\documentclass[preprint,12pt]{elsarticle}
\usepackage{amsmath,amssymb}   
\usepackage{latexsym}
\usepackage{graphicx} 
\usepackage{subfigure}   
\usepackage{color}
\usepackage{leftidx}
\usepackage{bm}
\journal{Physica D: Nonlinear Phenomena}

\begin{document}

\begin{frontmatter}



\title{A multi-domain model for microcirculation in optic nerve: blood flow and oxygen transport }
%A multi-domain model for blood microcirculation and oxygen transport in optic nerve
% 

\author[label1]{Zilong Song}
\address[label1]{Department of Mathematics and Statistics, Utah State University,3900 Old Main Hill,Logan,84322, UT,USA}

\author[label2]{Shixin Xu}
\address[label2]{Duke Kunshan University, 8 Duke Ave, Kunshan, Jiangsu, China}

\author[label3,label4]{Robert Eisenberg}
\address[label3]{Department of Applied Mathematics, Illinois Institute of Technology, Chicago, 60616, IL, USA}
\address[label4]{Department of Physiology and Biophysics, Rush University, Chicago, 60612, IL, USA}

\author[label5,label6,label7]{Huaxiong Huang}
\address[label5]{Research Centre for Mathematics, Advanced Institute of Natural Sciences, Beijing Normal University (Zhuhai), China}
\address[label6]{BNU-HKBU United International College, Zhuhai China}
\address[label7]{Department of Mathematics and Statistics, York University, Toronto, M3J 1P3, ON, Canada}

\begin{abstract}
Microcirculation of blood and oxygen transport play important roles in biological function of optic nerve and are directly affected by damages or pathologies. This work develops a multi-domain model for optic nerve, that includes important biological structures and various physical mechanisms in blood flow and oxygen delivery. The two sets of vasculature network are treated as five domains in the same geometric region, with various exchanges among them (such as Darcy's law for fluid flow) and with the tissue domain (such as water leak, diffusion). The numerical results of the coupled model for a uniform case of vasculature distribution show mechanisms and scales consistent with literature and intuition. The effects of various important model parameters (relevant to pathological conditions) are investigated to provide insights into the possible implications. The vasculature distribution (resting volume fractions here) has significant impacts on the blood circulation and could lead to insufficient blood supply in certain local region and in turn affect the oxygen delivery. The water leak across the capillary wall will have nontrivial effects after the leak coefficients pass a threshold.  The periodic arterial pressure conditions lead to expected periodic patterns and stable spatial profiles, and the uniform case is almost the averaged version of periodic case. The effects of viscosity, the stiffness of blood vessel wall, oxygen demand, etc. have also been analyzed. The framework can be extended to include ionic transport or to study the retina when more biological structural information is available. 
\end{abstract}



\end{frontmatter}


\section{Introduction}

Most of biology involves complex structures that are nearly machines designed to perform biological functions. Machines in our technology cannot be understood without their structure as well as the physics that they use and the engineering function they perform. 
A natural approach for these biological problems is to use physical laws in the form of conservation laws in three dimensions and time, together with structural information presented by biologists, and to compute physiological results required by biologists and clinicians. This approach has been used to study biological systems of some complexity, for example, the electrocytes of electrical eels \cite{cao2020,song2020}, the optic nerve of amphibians \cite{zhu2021}, the lens of the eye \cite{zhu2019}. The history of this approach is reviewed in \cite{bob2022}.

The eye is made of a series of tissue structures such as the lens, the retina, and the optical nerve that work together. The retina converts the light to electric signals, which travele to the brain through optic nerve. In each tissue structure, many physical mechanisms \cite{guidoboni2019,pittman2016,Hille2001} involving fluid flow and transport of ions and oxygen work together in well-designed biological structures to achieve specific goals and maintain homeostasis states. Many eye diseases such as glaucoma \cite{moyseyenko2023} and diabetic retinopathy \cite{ivanova2017,gerber2015} are due to the damage of optical nerve and retina related to the coupling of the above mechanisms. Cardiovascular pathologies such as hypertension and diabetes can induce changes in micro-circulations in the retina and optical nerve, and then are reflected by functional or structural changes. Non-invasive experimental/clinic data from the eye (say the retina) can provide an important ``window" into the cardiovascular pathologies \cite{lamb2007,guidoboni2020}. Therefore, mathematical modeling is needed to uncover the relationship between diseases/alterations in structures and the physical mechanisms. The challenges are the coupling of different mechanisms in multiple time and spatial scales  as well as  the incorporation of crucial biological structural information. On the other hand, too many scales and too much cellular detail in the model will make analysis/computation extremely tricky.

%between diseases/alterations in structures and the main contributing parameters/factors from the physical mechanism. 

There have been many modeling and experimental studies \cite{guidoboni2019, wang2009, prada_thesis,physiology2009,cessac2022} for both retina and optic nerve, particularly for retina since vascular and geometrical information is more accessible experimentally. The book \cite{guidoboni2019} includes an excellent review of different levels of mathematical models for blood flow in the eye as well as oxygen transport, see also \cite{secomb2004,popel1989,goldman2008,pittman2016}. The microcirculation and autoregulation have been studied in \cite{secomb2011, aletti2015, physiology2009,arciero2013,arciero2008}. A recent work \cite{causin2016} has studied the coupling between the blood microcirculation and oxygen transport  in retina, where a computer-generated vascular tree network is used for blood flow in vessels and delivery of oxygen to the tissue. A one-dimensional model of blood circulation in retina arterial network has been developed by utilizing clinic imaging \cite{julien2023}. A viscoelastic and porous-media model \cite{causin2014,prada_thesis} has been developed to study the mechanics and hemodynamics of optic nerve head. The blood flow and neurovascular coupling mechanisms in optic nerve have been reviewed in \cite{prada2016}.

Multi-domain modeling has shown success in the study of complex biological tissues, such as the lens \cite{zhu2019}, the optic nerve \cite{zhu2021,zhu2021b}, brain tissue \cite{mori2015}, and cardiac tissue \cite{henriquez1993,franzone2014}. At the tissue (coarse-grained) level, a given spatial point is present in every domain in the multidomain model, where exchanges occur between domains (to represent mechanisms at a refined cellular level). In particular,  a tridomain model \cite{zhu2021,zhu2021b} has been developed for fluid flow and electrodiffusion in optic nerve (without vasculature), and the clearance of potassium is studied. The multi-domain model for the vasculature here can be treated as generalization of compartment (vessel segment) models \cite{ye1993,guidoboni2019}, in the sense that more spatial and structural information could be incorporated.

The objective of this work is to develop multi-domain model for blood microcirculation and oxygen transport in optic nerve, with important biological structures and physical mechanisms incorporated. This work will focus on the vasculature and the exchanges between the vasculature and tissue in optic nerve, and can potentially be combined with the previous tri-domain model on fluid flow and electrodiffusion of ions in tissue \cite{zhu2021, zhu2021b} in future work.  This work studies the optic nerve because of its relative simple geometric structure, but the framework can be generalizaed to the retina. The multi-domain model includes five domains for the vasculature due to two complete sets of blood circulation pathways and one domain for the tissue. The part for blood circulation incorporates Darcy's law for in-domain flow, the leak to the tissue (the changes are related to diseased state), and volume fraction (equivalently blood vessel radius) changes due to force balance. The part for the oxygen transport considers both the dissolved and bound oxygen in hemoglobin and incorporates the mechanisms due to diffusion, convection and oxygen consumption.

The coupled model is solved by a finite-difference scheme and a Matlab solver (ode15s). The simulated results for the baseline case with a uniform resting volume fraction show reasonable mechanisms and consistent scales for important quantities with literature. More importantly, the effects of non-uniform resting volume fractions, periodic boundary condition, and various parameters are studied to provide insights into the consequences of biological structural changes and parameter changes due to diseases. For example, with non-uniform distribution of resting volume fractions, local regions could suffer insufficient capillary exchanges and oxygen supply. The leak coefficient could have have strong impact on blood circulation and oxygen delivery after it passes a threshold.  With periodic boundary arterial pressures, the results show almost the same averaged quantities as the baseline simulation with constant pressure conditions. The effects of viscosity, the stiffness of blood vessel wall, the demand for oxygen, the local partial blockage of vasculature etc. have also been analyzed.

The manuscript is organized as follows. Section 2 develops the mathematical model of blood circulation and oxygen transport with six domains. The model is simulated in Section 3 for a baseline case with uniform resting volume fractions and constant boundary pressures. The effects of various parameters and biological structural changes are analyzed in Section 4. Conclusions are provided in Section 5.

\section{Mathematical Model with Six Domains}

\begin{table}[h]
\begin{center}
\begin{tabular}{|c|c|c|c|}
%\hline
%Abbreviation &  Full name & Abbreviation &  Full name\\
\hline
 CRA & Central Retinal Artery & CRV & Central Retinal Vein\\
\hline
PCA & Posterior Ciliary Artery
& RBC & Red Blood Cell\\
\hline
MAP & Mean Arterial Pressure
& 1D & one-dimensional\\
\hline
\end{tabular}
\end{center}
\caption{Abbreviations in the paper}
\label{table1}
\end{table}

\begin{figure}
\begin{center}
\subfigure[Optic nerve]{\includegraphics[width=0.8\textwidth]{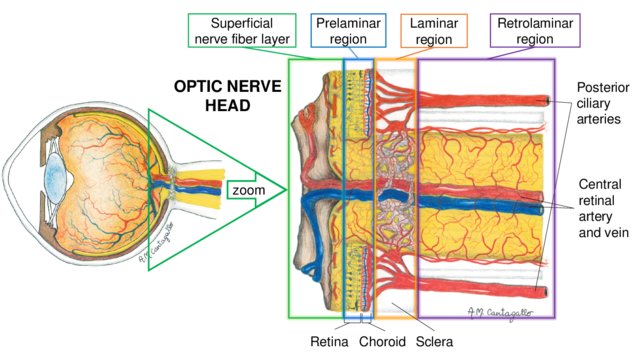}}
\subfigure[Six domains  and pathways]{\includegraphics[width=0.55\textwidth]{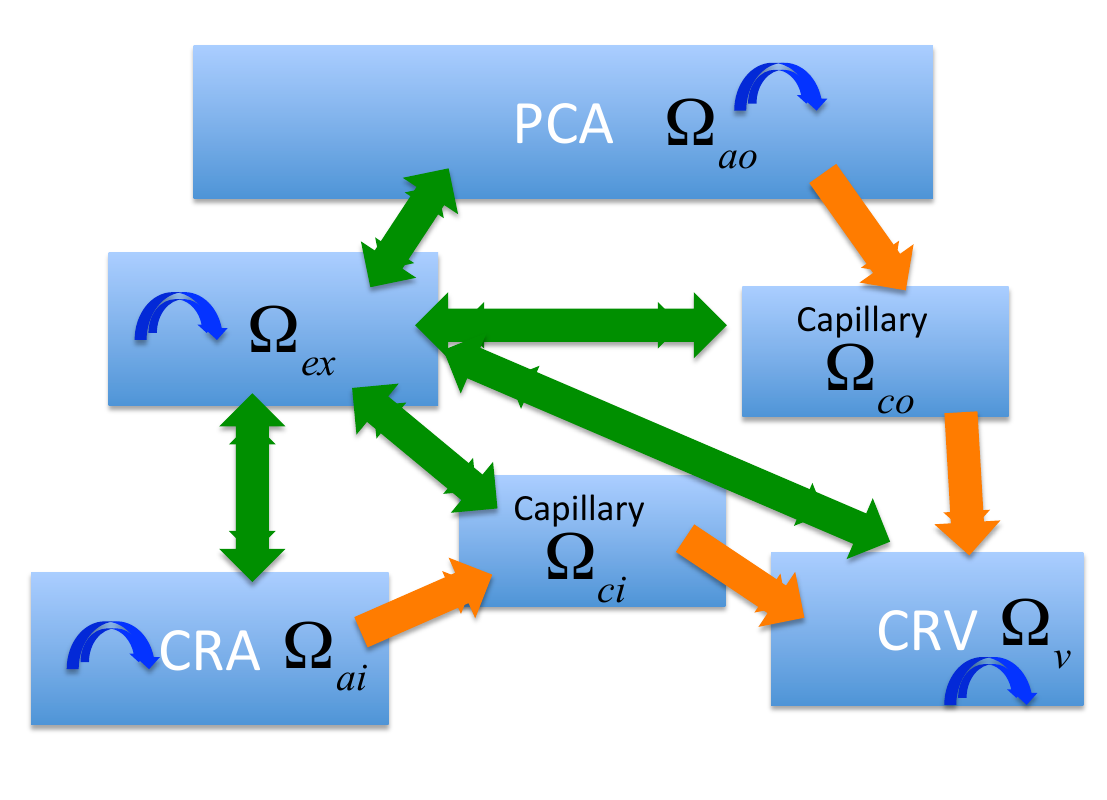}}
\caption{\label{fig1} (a) The optic nerve and retrolaminar region in an axial sectional view. The figure is taken from \cite{prada2016}, and we are grateful to the authors for making such a useful drawing.  (b) The six vascular and tissue domains. The green arrows denote the water exchanges between vascular domain and tissue (extra-vascular domain), the orange arrows denote the blood flow pathways, and the blue arrows denote the in-domain blood circulation.}
\end{center}
\end{figure}

Some abbreviations are listed in Table \ref{table1} for easy reading. Figure  \ref{fig1}(a) sketches the optic nerve, where the retrolaminar region is the major region with nearly radial symmetry. In this work, we focus on the retrolaminar region in the optic nerve \cite{prada_thesis,guidoboni2019,olver1994,onda1995} in a 1D setting. We consider six overlapping domains for the vasculature and tissue in the same geometric region $R_0<r<R_1$ and $0<z<L$ in cylindrical coordinates, but it is assumed to be uniform (or averaged) in the longitudinal $z$ direction so that it is a 1D problem in the radial direction and symmetric about $r=0$. The six domains are sketched in Figure \ref{fig1}(b) and are defined as:
\begin{description}
\item[$\Omega_{ai}$:] Arteries and arterioles starting from CRA at inner boundary $r=R_0$ 
\item[$\Omega_{ao}$:] Arteries and arterioles starting from PCA at outer boundary $r=R_1$
\item[$\Omega_{v}$:] Veins and venules connected to CRV at $r=R_0$
\item[$\Omega_{ci}$:] Capillaries   between $\Omega_{ai}$ and  $\Omega_{v}$
\item[$\Omega_{co}$:] Capillaries between $\Omega_{ao}$ and  $\Omega_{v}$
\item[$\Omega_{ex}$:] the extra-vascular domain, i.e, tissue domain including the axons and glial cells and extracellular space.
\end{description}

\subsection{Water/blood circulation}
\label{section2_1}

The water/blood circulation in optic nerve is based on porous-media type flows and exchanges among the domains. The circulation is both the blood flow among the vascular domains (the first 5 domains above) and the water flow (leak) for the exchange between vascular and extravascular domains. For the 1D geometric region $R_0<r<R_1$,  the governing equations from conservation laws are
\begin{equation}
\label{eq1}
\begin{aligned}
& \frac{\partial \eta_{ai}}{\partial t}  + \frac{1}{r} \frac{\partial \left( r {\eta}_{ai} u_{ai}^r\right)}{\partial r}  + Q_{ai,ex} +  Q_{ai,ci}=0, \\
& \frac{\partial \eta_{ao}}{\partial t}  + \frac{1}{r} \frac{\partial \left( r {\eta}_{ao} u_{ao}^r\right)}{\partial r}  + Q_{ao,ex} +  Q_{ao,co}=0, \\
& \frac{\partial \eta_{v}}{\partial t}  + \frac{1}{r} \frac{\partial \left( r {\eta}_v u_v^r\right)}{\partial r} + Q_{v,ex} + Q_{v,co} + Q_{v,ci} =0,\\
& \frac{\partial \eta_{ci}}{\partial t}  + Q_{ci,ex} - Q_{ai,ci} - Q_{v,ci} =0,\\
& \frac{\partial \eta_{co}}{\partial t}  + Q_{co,ex} - Q_{ao,co}- Q_{v,co}=0,\\
& \frac{1}{r} \frac{\partial }{\partial r} \left( \sum_{k=ai,ao,v,ex}{\eta}_k r u_k^r\right)=0,\\
& \eta_{ai} + \eta_{ao}  + \eta_{v} + \eta_{ex} + \eta_{ci} + \eta_{co}=1,
\end{aligned}
\end{equation}
where $\eta_j$ is the volume fraction of each domain $\Omega_j$ ($j=ai,ao,v,ci,co,ex$), the spatial derivative terms $\frac{1}{r} \frac{\partial}{\partial r}\left(r (\eta_k u_k^r)\right)$ are the in-domain flows in the polar coordinates with velocity $u_k^r$ ($k=ai,ao,v,ex$), all the $Q_{i,j}$ are the exchanges between different domains. There is in-domain flow in artery and vein domains, since they are connected with branching structures. There is also in-domain flow in extravascular domain since it is spatially connected.  But there are no in-domain flows for capillary domains at this tissue-scale modeling, because they are only connected to arteries and veins at a smaller spatial scale and different capillary networks do not exchange directly with each other.  The second last equation is derived by combining the dynamics of $\eta_{ex}$ and the algebraic constraint in the last equation. We will give the detailed models for all the terms in above equation and boundary conditions in the next subsections.

\subsubsection{In-domain flows and permeability}

The in-domain velocity in vascular domains follows the Darcy's law
\begin{equation}
\label{eq2}
\begin{aligned}
{u}_k^r = -\frac{\kappa_k (r) \tau_k }{\mu_b} \frac{\partial p_k}{\partial r},  \quad k = ai,ao,v,
\end{aligned}
\end{equation}
where $\kappa_k$ is the in-domain water permeability, $\mu_b$ is the viscosity of blood, and $\tau_k$ is the tortuosity.  The tortuosity $\tau_k$ includes the effect of vessel structure and orientation of blood vessels (in $r$-direction). From derivations in appendix and \cite{prada_thesis}, the permeability is given by
\begin{equation}
\label{eq3}
\begin{aligned}
\kappa_k (r) =\frac{1}{8}\beta_k(r)  {\eta}_k, \quad k = ai,ao,v,
\end{aligned}
\end{equation}
where $\beta_k(r)$ depends on structure of blood vessels, e.g., the distribution of branches, the level of branching, segment length etc.  But for simplicity, the coefficient $\beta_k(r)$ is assumed to be a constant here and its value is estimated in the appendix. The permeability $\kappa_k$ may depend on the concentration of oxygen through the changes in $\eta_k$ due to  changes in the vessel properties. For the extravascular domain $\Omega_{ex}$, we set
\begin{equation}
\label{eq4}
\begin{aligned}
{u}_{ex}^r = -\frac{\kappa_{ex} \tau_{ex} }{\mu_{ex}} \frac{\partial p_{ex}}{\partial r},
\end{aligned}
\end{equation}
where $k_{ex},\tau_{ex},\mu_{ex}$ are the permeability, tortuosity and viscosity in  $\Omega_{ex}$.

\subsubsection{Exchange through blood vessel wall}

Here, we consider the water exchange rates $Q_{j,ex}$ between the vascular domains $\Omega_j$ ($j= ai,ao, v, ci,co$) and the extravascular domain $\Omega_{ex}$, due the leak/exchange through blood vessel wall. The water exchange rate $Q_{j,ex}$ per unit control volume follows the form
\begin{equation}
\label{eq5}
\begin{aligned}
Q_{j,ex} =M_{j,ex} U_{j,ex}, \quad j= ai,ao, v, ci,co,
\end{aligned}
\end{equation}
where $M_{j,ex}$ is the area of blood vessel wall for domain $\Omega_j$ per unit control volume and $U_{j,ex}$ is the water velocity from $\Omega_j$ to $\Omega_{ex}$.  The quantity $M_{j,ex}$ is modeled by (see appendix for details)
\begin{equation}
\label{eq6}
\begin{aligned}
M_{j,ex} =M_{j,ex}^0 \sqrt{\eta_j},
\end{aligned}
\end{equation}
where $M_{j,ex}^0$ is a constant depending on blood vessel structures and estimated in appendix (related to $\beta_{j}$).  The dependence of $M_{j,ex}$ on the radius of blood vessel is reflected through $\eta_j$. The water velocity across the blood vessel wall is modeled as
\begin{equation}
\label{eq7}
\begin{aligned}
U_{j,ex} = L_{j,ex} (p_j - p_{ex}-  (\pi_j-\pi_{ex})), \quad j= ai,ao, v, ci,co
\end{aligned}
\end{equation}
where $L_{j,ex}$ is the leak coefficient  \cite{harrer2004,physiology2009} (for capillary, see also \cite{bentzer2001} \cite{taylor1981capillary} \cite{antcliff2001}), and $\pi_j,\pi_{ex} $ are the colloidal osmotic pressures \cite{physiology2009} (e.g., due to ions, proteins etc.). The leak coefficient in artery and vein domains is much smaller than that in capillary domains  under normal conditions.

\subsubsection{Exchange between vascular domains}

The blood flow rates from artery to capillary or from capillary to vein are modeled by 
\begin{equation}
\label{eq8}
\begin{aligned}
Q_{ai,ci} = K_{ai,ci} (p_{ai} - p_{ci}), \quad Q_{v,ci} = K_{v,ci} (p_{v} - p_{ci}),\\
Q_{ao,co} = K_{ao,co} (p_{ao} - p_{co}), \quad Q_{v,co} = K_{v,co} (p_{v} - p_{co}),
\end{aligned}
\end{equation}
where $K_{j,k}$ is the effective conductance (or the inverse of resistance). We take $K_{ai,ci}$ from artery domain $\Omega_{ai}$ to capillary domain $\Omega_{ci}$ for example. It relates to both artery and capillary domains since it flows in the small arterioles for some distance before reaching the capillary. The conductance is modeled as the harmonic average
\begin{equation}
\label{eq9}
\begin{aligned}
& K_{ai,ci} = \frac{K_{ai} K_{ci}}{K_{ai} + K_{ci}}, \quad K_{ai}  =\delta_{ai} \eta_{ai}^2 \frac{\epsilon_{ai}}{\tanh(\epsilon_{ai})}, \quad K_{ci} = \delta_{ci} \eta_{ci}^2\frac{\epsilon_{ci}}{\tanh(\epsilon_{ci})},
\end{aligned}
\end{equation}
where $K_{j}$ ($j=ai,ci$) are the effective conductance in the two domains with parameter $\delta_{j}$ estimated in the appendix (related to $\beta_j$ and $\mu_b$). The dimensionless ratio ${\epsilon_{j}}/{\tanh(\epsilon_{j})}$ (with $j=ai,ci$) is a  correction factor due to the water leak \cite{physiology2009}
\begin{equation}
\label{eq10}
\begin{aligned}
\epsilon_{j} = \frac{1}{2}\sqrt{(M_{j,ex}L_{j,ex})/(\delta_{j} \eta_{j}^2)}, \quad j=ai,ci.
\end{aligned}
\end{equation} 
In the limit when there is no water leak through blood vessel wall  $L_{j,ex} \rightarrow 0$, we get $\epsilon_{ai} \rightarrow 0$ and  ${\epsilon_{j}}/{\tanh(\epsilon_{j})} \rightarrow 1$, so there is no effect from correction factor. The presence of leak increases the conductance. In a similar way, we can define $K_{ao, co}$ $K_{v,ci}$, $K_{v,co}$, as well as the parameters $\delta_{j}, \epsilon_{j}$ ($j=ao,co,v$). 

%Remark: maybe we can remove the $\epsilon_{ai},\epsilon_{v}$ part, because the leak is considered in $Q_{ai,ex}$ and $Q_{v,ex}$ (that inner loop). But the inner circulation for $ci$ is not in the model, could be treated in the  $\epsilon_{ai}$, i.e., the permeability is increased.

\subsubsection{Force balance on blood vessel wall}

The force balance on the blood vessel wall of each vascular domain is modeled by \cite{physiology2009,zhu2021}
\begin{equation}
\label{eq11}
\begin{aligned}
\lambda_j (\eta_j - \eta_j^{re}) =p_j  - p_{ex} - (P_j^{re}-P_{ex}^{re}),\quad j = ai,ao, v, ci, co,
\end{aligned}
\end{equation}
where $\lambda_j$ is the elastic modulus of the blood vessel wall \cite{prada_thesis,camasao2021}, $P_j^{re}, P_{ex}^{re}$ are resting or reference pressures, and $\eta_j^{re}$ are the resting volume fractions. The modulus $\lambda_j$ is set as constant here, but it could be affected by the oxygen concentration and in turn leads to changes in volume fractions (or the radius of blood vessels). The quantity $\eta_j^{re}(r)$ could depend on the spatial variable $r$ and the profile is related to the structural information. We investigate both uniform and Gaussian profiles for $\eta_j^{re}(r)$ in later examples. An alternative model of force balance \cite{julien2023, bartolo2022} is to use $\sqrt{\eta_j}$ and $\sqrt{ \eta_j^{re}}$ in the formula (\ref{eq11}). If $\eta_j$ is not far from  $\eta_j^{re}$, the two models will be similar up to first-order approximations with slightly different definitions for the modulus.

%Effect of oxygen: the oxygen will affect the dilation of blood vessel and hence the effective resistance/permeability. Two ways of incorporating the effect of oxygen: (1) put it in $\beta(C_k)$, since $\beta$ has some information of initial diameter of vessel? but diameter info is already in $\eta$? (2) put it in the modulus $K_j(C_k)$ (or resting state $\eta_j^{re}$) of force balance, so local volume $\eta$ or local diameter changes with effect of oxygen, then in turn affect $\kappa$ since $\kappa$ depends on $\eta$. Depend on only $C_{ex}$ in the tissue? Oxygen could  affect the smooth muscle tone, and then the wall tension. 

\subsubsection{Boundary conditions}
\label{section2_1_5}

As there are in-domain flows (i.e., spatial derivatives) for the four domains $\Omega_{j}$, $j=ai,ao,v,ex$, we need to propose boundary conditions at $r=R_0,R_1$ for these domains.\\
(1) For the artery domain $\Omega_{ai}$, we set
\begin{equation}
\label{eq12}
\begin{aligned}
p_{ai} (R_0,t) = P_{ai,0}(t),\quad \frac{\partial p_{ai}}{\partial r} (R_1,t) = 0,
\end{aligned}
\end{equation}
where $P_{ai,0}$ is the given CRA pressure at the start of that artery domain $\Omega_{ai}$. The condition at the outer boundary $r=R_1$ means no blood flow out of the region there. Alternatively, if $P_{ai,0}(t)$ is not given, we can use the prescribed the blood flow rates at boundary
\begin{equation}
\label{eq13}
\begin{aligned}
\eta_{ai} R_0 u_{ai}^r= Q_{ai}^\ast.
\end{aligned}
\end{equation}
(2) For the other artery domain $\Omega_{ao}$, we set
\begin{equation}
\label{eq14}
\begin{aligned}
\frac{\partial p_{ao}}{\partial r} (R_0,t) = 0, \quad p_{ao} (R_1,t) = P_{ao,1}(t),
\end{aligned}
\end{equation}
where $P_{ao,1}$ is the given PCA pressure at the start of artery domain $\Omega_{ao}$ and there is no blood flow on the other end. Alternatively,  if $P_{ao,1}(t)$ is not given, we can use the flow rates condition
\begin{equation}
\label{eq15}
\begin{aligned}
\eta_{ao} R_1 u_{ao}^r= -Q_{ao}^\ast.
\end{aligned}
\end{equation}
(3) For the vein domain $\Omega_{v}$, we set
\begin{equation}
\label{eq16}
\begin{aligned}
 p_{v} (R_0,t) = P_{v,0}(t),\quad \frac{\partial p_{v}}{\partial r} (R_1,t) = 0.
\end{aligned}
\end{equation}
where $P_{v,0}(t)$ is the given CRV pressure at inner boundary where the blood drains out of this optic nerve region.\\
(4) For extravascular domain $\Omega_{ex}$, we set
\begin{equation}
\label{eq17}
\begin{aligned}
\frac{\partial p_{ex}}{\partial r} (R_0,t) = 0,\quad \frac{\partial p_{ex}}{\partial r} (R_1,t) = 0,
\end{aligned}
\end{equation}
which means no water flow out of the region through domain $\Omega_{ex}$.

\subsection{Oxygen transport}  

We start by introducing some background knowledge and use of notations here. The total oxygen concentration  $\bar{C}_{O_2}$ (per unit blood volume) in vascular domains consists of two parts: the dissolved oxygen and the oxygen bound to hemoglobin in red blood cells (RBC) \cite{guidoboni2019,popel1989,physiology2008}. Mathematically, we have 
\begin{equation}
\label{eq18}
\begin{aligned}
& \bar{C}_{O_2}= C_{O_2} + H \, S_{O_2}
\end{aligned}
\end{equation}
where $C_{O_2}$ the dissolved oxygen, $H$ is the oxygen-binding capacity  of blood (per unit blood volume),  and $S_{O_2}$ is the oxygen saturation of Hemoglobin. In many works \cite{pittman2016,causin2016}, the dissolved oxygen concentration is often represented by oxygen partial pressure $P_{O_2}$ with the formula
\begin{equation}
\label{eq19}
\begin{aligned}
C_{O_2}= \alpha_{O_2} P_{O_2}
\end{aligned}
\end{equation}
where $\alpha_{O_2}$ is the solubility coefficient of oxygen in blood. But in order to distinguish the pressure $p_k$ for water/blood flows and the oxygen partial pressure here, we will directly use the oxygen concentration $C_{O_2}$ in our model. The oxygen saturation $S_{O_2}$ is given by Hill's equation \cite{causin2016,popel1989}
\begin{equation}
\label{eq20}
\begin{aligned}
S_{O_2} = \frac{P_{O_2}^{n}}{P_{O_2}^{n} + P_{50}^{n}}= \frac{C_{O_2}^{n}}{C_{O_2}^{n} + C_{50}^{n}} 
\end{aligned}
\end{equation}
where we have multiplied $\alpha_{O_2}$ before $P_{O_2}$ and $P_{50}$ to get the last equality,  $C_{50}= \alpha_{O_2} P_{50}$ is the half-saturation constant, and $n$ is the Hill's  exponent parameter. To estimate $H$, it is often written as \cite{pittman2016}
\begin{equation}
\label{eq21}
\begin{aligned}
H =   [Hb] C_{Hb},
\end{aligned}
\end{equation}
where $[Hb]$ is the hemoglobin concentration per unit volume of blood, $C_{Hb}$ is the oxygen-binding capacity of hemoglobin.  Under normal conditions, the quantities $H, [Hb], C_{Hb}$ can be assumed as constants, see table in appendix. But here we will also consider abnormal cases with large water leak through blood vessels, so the hemoglobin concentration and hence the quantity $H$ can vary spatially, since hemoglobin can not leak out with water.

In each of five vascular domains, we have the above quantities and  relations. To simplify the notations, we will omit the subscript $O_2$ but add the domain subscript. In summary, we have
\begin{equation}
\label{eq22}
\begin{aligned}
& \bar{C}_{k} (C_{k}, H_k)=C_{k} +  H_k  \frac{C_{k}^{n}}{C_{k}^{n}+ C_{50}^{n}}, \quad k = ai, ao, v, ci, co,
\end{aligned}
\end{equation}
where $\bar{C}_{k}$, $C_{k}$ and $H_k$ are the total oxygen concentration, dissolved oxygen concentration, and oxygen-binding capacity of blood in the domain $\Omega_k$.  For the extravascular domain $\Omega_{ex}$, only the dissolved oxygen concentration $C_{ex}(r,t)$ is needed and well-defined.

By conservation laws, the dynamics of $H_k$ ($k = ai, ao, v, ci, co$) is given by 
\begin{equation}
\label{eq23}
\begin{aligned}
& \frac{\partial (\eta_{ai}  H_{ai})}{\partial t}  - \frac{1}{r} \frac{\partial }{\partial r} \left(r {\eta}_{ai} u_{ai}^r  H_{ai} \right) + Q_{ai,ci} H_{ai}=0, \\
& \frac{\partial (\eta_{ao}  H_{ao}) }{\partial t}  - \frac{1}{r} \frac{\partial }{\partial r} \left( r {\eta}_{ao} u_{ao}^r H_{ao} \right) +   Q_{ao,co} H_{ao}=0, \\
& \frac{\partial (\eta_{v}   H_{v} )}{\partial t}  - \frac{1}{r} \frac{\partial }{\partial r} \left(r {\eta}_{v} u_{v}^r H_{v} \right) +  Q_{v,ci} H_{ci} +  Q_{v,co} H_{co}=0, \\
& \frac{\partial (\eta_{ci}  H_{ci}) }{\partial t}  -Q_{ai,ci} H_{ai}-  Q_{v,ci} H_{ci}=0,\\
& \frac{\partial (\eta_{co}  H_{co} )}{\partial t}  -Q_{ao,co} H_{ao} - Q_{v,co} H_{co}=0,
\end{aligned}
\end{equation}
Under normal conditions when all the water leak terms $Q_{j,ex}$ ($j=ai,ao,v,ci,co$) through blood vessel walls are negligible small compared with other term $Q_{i,j}$ in blood flow equations in (\ref{eq1}), this system is equivalent to the first five equations in (\ref{eq1}) with constant solutions $H_k=H_0$ ($k = ai, ao, v, ci, co$) (verified in simulations). Next, the equations for oxygen exchange are given by
\begin{equation}
\label{eq24}
\begin{aligned}
& \frac{\partial}{\partial t}(\eta_{ai} \bar{C}_{ai}) +\frac{1}{r} \frac{\partial \left( {\eta}_{ai} r J_{ai}^r\right)}{\partial r} + S_{ai,ex} +  S_{ai,ci}=0,\\
& \frac{\partial}{\partial t}(\eta_{ao} \bar{C}_{ao}) +\frac{1}{r} \frac{\partial \left({\eta}_{ao} r J_{ao}^r\right)}{\partial r} + S_{ao,ex} +  S_{ao,co} =0,\\
& \frac{\partial}{\partial t}(\eta_{v} \bar{C}_{v}) +\frac{1}{r} \frac{\partial \left({\eta}_v r J_v^r\right)}{\partial r} + S_{v,ex} + S_{v,ci} + S_{v,co} =0, \\
& \frac{\partial}{\partial t}(\eta_{ci} \bar{C}_{ci}) + S_{ci,ex} -S_{ai,ci} - S_{v,ci} =0,\\
& \frac{\partial}{\partial t}(\eta_{co} \bar{C}_{co}) + S_{co,ex} -S_{ao,co} - S_{v,co} =0,\\
& \frac{\partial}{\partial t}(\eta_{ex} C_{ex}) +\frac{1}{r} \frac{\partial \left( \eta_{ex} r J_{ex}^r\right)}{\partial r} + S_{ex}  - \sum_{j=ai,ao,v,ci,co }S_{j,ex} =0,\\
\end{aligned}
\end{equation}
where $J_{j}^r$ ($j=ai,ao,v,ex$) are the in-domain oxygen fluxes in the $r$-direction, $S_{i,j}$ are the oxygen exchange rates between different domains, and $S_{ex}$  is the consumption rate of oxygen (e.g., for pumps of ions on axon \cite{Hille2001}) given by the classical Michaelis-Menten kinetics \cite{causin2016,arciero2008,secomb2004} 
\begin{equation}
\label{eq25}
\begin{aligned}
S_{ex}(C_{ex}) =S_{ex}^{max} \frac{C_{ex}}{C_{ex} + C_{1/2}},
\end{aligned}
\end{equation}
where $S_{ex}^{max}$ and $C_{1/2}$ are two parameters for the maximum consumption rate and concentration at half-max consumption. Detailed models for in-domain and inter-domain fluxes $J_{j}^r$ and $S_{i,j}$, as well as the boundary conditions, will be given below.

\subsubsection{In-domain oxygen flux}

The in-domain oxygen fluxes in the four domains $\Omega_j$ ($j=ai,ao,v,ex$) consist of convection and diffusion terms
\begin{equation}
\label{eq26}
\begin{aligned}
&J_k^r = \bar{C}_k u_k^r - D_k \tau_k \frac{\partial C_k}{\partial r}, \quad k = ai,ao,v,\\
& J_{ex}^r = {C}_{ex} u_{ex}^r - D_{ex} \tau_{ex} \frac{\partial C_{ex}}{\partial r},
\end{aligned}
\end{equation}
where $D_j,\tau_j,u_j^r$ ($j=ai,ao,v,ex$) are the diffusion constant of oxygen, the tortuosity, and the blood/water velocity defined in (\ref{eq2},\ref{eq4}).

\subsubsection{Oxygen exchange through blood vessel wall}

The oxygen exchange rate per unit volume from $\Omega_j$ to $\Omega_{ex}$ follows 
\begin{equation}
\label{eq27}
\begin{aligned}
S_{j,ex} =M_{j,ex} J_{j,ex}, \quad j= ai,ao, v, ci,co,
\end{aligned}
\end{equation}
where $M_{j,ex}$ is defined in (\ref{eq6}).  The oxygen flux $J_{j,ex}$ per unit area is modeled by
\begin{equation}
\label{eq28}
\begin{aligned}
J_{j,ex} =l_{j,ex} (C_j - C_{ex}) + U_{j,ex} C_{j,ex}^{upwind}, \quad j =  ai,ao,v, ci, co,
\end{aligned}
\end{equation}
where $U_{j,ex}$ is defined in (\ref{eq7}), $l_{j,ex}$ is the oxygen permeability through the blood vessel wall \cite{causin2016,goldman2008}, and $C_{j,ex}^{upwind}$ is upwind concentration determined by the sign of $U_{j,ex}$.  Combining (\ref{eq27},\ref{eq28}) with the use of (\ref{eq5}), we obtain
\begin{equation}
\label{eq29}
\begin{aligned}
S_{j,ex} =M_{j,ex} l_{j,ex} (C_j - C_{ex}) +Q_{j,ex} C_{j,ex}^{upwind} , \quad j= ai,ao, v, ci,co,
\end{aligned}
\end{equation}
and $C_{j,ex}^{upwind}$ is given by 
\begin{equation}
\label{eq30}
\begin{aligned}
C_{j,ex}^{upwind} = \begin{cases} C_j & \mathrm{if} ~Q_{j,ex}>0,\\
C_{ex} & \mathrm{if} ~Q_{j,ex}<0.\\
\end{cases}
\end{aligned}
\end{equation}

\subsubsection{Oxygen exchange between vascular domains}

The oxygen exchange rate between capillaries and other vascular domains are
\begin{equation}
\label{eq31}
\begin{aligned}
& S_{ai,ci} =Q_{ai,ci} \bar{C}_{ai} + D_{ai,ci} (C_{ai}- C_{ci}),\quad S_{ao,co} = Q_{ao,co} \bar{C}_{ao} + D_{ao,co} (C_{ao}- C_{co}),\\
& S_{v,ci} = Q_{v,ci} \bar{C}_{ci,v}^{upwind} + D_{v,ci} (C_{v}- C_{ci}),\quad S_{v,co} = Q_{v,co} \bar{C}_{co,v}^{upwind} + D_{v,co} (C_{v}- C_{co}),
\end{aligned}
\end{equation}
where $D_{i,j}$ are effective diffusion constant between domains (estimated in appendix), the $Q_{i,j}$ terms are defined in (\ref{eq8}), and $\bar{C}_{k,v}^{upwind}$ ($k=ci,co$) is the total oxygen concentration before entering to $\Omega_v$. We take $\bar{C}_{ci,v}^{upwind}$ for example. 
One simple option is $\bar{C}_{ci,v}^{upwind}  = \bar{C}_{ci}$, but this does not consider the possible gradient inside the capillary network point.  If there is more than sufficient oxygen supply for exchange to extravascular domain, we assume a linear decreasing profile for dissolved oxygen concentration ${C}_{ci}$ within a point (network) in capillary domain, which implies
\begin{equation}
\label{eq32}
\begin{aligned}
{C}_{ci,v}^{upwind} = 2 {C}_{ci} - {C}_{ai}, \quad \mathrm{if} \quad {C}_{ci,v}^{upwind}>C_{ex}.
\end{aligned}
\end{equation}
When  the above ${C}_{ci,v}^{upwind}\le C_{ex}$, the capillary domain could not supply oxygen through the permeability term in (\ref{eq28}) so the concentration will not decrease further. In summary, we adopt the model
 \begin{equation}
 \label{eq33}
\begin{aligned}
& {C}_{ci,v}^{upwind} = \mathrm{max}\{ 2 {C}_{ci} - {C}_{ai}, C_{ex}\},
\end{aligned}
\end{equation}
and define
\begin{equation}
\label{eq34}
\begin{aligned}
\bar{C}_{ci,v}^{upwind} = \bar{C} ({C}_{ci,v}^{upwind}, H_{ci}),
\end{aligned}
\end{equation}
by using (\ref{eq22}). Similarly, $\bar{C}_{co,v}^{upwind}$ can be defined by replacing subscripts $ci,ai$ by $co,ao$.

\subsubsection{Boundary conditions}

Now we propose boundary conditions for both $H_k$ ($k=ai,ao,v$) and $C_j$ ($j=ai,ao,v,ex$) at $r=R_0,R_1$.  For the $H_k$, we have
\begin{equation}
\label{eq35}
\begin{aligned}
&H_{ai} (R_0,t) = H_{0}, \quad \frac{\partial H_{ai}}{\partial r} (R_1,t) = 0,\\
& \frac{\partial H_{ao}}{\partial r} (R_0,t) = 0,\quad H_{ao} (R_1,t) = H_0,\\
& \frac{\partial H_{v}}{\partial r} (R_0,t) = 0, \quad \frac{\partial H_{v}}{\partial r} (R_1,t) = 0,
\end{aligned}
\end{equation}
where $H_0$ is averaged oxygen-binding capacity of blood under normal conditions, and all the Neumann conditions imply that hemoglobins (or RBC) can not flow out these boundaries consistent with equations (\ref{eq12},\ref{eq14},\ref{eq16}). For the concentrations, we have
\begin{equation}
\label{eq36}
\begin{aligned}
&C_{ai} (R_0,t) = C_{ai,0}, \quad \frac{\partial C_{ai}}{\partial r} (R_1,t) = 0,\\
& \frac{\partial C_{ao}}{\partial r} (R_0,t) = 0,\quad C_{ao} (R_1,t) = C_{ao,1},\\
& \frac{\partial C_{v}}{\partial r} (R_0,t) = 0, \quad \frac{\partial C_{v}}{\partial r} (R_1,t) = 0,\\
&\frac{\partial C_{ex}}{\partial r} (R_0,t) = 0,\quad \frac{\partial C_{ex}}{\partial r} (R_1,t) = 0,
\end{aligned}
\end{equation}
where $C_{ai,0}$ and $C_{ao,1}$ are given dissolved oxygen concentrations at arteries CRA and PCA. Neumann conditions are adopted for other conditions, which will not influence much on the numerical results, since the in-domain diffusion terms are very small compared with the convection terms by fluid flow.

\subsection{Nondimensionalization}

\subsubsection{The water/blood circulation}

We adopt the scalings
\begin{equation}
\label{eq37}
\begin{aligned}
& \tilde{t} = \frac{t}{t_0}, \quad \tilde{r} = \frac{r}{R_1}, \quad \tilde{p}_k = \frac{p_k}{P_{v,0}},\quad \tilde{\pi}_k = \frac{\pi_k}{P_{v,0}},\\
&  \tilde{P}_k^{re} = \frac{P_k^{re}}{P_{v,0}}, \quad \tilde{Q}_{j,k} = {Q}_{j,k} t_0, \quad \tilde{\lambda}_j = \frac{\lambda_j}{P_{v,0}}.
\end{aligned}
\end{equation}
The time scale $t_0$ is chosen to be 
\begin{equation}
\label{eq38}
\begin{aligned}
t_0 = 1\mathrm{mm/(1 cm/s)} = 0.1 \mathrm{s}
\end{aligned}
\end{equation}
which is consistent with  some typical blood flow velocity \cite{physiology2009} and will give $O(1)$ dimensionless velocity for the in-domain blood flows. Substituting the above scalings in the system in Section \ref{section2_1} and after removing the tilde for the unknown variables $\tilde{p}_k, \tilde{Q}$ and independent variables $\tilde{r}, \tilde{t}$ for mathematical simplicity, we have the dimensionless system
\begin{equation}
\label{eq39}
\begin{aligned}
& \frac{\partial \eta_{ai}}{\partial t}  - \frac{1}{r} \frac{\partial }{\partial r} \left(\bar{\kappa}_{ai}  r \frac{\partial p_{ai}}{\partial r} \right) + Q_{ai,ex}+ Q_{ai,ci}=0, \\
& \frac{\partial \eta_{ao}}{\partial t}  - \frac{1}{r} \frac{\partial }{\partial r} \left(\bar{\kappa}_{ao}  r \frac{\partial p_{ao}}{\partial r} \right) +  Q_{ao,ex}+  Q_{ao,co}=0, \\
& \frac{\partial \eta_{v}}{\partial t}  - \frac{1}{r} \frac{\partial }{\partial r} \left(\bar{\kappa}_{v}  r \frac{\partial p_{v}}{\partial r} \right) + Q_{v,ex} +  Q_{v,ci}+  Q_{v,co}=0, \\
& \frac{\partial \eta_{ci}}{\partial t}  +  Q_{ci,ex} -Q_{ai,ci}-  Q_{v,ci}=0,\\
& \frac{\partial \eta_{co}}{\partial t}  +  Q_{co,ex} -Q_{ao,co}- Q_{v,co}=0,\\
& \frac{\partial }{\partial r} \left( \sum_{k=v,ai,ao,ex} \bar{\kappa}_{k}  r \frac{\partial p_{k}}{\partial r} \right) =0, \\
\end{aligned}
\end{equation}
where the water/blood flow rates are 
\begin{equation}
\label{eq40}
\begin{aligned}
& Q_{j,ex}=\bar{L}_{j,ex} \left[p_{j} - p_{ex} - (\tilde{\pi}_{j} - \tilde{\pi}_{ex})\right],\quad j  = ai, ao, v, ci, co\\
& Q_{k,ci}= \bar{K}_{k,ci}(p_{k} - p_{ci}), \quad k= ai,v,\\
& Q_{k,co}= \bar{K}_{k,co}(p_{k} - p_{co}), \quad k= ao,v,\\
\end{aligned}
\end{equation}
and the parameters are
\begin{equation}
\label{eq41}
\begin{aligned}
& \bar{L}_{j,ex} = \tilde{L}_{j,ex} \sqrt{\eta_{j}}, \quad \tilde{L}_{j,ex} = M_{j,ex}^0 L_{j,ex} P_{v,0} t_0, \quad j  = ai, ao, v, ci, co,\\
& \bar{\kappa}_{j} =\tilde{\beta}_{j} \eta_{j}^2, \quad  \tilde{\beta}_{j} = \frac{ \beta_{j} \tau_{j} P_{v,0} t_0}{8 R_1^2 \mu_b}, \quad j= ai, ao, v,\\
& \bar{\kappa}_{ex} =\tilde{\kappa}_{ex} \eta_{ex},\quad \tilde{\kappa}_{ex}= \frac{P_{v,0} \kappa_{ex} \tau_{ex} t_0}{R_1^2 \mu_{ex}},\\
& \bar{K}_{j,ci} =\frac{\bar{K}_{j} \bar{K}_{ci}}{\bar{K}_{j}+\bar{K}_{ci}},\quad j= ai, v, \quad  \bar{K}_{j,co} =\frac{\bar{K}_{j} \bar{K}_{co}}{\bar{K}_{j}+\bar{K}_{co}},\quad j= ao, v\\
& \bar{K}_{j}  =\tilde{\delta}_{j} \eta_{j}^2 \frac{\bar{\epsilon}_{j}}{\tanh(\bar{\epsilon}_{j})},\quad \bar{\epsilon}_{j} = \frac{1}{2} \eta_{j}^{-3/4} \sqrt{\tilde{L}_{j,ex}/\tilde{\delta}_{j} } , \quad  \tilde{\delta}_{j} = \delta_{j} P_{v,0} t_0, \quad j  = ai, ao, v, ci, co.
\end{aligned}
\end{equation}
In the above notations (and in the next subsection), the quantities with a bar (e.g. $\bar{\kappa}_{j}$) are effective coefficients and depend on the unknown variables (e.g., $\eta_j$), whereas quantities with a tilde (e.g., $\tilde{\beta}_j$) are dimensionless parameters that does not depend the unknowns. The algebraic constraint for volume fractions is 
\begin{equation}
\label{eq42}
\begin{aligned}
\eta_{ai} + \eta_{ao}  + \eta_{v} + \eta_{ex} + \eta_{ci} + \eta_{co}=1,
\end{aligned}
\end{equation}
and the force balance constraint is given by 
\begin{equation}
\label{eq43}
\begin{aligned}
\tilde{\lambda}_j (\eta_j - \eta_j^{re}) =p_j  - p_{ex} - (\tilde{P}_j^{re}-\tilde{P}_{ex}^{re}),\quad j = ai,ao, v, ci, co.
\end{aligned}
\end{equation}
The dimensionless form for the boundary conditions in Section  \ref{section2_1_5} will have the same form except that the given pressures are nondimentionalized. The dimensionless flow rates conditions in (\ref{eq13},\ref{eq15}) have the form
\begin{equation}
\label{eq43_1}
\begin{aligned}
-\left.\bar{\kappa}_{ao} \frac{\partial p_{ao}}{\partial r}\right|_{r=R_1=1} = -Q_{ao}^\ast,\quad -\left.\bar{\kappa}_{ai} r \frac{\partial p_{ai}}{\partial r}\right|_{r=R_0} = Q_{ai}^\ast.
\end{aligned}
\end{equation}

\subsubsection{The oxygen transport}

We adopt the scalings
\begin{equation}
\label{eq44}
\begin{aligned}
& \tilde{t} = \frac{t}{t_0}, \quad \tilde{r} = \frac{r}{R_1}, \quad \tilde{C}_k = \frac{C_k}{C_{ai,0}},\quad \tilde{\bar{C}}_k = \frac{\bar{C}_k}{C_{ai,0}}, \quad \tilde{H}_{k} = \frac{H_k }{C_{ai,0}},\\
&\tilde{S}_{ex}^{max} = \frac{{S}_{ex}^{max} t_0}{C_{ai,0}}, \quad \tilde{C}_{50}= \frac{C_{50}}{C_{ai,0}}, \quad \tilde{C}_{1/2}= \frac{C_{1/2}}{C_{ai,0}}, 
\end{aligned}
\end{equation}
The timescale $t_0$ is the same as in (\ref{eq38}), since the in-domain diffusion timescale $R_1^2/D_{k} \sim 10^3$s is much larger. Substituting the scalings and after removing the tilde for the unknowns $\tilde{H}_k,\tilde{C}_k, \tilde{\bar{C}}_k$ and independent variables $\tilde{t},\tilde{r}$, we have  the equations for $H_k$
\begin{equation}
\label{eq45}
\begin{aligned}
& \frac{\partial (\eta_{ai}  H_{ai})}{\partial t}  - \frac{1}{r} \frac{\partial }{\partial r} \left(\bar{\kappa}_{ai}  r \frac{\partial p_{ai}}{\partial r}  H_{ai} \right) + Q_{ai,ci} H_{ai}=0, \\
& \frac{\partial (\eta_{ao}  H_{ao}) }{\partial t}  - \frac{1}{r} \frac{\partial }{\partial r} \left(\bar{\kappa}_{ao}  r \frac{\partial p_{ao}}{\partial r} H_{ao} \right) +   Q_{ao,co} H_{ao}=0, \\
& \frac{\partial (\eta_{v}   H_{v} )}{\partial t}  - \frac{1}{r} \frac{\partial }{\partial r} \left(\bar{\kappa}_{v}  r \frac{\partial p_{v}}{\partial r} H_{v} \right) +  Q_{v,ci} H_{ci} +  Q_{v,co} H_{co}=0, \\
& \frac{\partial (\eta_{ci}  H_{ci}) }{\partial t}  -Q_{ai,ci} H_{ai}-  Q_{v,ci} H_{ci}=0,\\
& \frac{\partial (\eta_{co}  H_{co} )}{\partial t}  -Q_{ao,co} H_{ao} - Q_{v,co} H_{co}=0,
\end{aligned}
\end{equation}
and the equations for $\bar{C}_k$ and $C_{ex}$
\begin{equation}
\label{eq46}
\begin{aligned}
& \frac{\partial}{\partial t}(\eta_{ai} \bar{C}_{ai})  - \frac{1}{r} \frac{\partial }{\partial r} \left(\bar{\kappa}_{ai} \bar{C}_{ai} r \frac{\partial p_{ai}}{\partial r}  + \bar{D}_{ai}  r \frac{\partial C_{ai}}{\partial r}  \right) + S_{ai,ex}   + S_{ai,ci}  =0,\\
& \frac{\partial}{\partial t}(\eta_{ao} \bar{C}_{ao})  - \frac{1}{r} \frac{\partial }{\partial r} \left(\bar{\kappa}_{ao} \bar{C}_{ao} r \frac{\partial p_{ao}}{\partial r}  + \bar{D}_{ao} r \frac{\partial C_{ao}}{\partial r}  \right) + S_{ao,ex}  + S_{ao,co} =0,\\
& \frac{\partial}{\partial t}(\eta_{v} \bar{C}_{v})  - \frac{1}{r} \frac{\partial }{\partial r} \left(\bar{\kappa}_{v} \bar{C}_{v} r \frac{\partial p_{v}}{\partial r}  + \bar{D}_{v}  r \frac{\partial C_{v}}{\partial r}  \right) + S_{v,ex}  + S_{v,ci} +S_{v,co}=0,\\
& \frac{\partial}{\partial t}(\eta_{ci} \bar{C}_{ci}) + S_{ci,ex} -S_{ai,ci} - S_{v,ci} =0,\\
& \frac{\partial}{\partial t}(\eta_{co} \bar{C}_{co}) + S_{co,ex} -S_{ao,co} - S_{v,co} =0,\\
& \frac{\partial}{\partial t}(\eta_{ex} C_{ex})  - \frac{1}{r} \frac{\partial }{\partial r} \left(\bar{\kappa}_{ex} {C}_{ex} r \frac{\partial p_{ex}}{\partial r}  + \bar{D}_{ex} r \frac{\partial C_{ex}}{\partial r}  \right)  + S_{ex}  - \sum_{j=ai,ao,v,ci,co} S_{j,ex}   =0,\\
\end{aligned}
\end{equation}
where total oxygen concentration $\bar{C}_k$ is 
\begin{equation}
\label{eq47}
\begin{aligned}
& \bar{C}_{k} (C_{k}, H_k)=C_{k} +   {H}_{k} \frac{C_{k}^{n}}{C_{k}^{n}+ \tilde{C}_{50}^{n}},\quad k = ai, ao, v, ci, co,
\end{aligned}
\end{equation}
the oxygen exchange and consumption rates  are given by 
\begin{equation}
\label{eq48}
\begin{aligned}
& S_{j,ex} =  Q_{j,ex} C_{j,ex}^{upwind} + \bar{l}_{j,ex} (C_{j} - C_{ex}),\quad j= ai, ao, v, ci, co,\\
& S_{ai,ci} = Q_{ai,ci} \bar{C}_{ai} +  \tilde{D}_{ai,ci}  (C_{ai} - C_{ci}),\\
& S_{ao,co} = Q_{ao,co} \bar{C}_{ao} +  \tilde{D}_{ao,co}  (C_{ao} - C_{co}),\\
& S_{v,k} = Q_{v,k} \bar{C}_{k,v}^{upwind} +  \tilde{D}_{v,k}  (C_{v} - C_{k}), \quad k = ci,co,\\
& S_{ex} =\tilde{S}_{ex}^{max} \frac{C_{ex}}{C_{ex} + \tilde{C}_{1/2}},
\end{aligned}
\end{equation}
and the coefficients are defined by
\begin{equation}
\label{eq49}
\begin{aligned}
&\bar{D}_j = \tilde{D}_{j} \eta_j,\quad \tilde{D}_{j} = \frac{D_{j} \tau_{j} t_0}{R_1^2}, \quad j= ai, ao, v, ex,\\
&\bar{l}_{j,ex} = \tilde{l}_{j,ex}  \sqrt{\eta_{j}}, \quad \tilde{l}_{j,ex} = M_{j,ex}^0 l_{j,ex} t_0, \quad j= ai, ao, v, ci, co,\\
 & \tilde{D}_{j,ci} = D_{j,ci} t_0, \quad j= ai, v,\quad  \tilde{D}_{j,co} = D_{j,co} t_0, \quad j= ao, v.
\end{aligned}
\end{equation}
The boundary conditions in (\ref{eq35},\ref{eq36}) will take the same form except that $C_{ai,0},C_{ao,1},H_0$ are nondimensionalized by $C_{ai,0}$.

\section{Numerical Results}

In our implementation, the partial differential equations in  (\ref{eq39},\ref{eq45},\ref{eq46}) with  boundary conditions are converted to dynamic systems by using finite-difference method for the spatial variable $r$. In the finite-difference scheme, central discretization is adopted and upwind scheme is used for the convection terms in (\ref{eq45},\ref{eq46}) (otherwise it is unstable). Then, they are solved as a whole system of differential-algebraic equations (DAE) by combining with the algebraic constraints (\ref{eq42},\ref{eq43},\ref{eq47}). The DAE system is solved in Matlab using the built-in solver ode15s.  All the 28 unknowns $p_k,\eta_k,C_k$ ($k=ai,ao,v,ci,co,ex$) and $\bar{C}_j,H_j$  ($j=ai,ao,v,ci,co$) at all discrete spatial points are solved simultaneously. In this section, we consider a uniform case that resting volume fractions $\eta_j^{re} (r)$ ($j=ai,ao,v,ci,co,ex$) follow uniform distribution, i.e., they are assumed to be constants. All the chosen parameters are given/estimated in appendix.  This case will serve as a reference case for the study of the effects of resting volume fractions and other parameters in the next section.

\subsection{Water/blood circulation}
\label{section3_1}

\begin{figure}[h]
\begin{center}
\includegraphics[width=0.45 \textwidth]{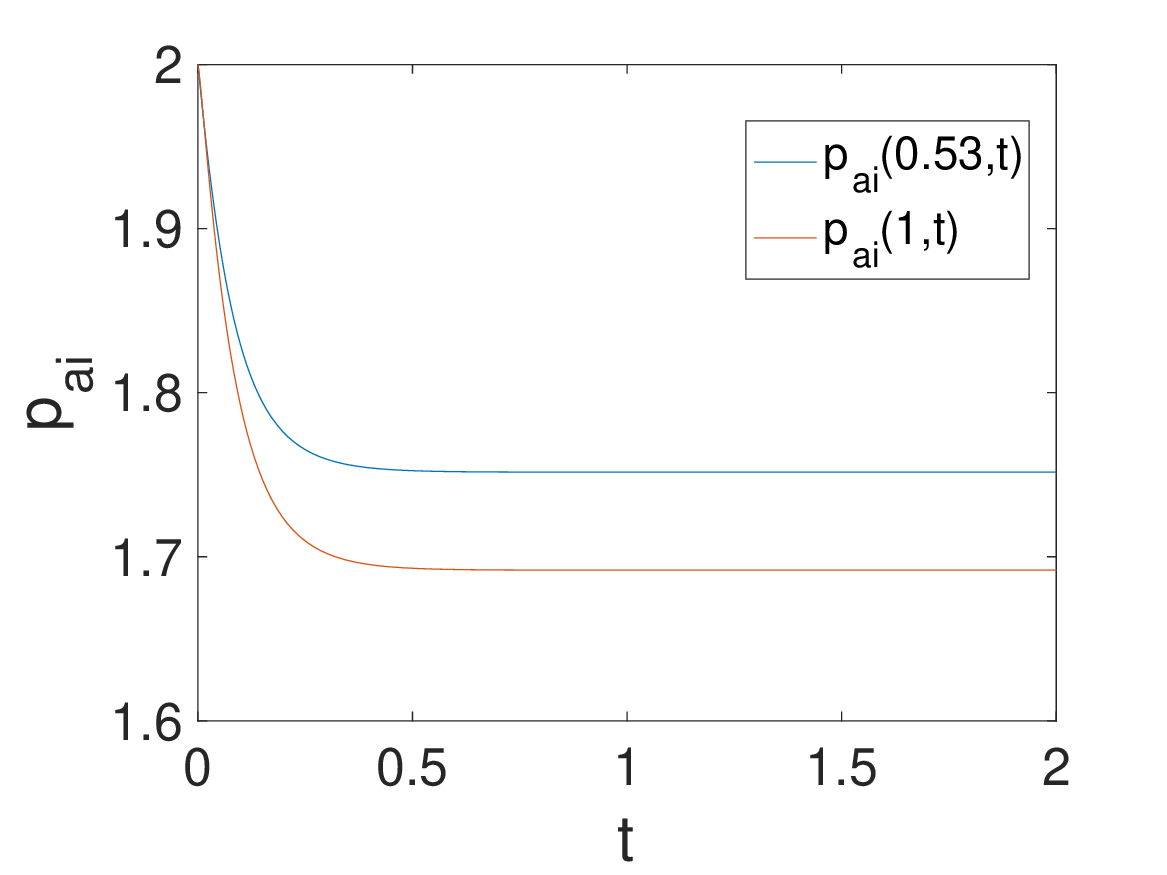}
\includegraphics[width=0.45 \textwidth]{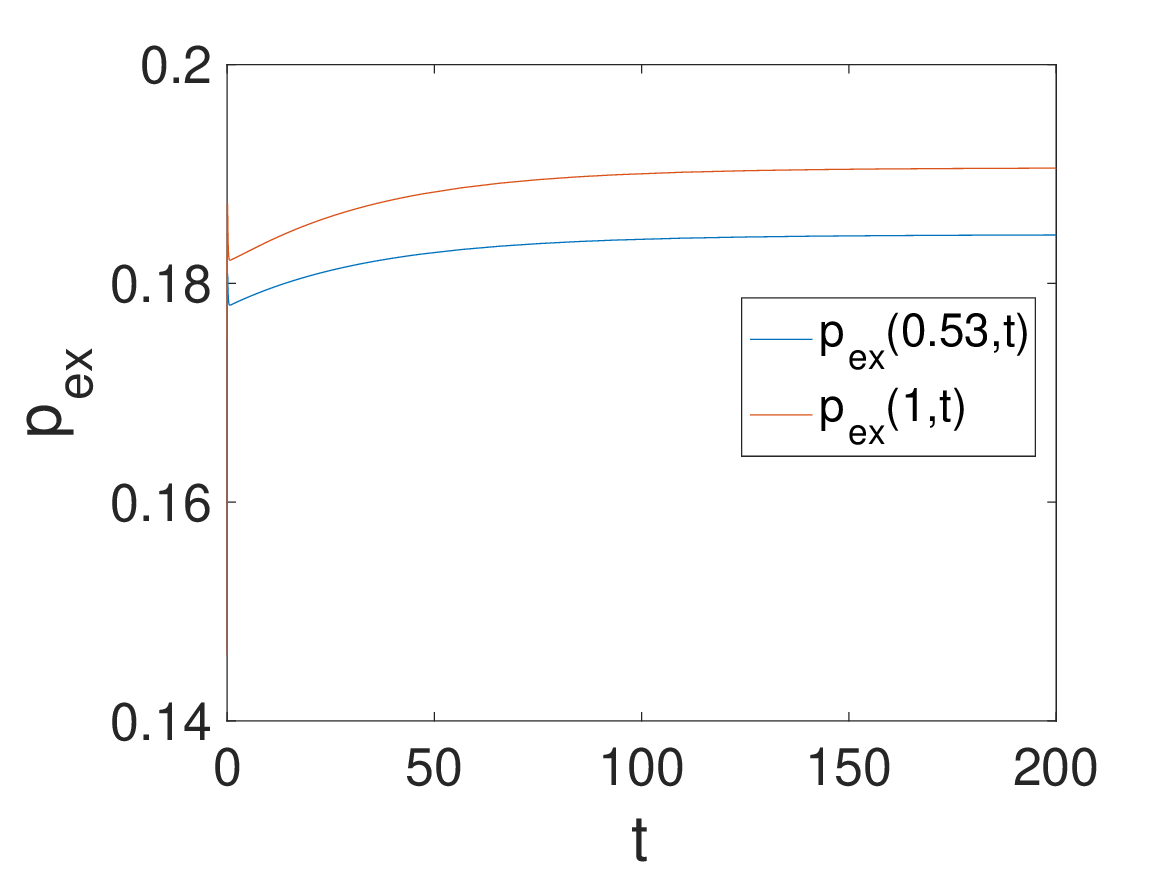}
\caption{The dynamics of $p_{ai}$ and $p_{ex}$ at two points $r=0.53,1$.}
\label{fig2}
\end{center}
\end{figure}

The values used in the initial conditions at $t=0$ are set to be consistent with the algebraic constraints and the boundary conditions for pressures. At $t=2$, most of unknowns have reached the steady state. Figure \ref{fig2}(a)  for the dynamics of artery pressure $p_{ai}$ at two locations $r=0.53, 1$.  The pressure $p_{ex}$ in Figure \ref{fig2}(b) reaches the steady state slower (at about $t=200$), because it is more sensitive to the small variations of volume fractions $\eta_k$ during the dynamics via the effective permeability $\bar{\kappa}_k$ in $(\ref{eq39})_6$.

\begin{figure}[h]
\begin{center}
\includegraphics[width=0.45 \textwidth]{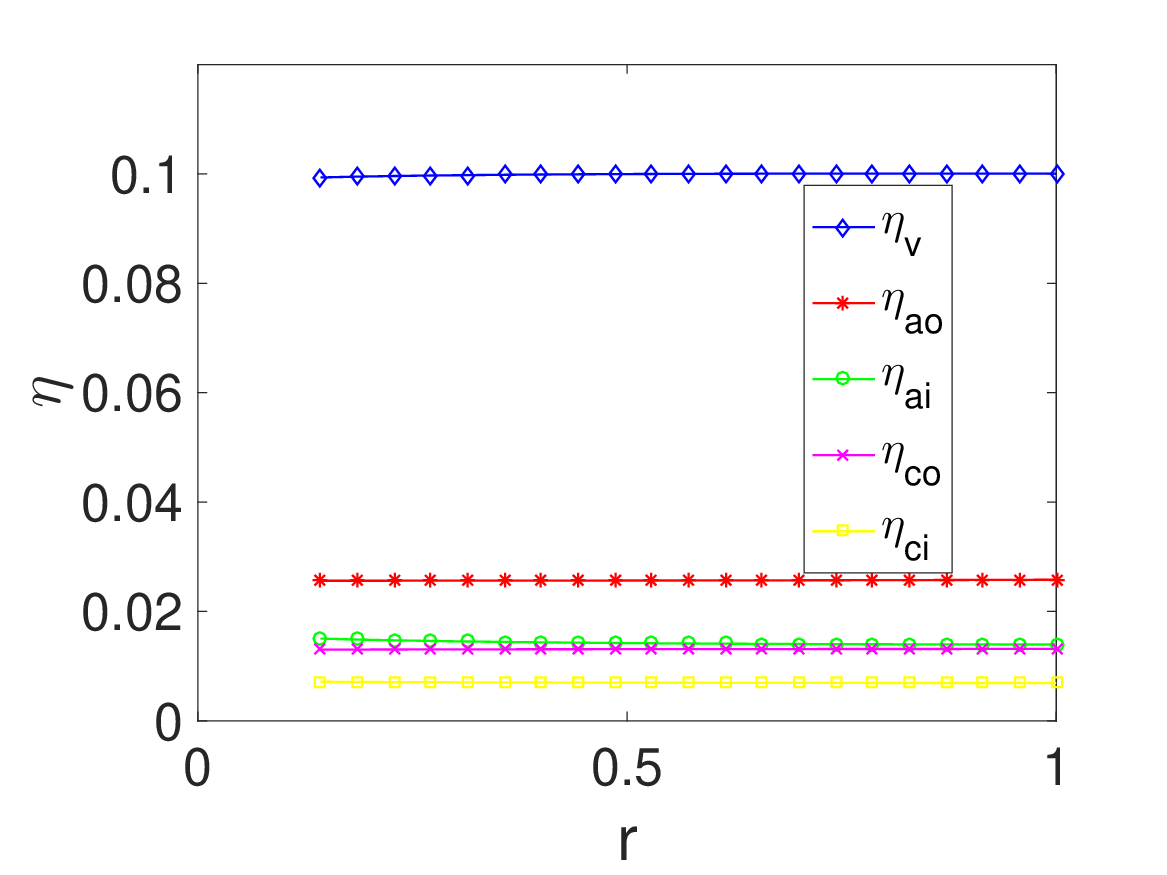}
\includegraphics[width=0.45 \textwidth]{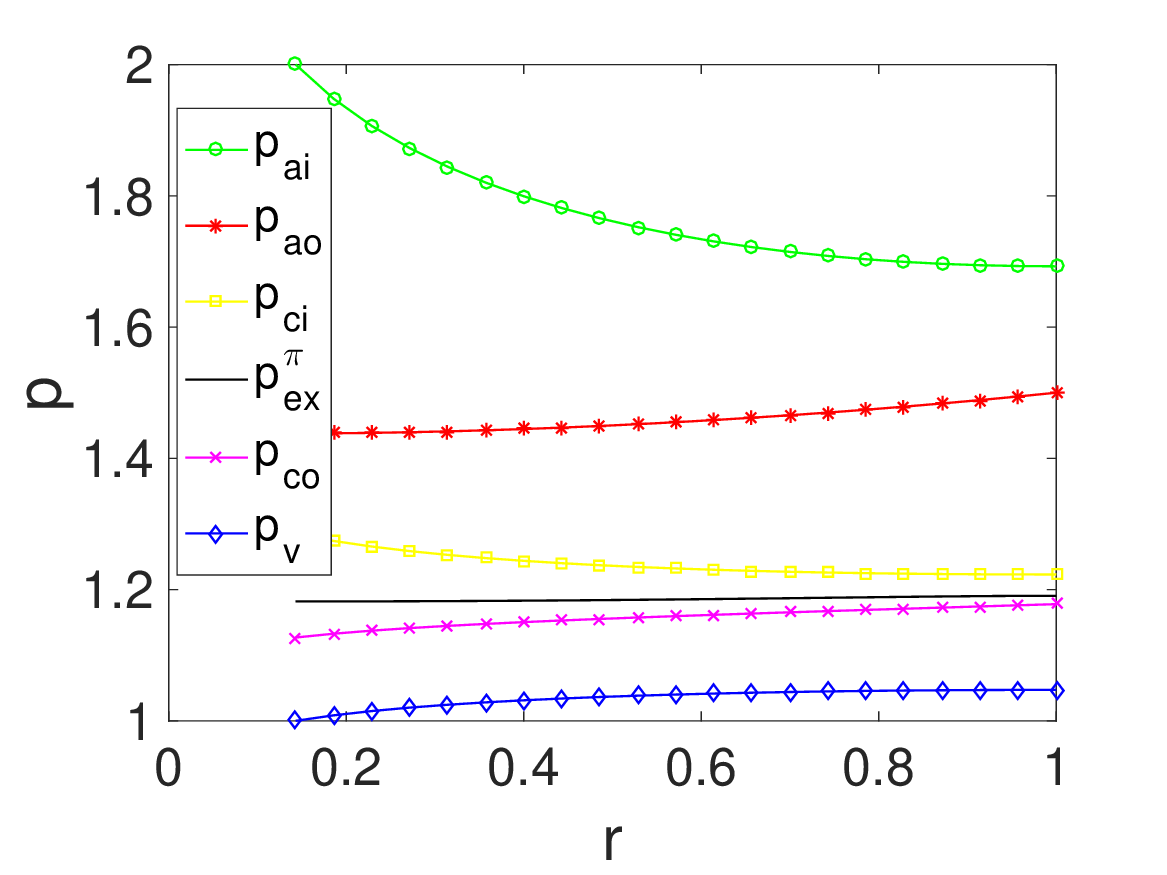}
\caption{The spatial profiles of volume fractions $\eta_{j}$ and pressures $p_{j}$ in domain $\Omega_j$ at steady state. }
\label{fig3}
\end{center}
\end{figure}

% The legend follows the same order of the curves from top to bottom.

Next, we illustrate the steady-state profiles with $t=200$ in the dimensionless geometric region $0.14<r<1$. Figure \ref{fig3}(a) shows profiles of volume fractions in the five vascular domains, which are almost uniform since the difference $\eta_k-\eta_k^{re}$ is quite small (with max around $2.5*10^{-3}$) due to the large moduli $\tilde{\lambda}_k$.  The remaining percentage of about 85\% is for $\eta_{ex}$ (not shown in the figure).  Figure \ref{fig3}(b) shows the pressures in the six domains, where for convenience of comparison we have defined 
\begin{equation}
\label{eq50}
\begin{aligned}
p_{ex}^\pi = p_{ex} + \Delta \tilde{\pi} =  p_{ex}  + \tilde{\pi}_v- \tilde{\pi}_{ex}.
\end{aligned}
\end{equation}
It shows that the pressures in the two artery domains drop along the direction of blood flow, and the pressure in capillary domain lies between the connected artery and vein domains.  The modified extravascular pressure $p_{ex}^\pi$ lies between the artery and vein pressures, and the sign of the difference between $p_{ex}^\pi$ and vascular pressures will determine the direction of water leak through blood vessel wall.

\begin{figure}[h]
\begin{center}
\includegraphics[width=0.45 \textwidth]{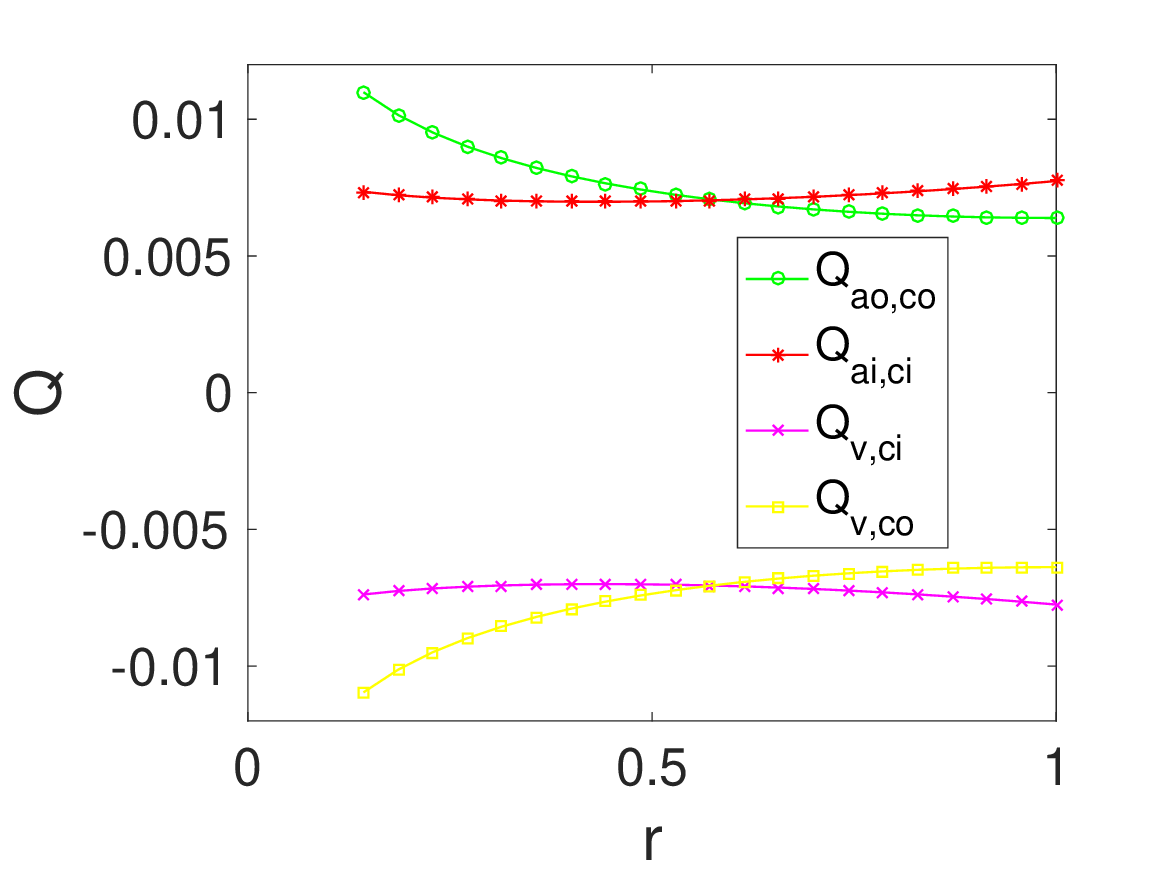}
\includegraphics[width=0.45 \textwidth]{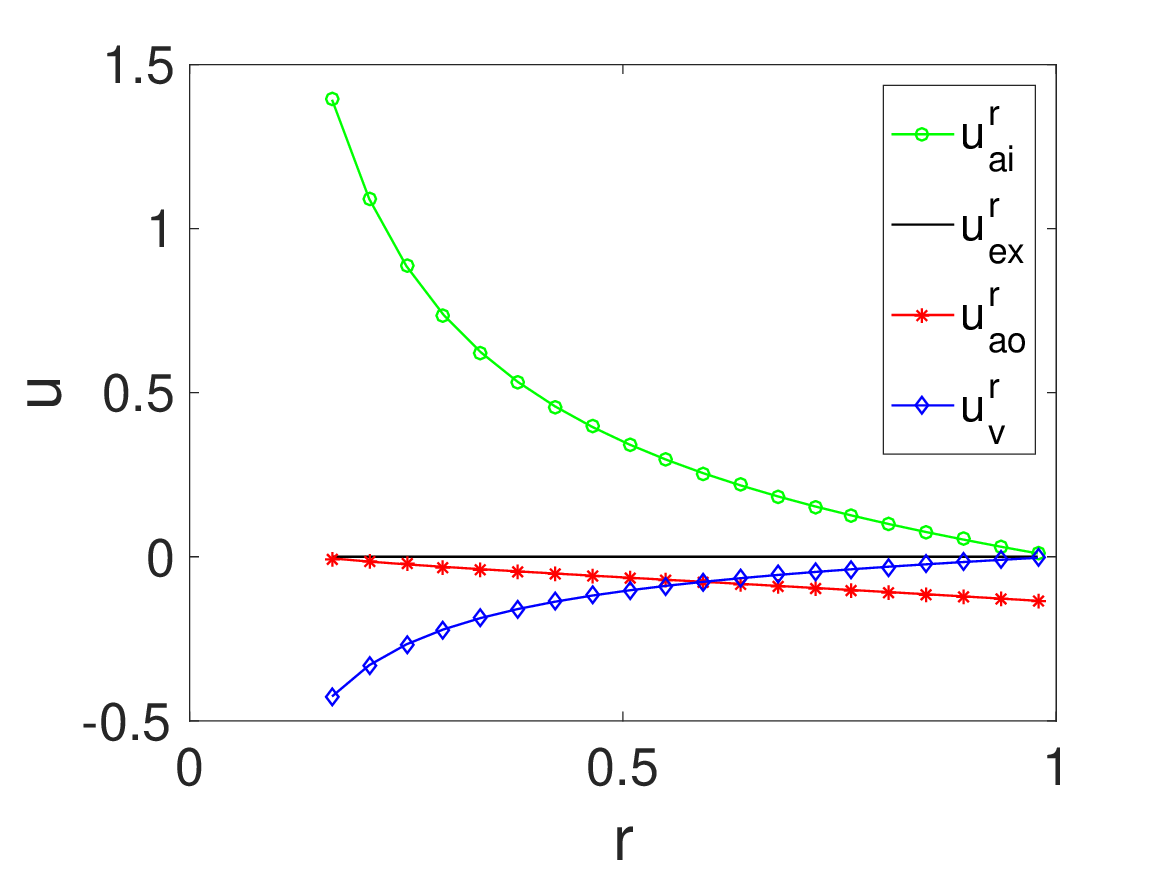}
\caption{Blood flow rates between vascular domains, and the in-domain velocities.}
\label{fig4}
\end{center}
\end{figure}

\begin{figure}[h]
\begin{center}
\includegraphics[width=0.45 \textwidth]{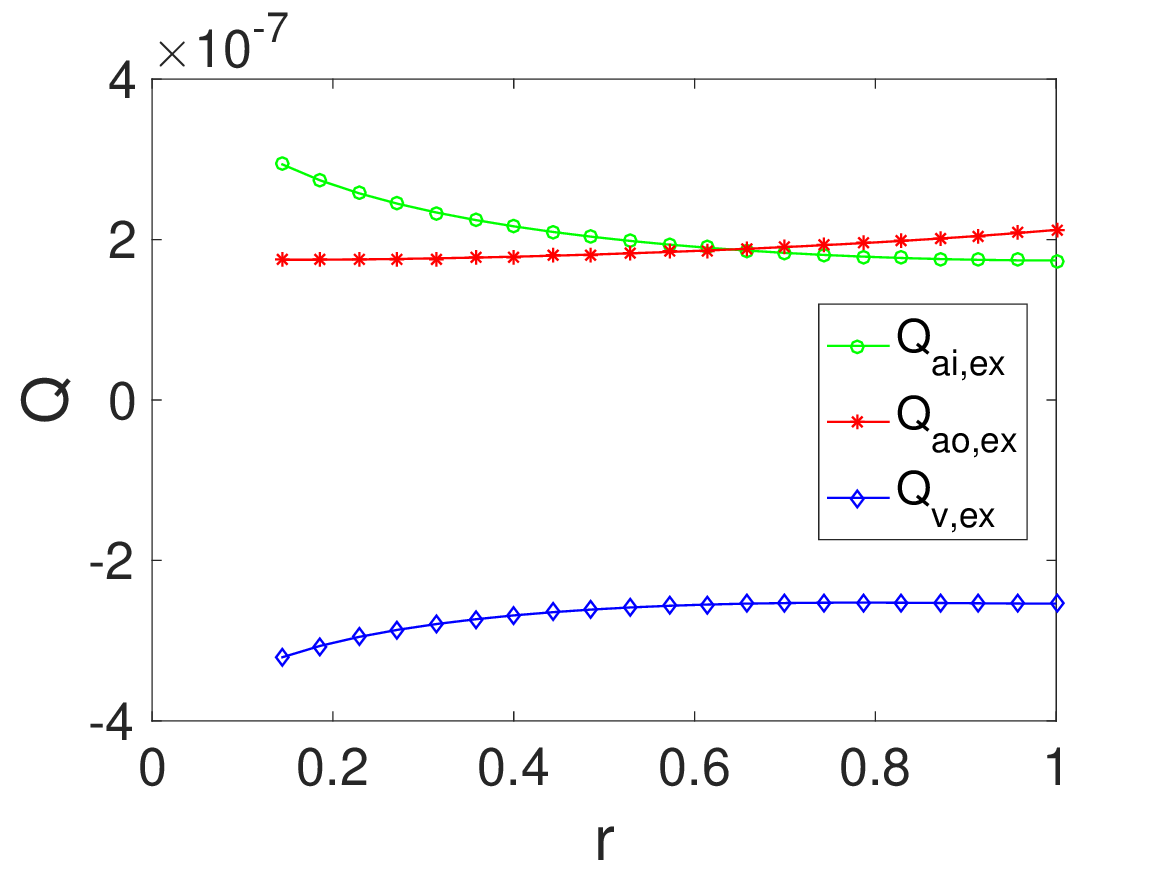}
\includegraphics[width=0.45 \textwidth]{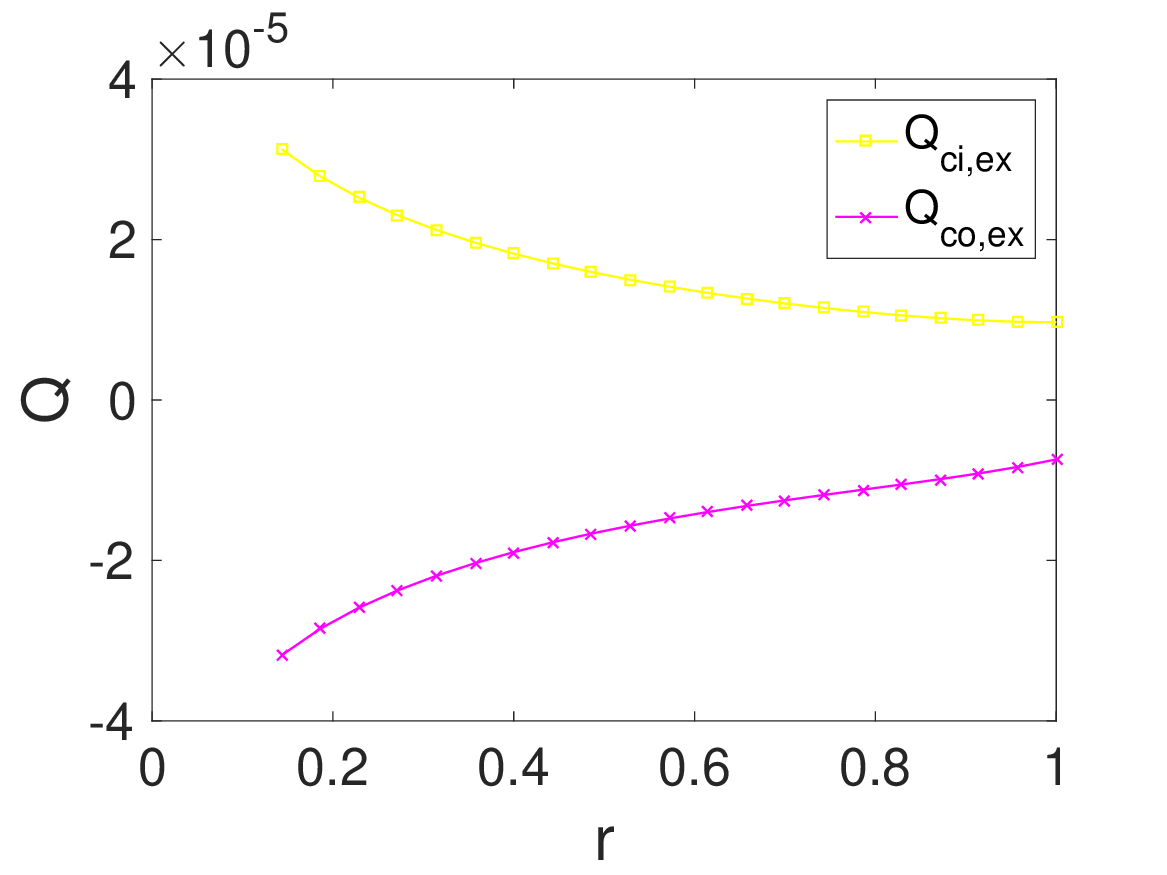}
\caption{The water flow rates across the blood vessel wall  between the vascular domains and the extravascular domain.}
\label{fig5}
\end{center}
\end{figure}

Figure \ref{fig4}(a) shows the blood flow rates going through capillaries between the vascular domains.These are the dominant terms in the governing equations and are to be balanced with the in-domain flows through arteries and veins. The signs of $Q_{i,j}$ in Figure \ref{fig4}(a) indicate that blood flows from artery to capillary and then from capillary to vein. Figure \ref{fig4}(b) shows the dimensionless in-domain blood/water velocities, which are defined as 
\begin{equation}
\label{eq51}
\begin{aligned}
&u_j^r= - \tilde{\beta}_j \eta_j  \frac{\partial p_j}{\partial r}, \quad j =ai,ao,v,\quad u_{ex}^r = -\tilde{\kappa}_{ex} \frac{\partial p_{ex}}{\partial r}.
\end{aligned}
\end{equation}
The sign of velocity in Figure \ref{fig4}(b) is consistent with the pressure drops in Figure \ref{fig3}(b) and we have the following expected observations
\begin{itemize}
\item for $\Omega_{ai}$, blood flows from inner boundary $r=R_0$ (i.e., the CRA) to outer boundary $r=R_1$ with decreasing velocity;
\item for $\Omega_{ao}$, blood flows from outer boundary (i.e., the PCA) to the inner boundary  with decreasing velocity; 
\item for $\Omega_{v}$, blood flows from the outer boundary to the inner boundary (i.e., the CRV) with increasing velocity; 
\item the water flow in $\Omega_{ex}$ is negligibly small compared with scale of blood flow.
\end{itemize}
The maximum velocity occurs at the start of artery (CRA) in $\Omega_{ai}$, where the pressure drop is most significant. The velocity at inner boundary in $\Omega_v$ is also relatively large, since the blood will eventually merge and drain from the system through the CRV. The scaling factor for the velocity is 
\begin{equation}
\label{eq52}
\begin{aligned}
&\frac{R_1}{t_0} = 0.79  \mathrm{cm/s},
\end{aligned}
\end{equation}
and the maximum velocity with units in this case is 1.1 cm/s, consistent with the scales in \cite{wang2009,julien2023,causin2016,nagaoka2006}. Figure \ref{fig5} shows the water flow rates across the blood vessel wall, which are much smaller than the blood flow rates in Figure \ref{fig4}(a) and expected for the normal parameters. But the water flow rates could change significantly under pathological conditions.

\subsection{Oxygen transport}

\begin{figure}[h]
\begin{center}
\includegraphics[width=0.45 \textwidth]{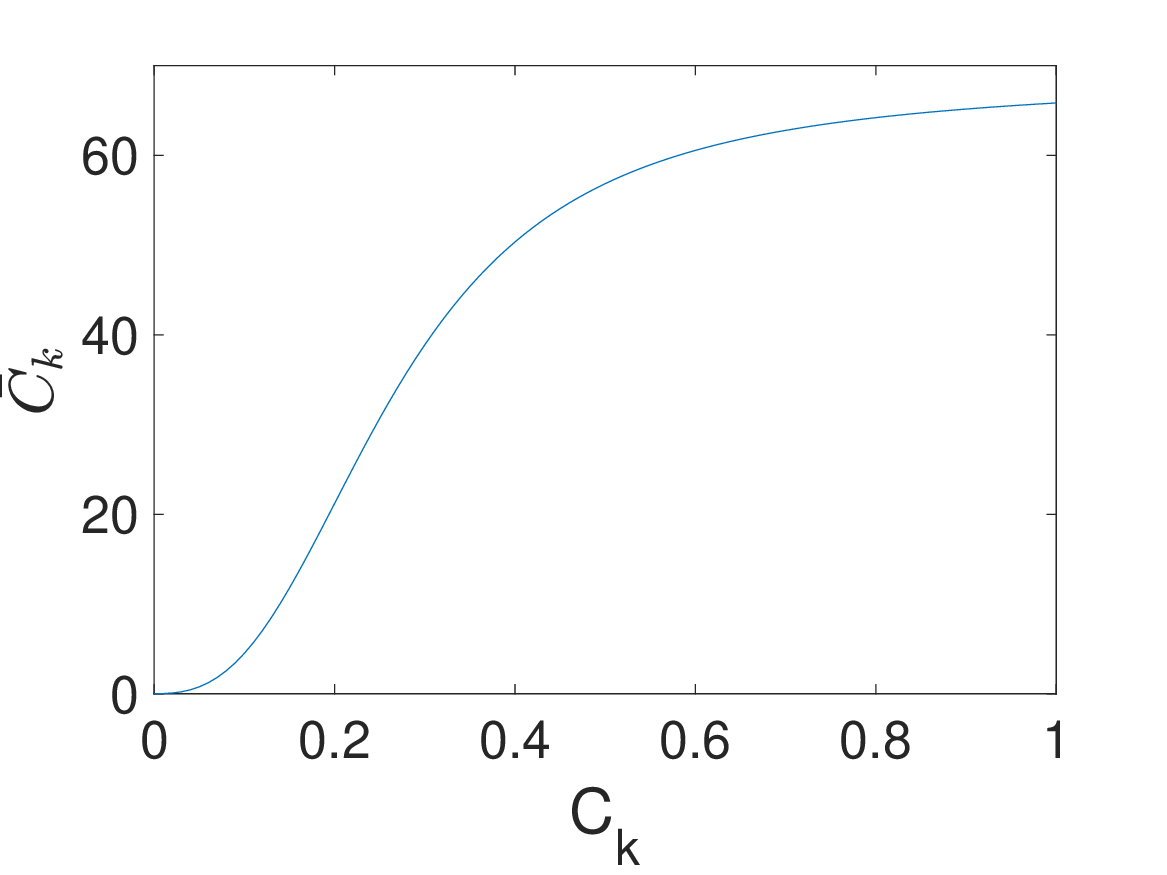}
\includegraphics[width=0.45 \textwidth]{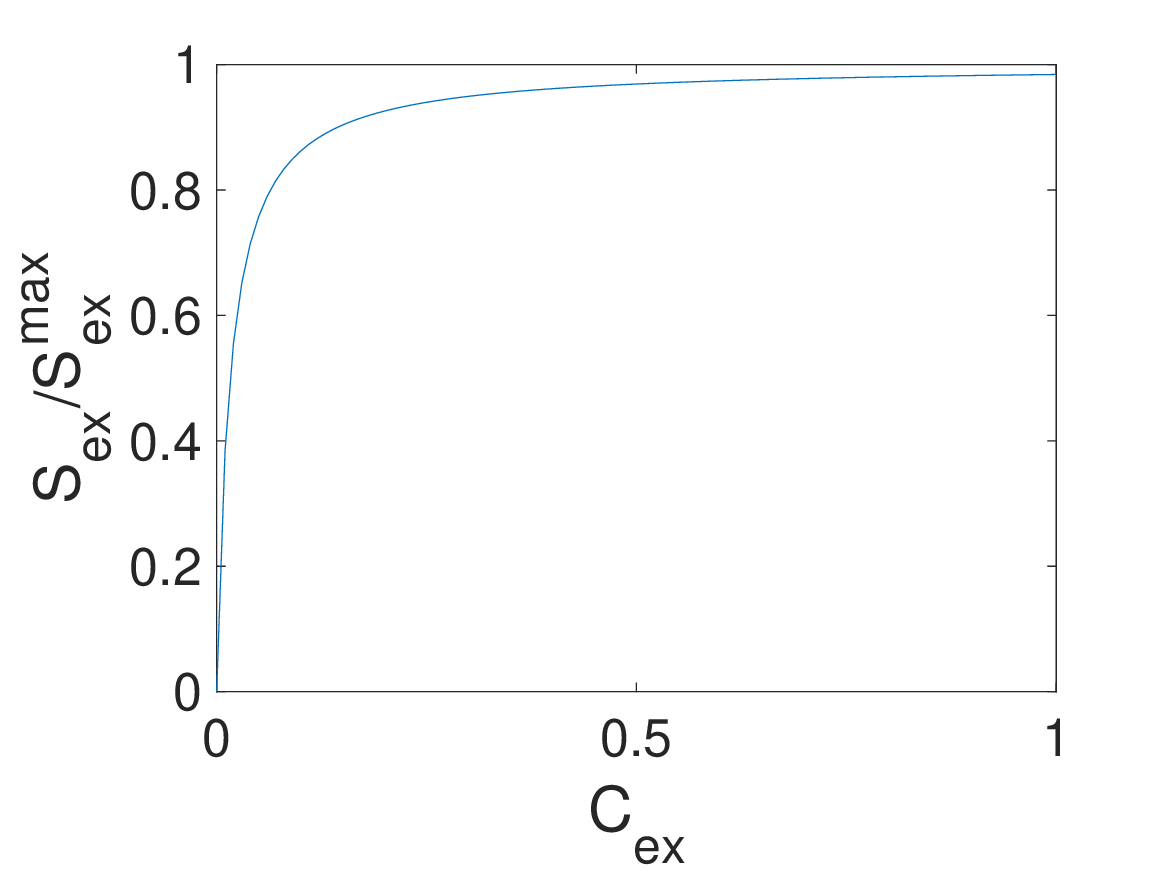}
\caption{Relation between total concentration $\bar{C}_k$ and dissolved concentration $C_k$, and the relation between consumption rate $S_{ex}/S_{ex}^{max}$ and dissolved concentration $C_{ex}$ in tissue.}
\label{fig6}
\end{center}
\end{figure}

To better understand the numerical results, we first show the two functions $\bar{C}_k(C_k)$ in the vascular domains and $S_{ex}(C_{ex})$ in the extravascular domain. Figure \ref{fig6}(a) shows relation between the total concentration $\bar{C}_k$ and dissolved concentration $C_k$ with $H_k=H_0=66.7$ in vascular domains, which implies that the majority of the oxygen is stored in the RBC. With $C_k=1$, we have $\bar{C} \approx 66$, so the dissolved oxygen is only about 1.5\% of the total oxygen while the oxygen stored in RBC is about $98.5\%$ as expected \cite{pittman2016}. Figure \ref{fig6}(b) shows the relation between the normalized consumption rate  $S_{ex}/S_{ex}^{max}$ and the oxygen concentration $C_{ex}$ with $\tilde{C}_{1/2}=0.016$, showing that the consumption rate is maintained at relatively high level (e.g., above $95\%$ of maximum) if $C_{ex}$ is maintained at reasonable values (e.g., $C_{ex}>0.3$).

\begin{figure}[h]
\begin{center}
\includegraphics[width=0.45 \textwidth]{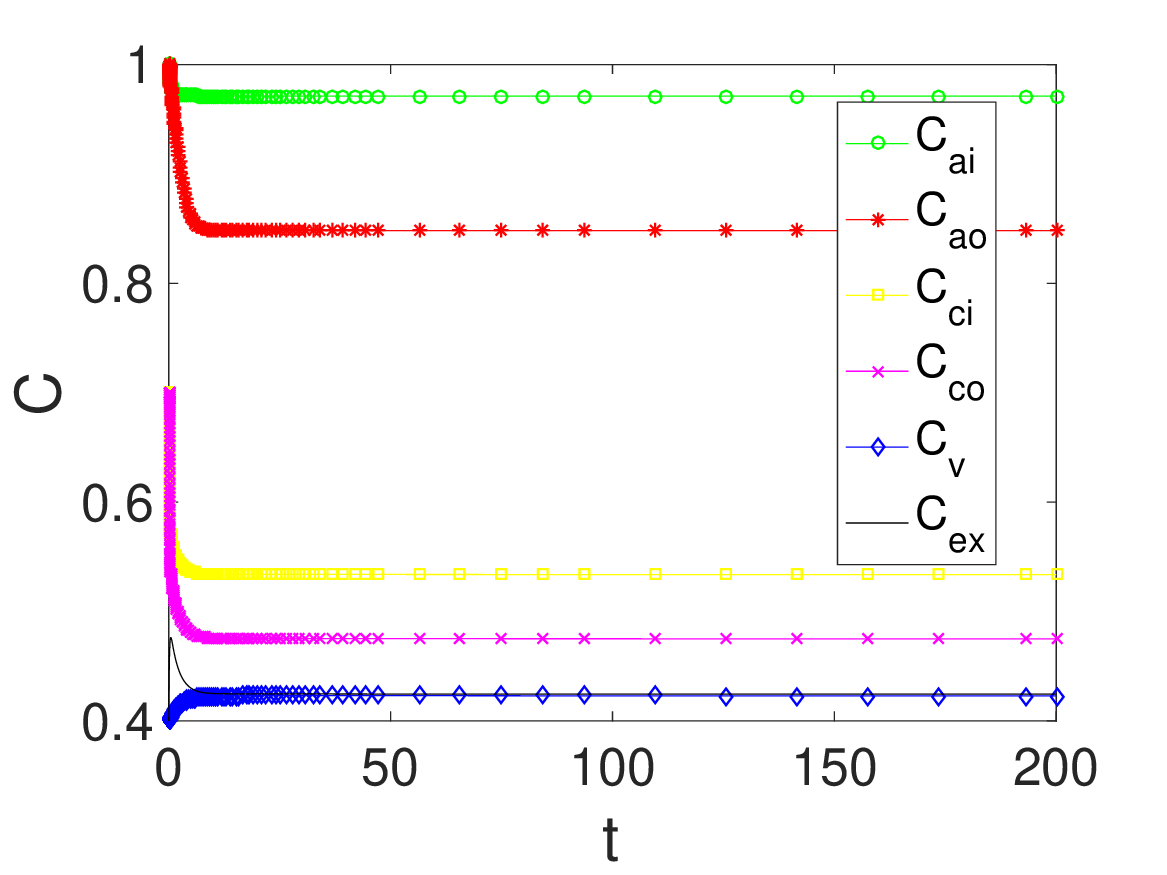}
\includegraphics[width=0.45 \textwidth]{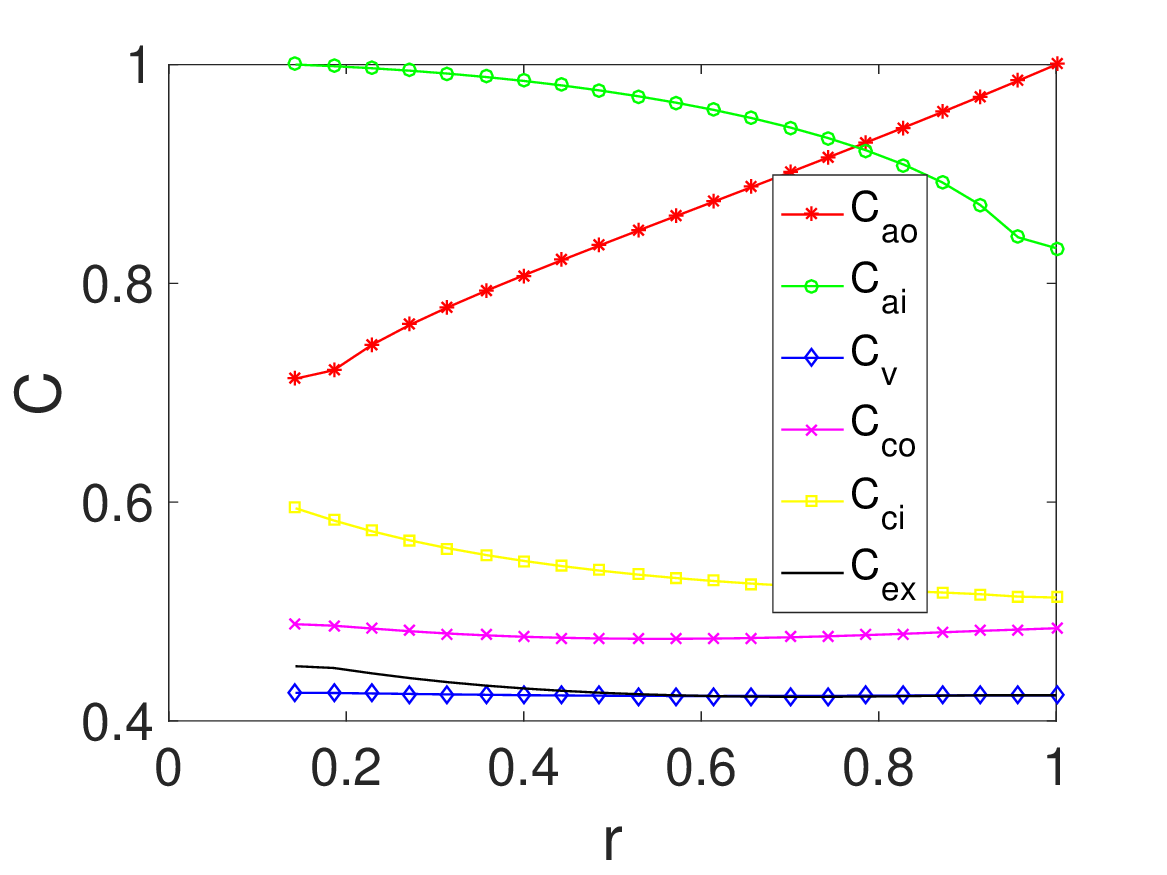}
\includegraphics[width=0.45 \textwidth]{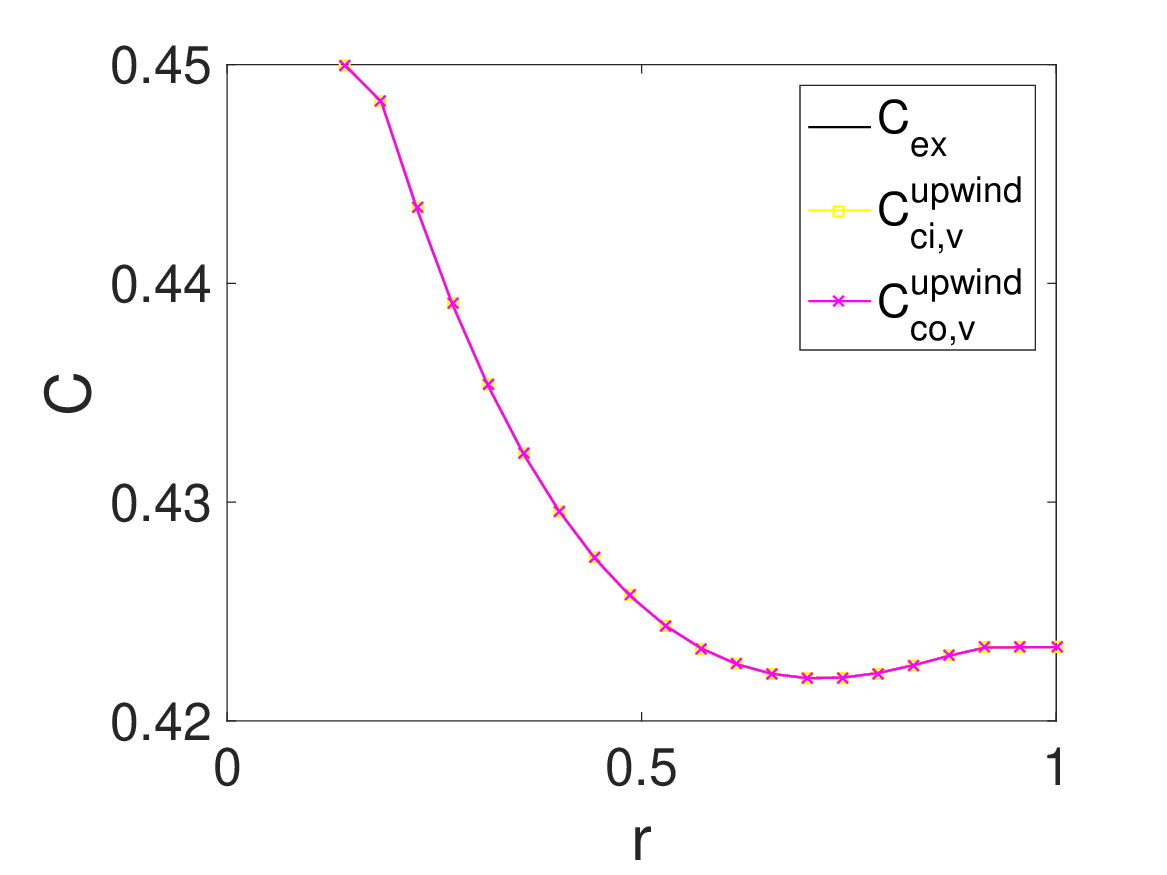}
\includegraphics[width=0.45 \textwidth]{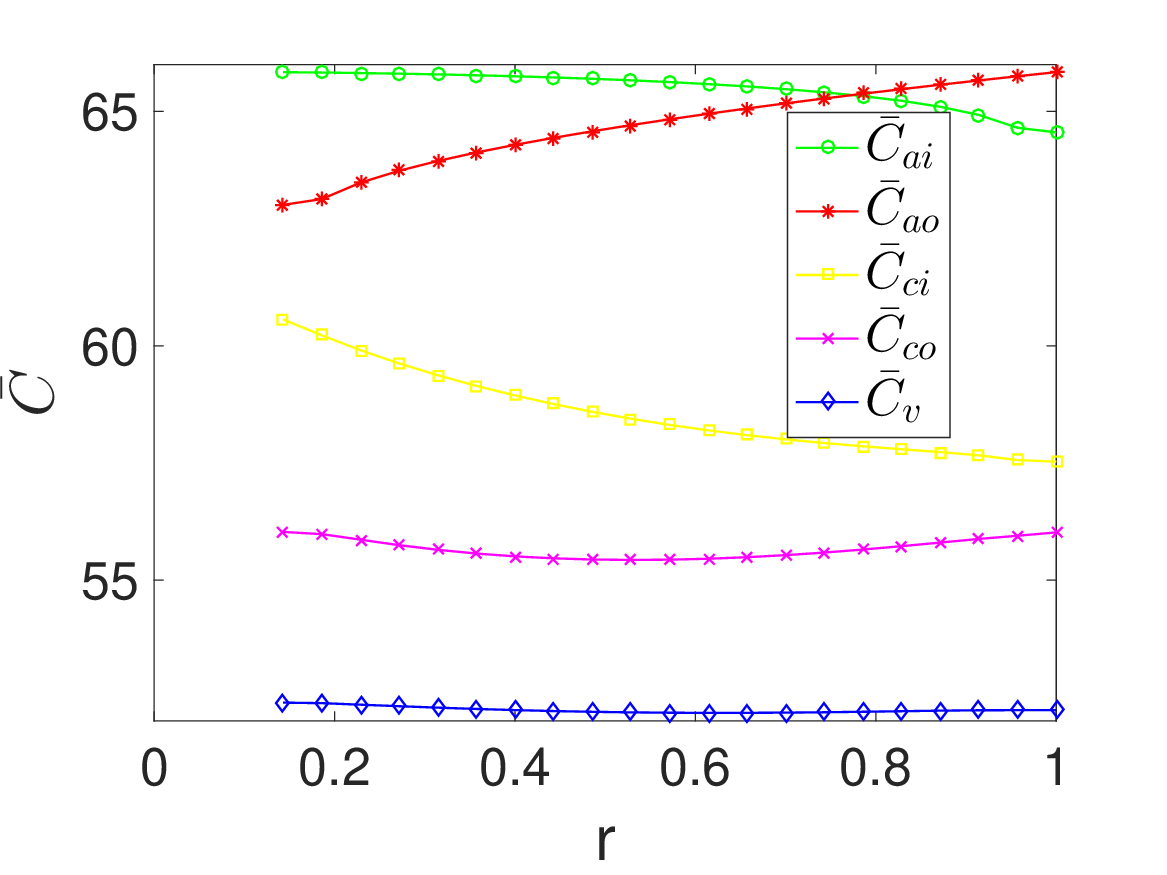}
\caption{Dynamics of concentrations at  $r=0.53$ and the steady-state profiles of concentrations.}
\label{fig7}
\end{center}
\end{figure}

Figure \ref{fig7}(a) shows the dynamics of concentrations at a middle point $r=0.53$, which implies that they already reach the steady state at $t=200$. Figure \ref{fig7}(b) shows the profiles of dissolved concentrations $C_k$ in six domains at steady state (at $t=200$). The highest two curves are for the two artery domains, and the concentrations decrease along the direction of blood flow. The middle two curves for the capillary domains lie between those for the artery and vein domains. This is expected since the major exchange and supply of oxygen occur via the capillary domains. The oxygen concentrations in the vein and extravascular domains have similar values and are the lowest among the domains.  Figure \ref{fig7}(c) shows the upwind concentrations $C_{co,v}^{upwind},C_{ci,v}^{upwind}$ before entering into the vein domain (see (\ref{eq33})), and in this reference case they are equal to $C_{ex}$. The total oxygen concentrations in Figure \ref{fig7}(d) follow similar shape as in Figure \ref{fig7}(b), but at a much larger order of magnitude.

\begin{figure}[h]
\begin{center}
\includegraphics[width=0.45 \textwidth]{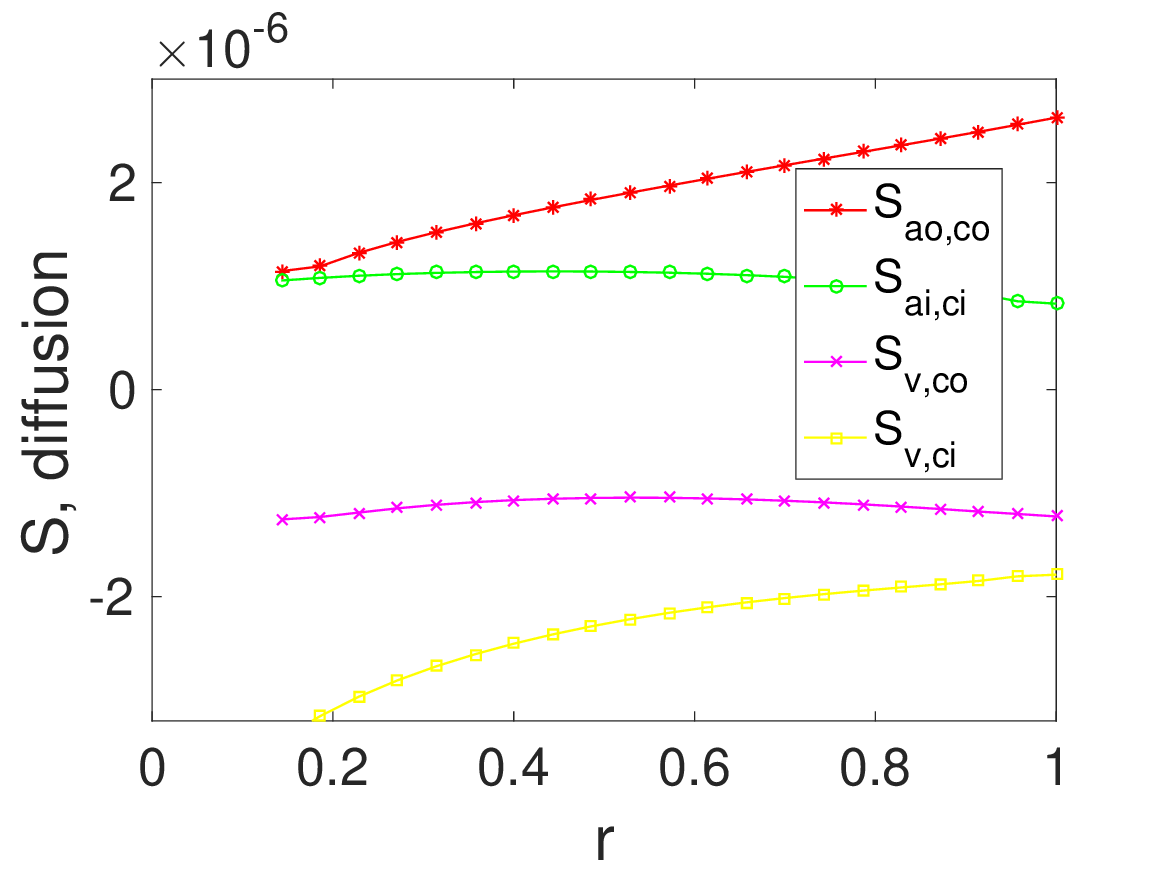}
\includegraphics[width=0.45 \textwidth]{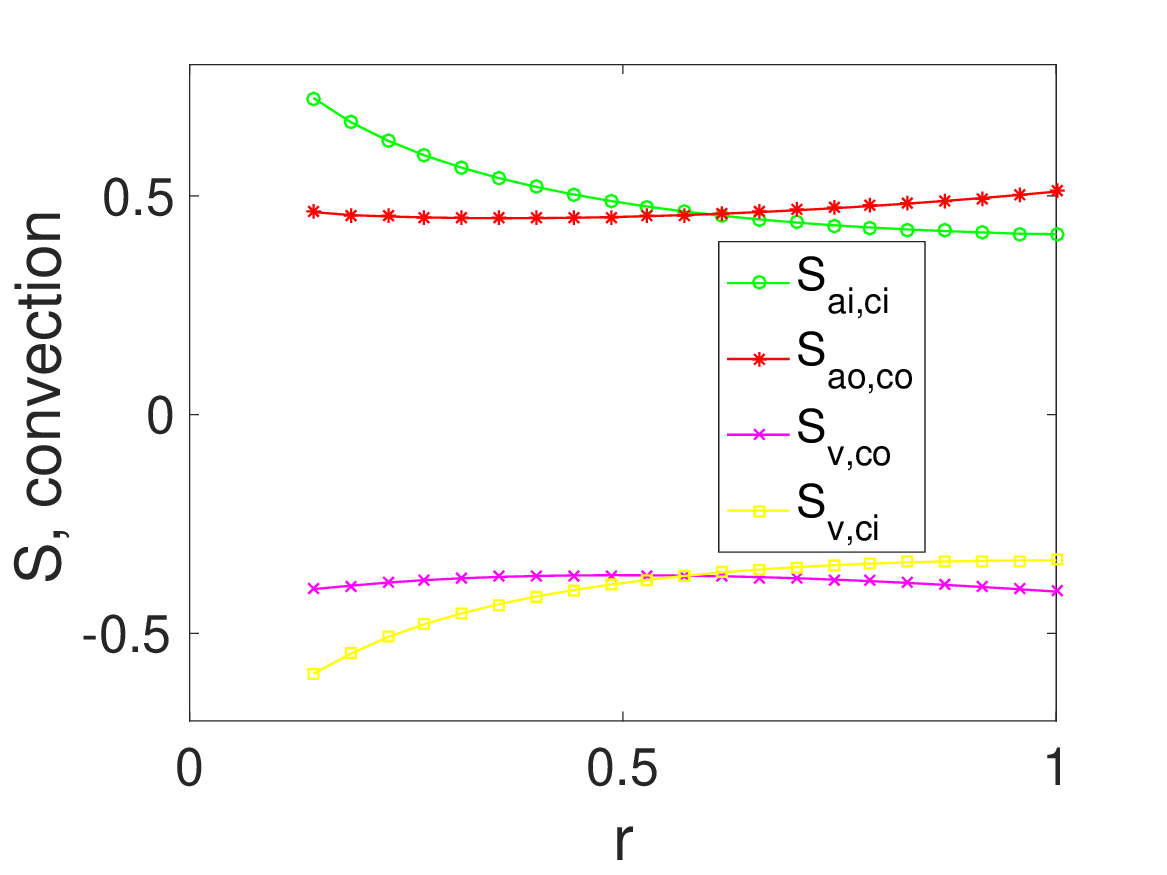}
\includegraphics[width=0.45 \textwidth]{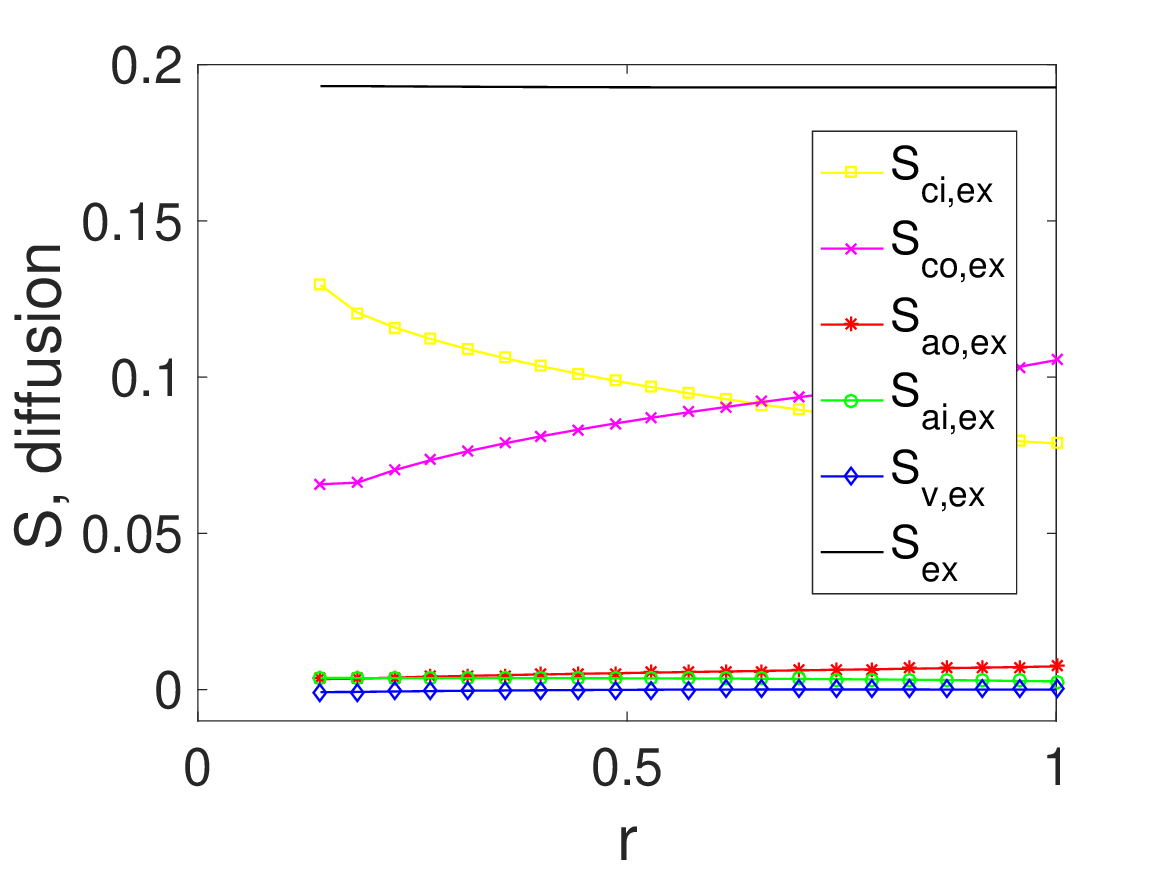}
\includegraphics[width=0.45 \textwidth]{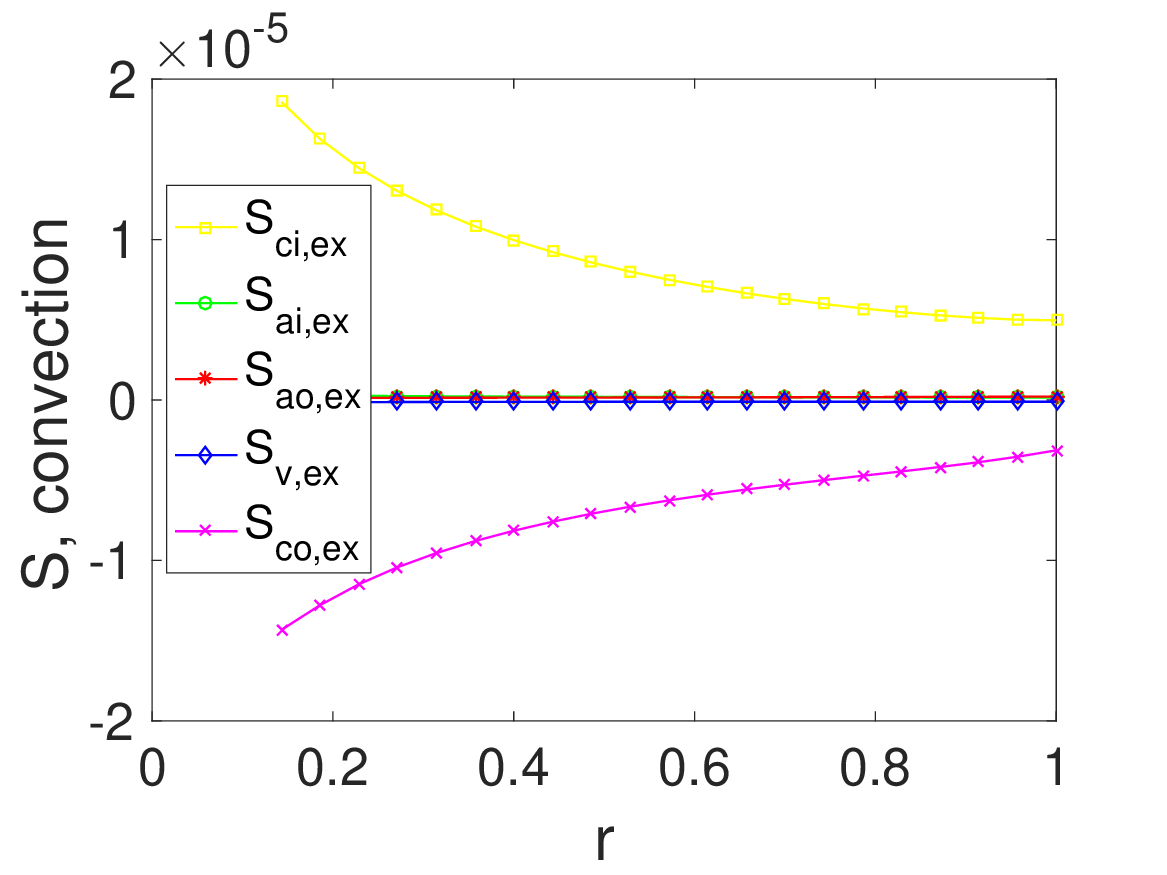}
\caption{Oxygen exchange rates between different domains due to diffusion and convection mechanisms.}
\label{fig8}
\end{center}
\end{figure}

Figure \ref{fig8} shows the oxygen exchange rates between different domains, including the diffusion mechanisms (related to $\tilde{D}_{i,j}$ and $\bar{l}_{i,j}$) and the convection mechanisms (related to blood/water flows $Q_{i,j}$). In the vascular domains, the oxygen exchange follows the direction of blood flow from artery to capillary and then to vein by the signs of $S_{i,j}$ in Figure \ref{fig8}(a,b), and the convection dominates the exchange process by the magnitude in Figure \ref{fig8}(a,b). The supply of oxygen to extravascular domain is mainly through diffusion across blood vessel wall compared with convection, shown in Figure \ref{fig8}(c,d). Figure \ref{fig8}(c) also shows that the two capillaries domains provide the major supply of oxygen to extravascular domain, and the total consumption rate $S_{ex}$ per unit volume is relatively stable. The drop for $C_k$ ($k=ai,ao$) in artery domains in Figure \ref{fig7}(b) is also due to the diffusion of oxygen through blood vessel wall. In summary, in the present model, the pathway of supply of oxygen is mainly from artery to the capillary by convection and to the extravascular domain (tissue) by diffusion, consistent with common sense.

\begin{figure}[h]
\begin{center}
\includegraphics[width=0.45 \textwidth]{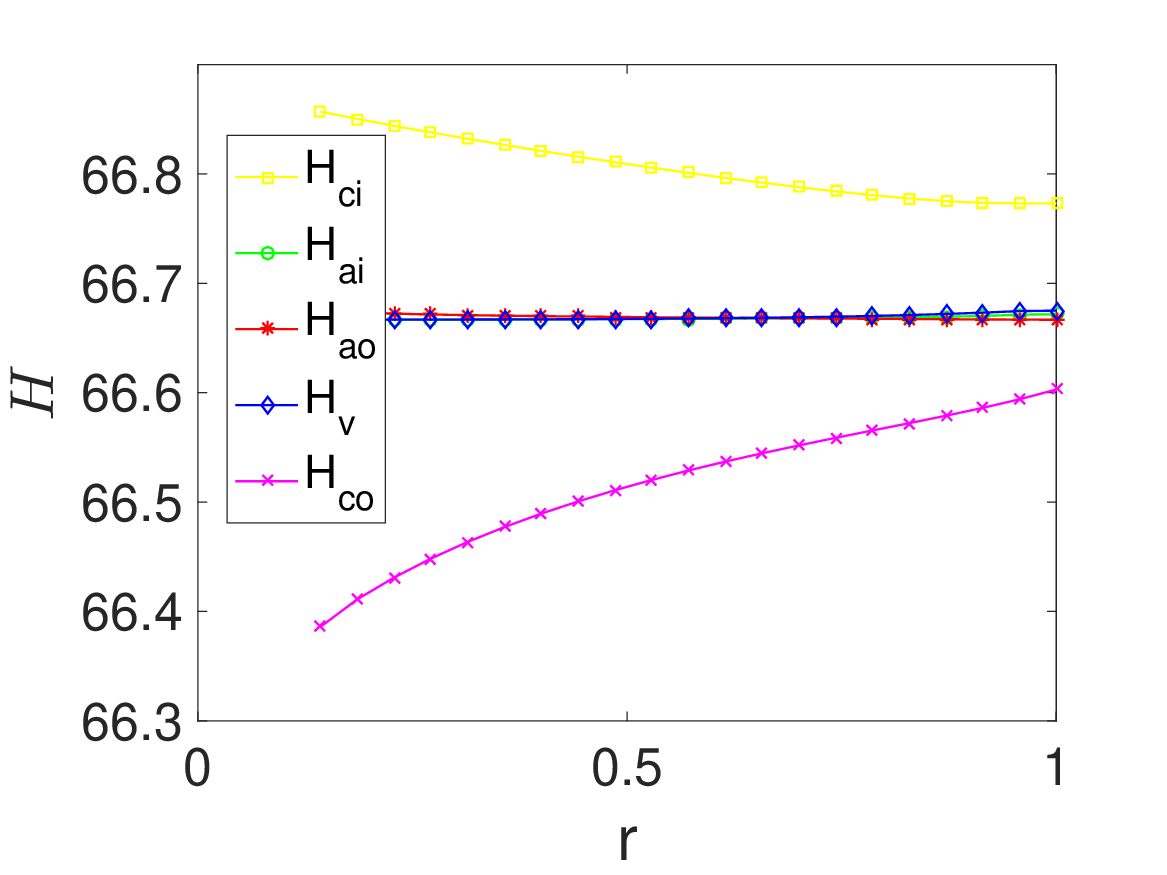}
\includegraphics[width=0.45 \textwidth]{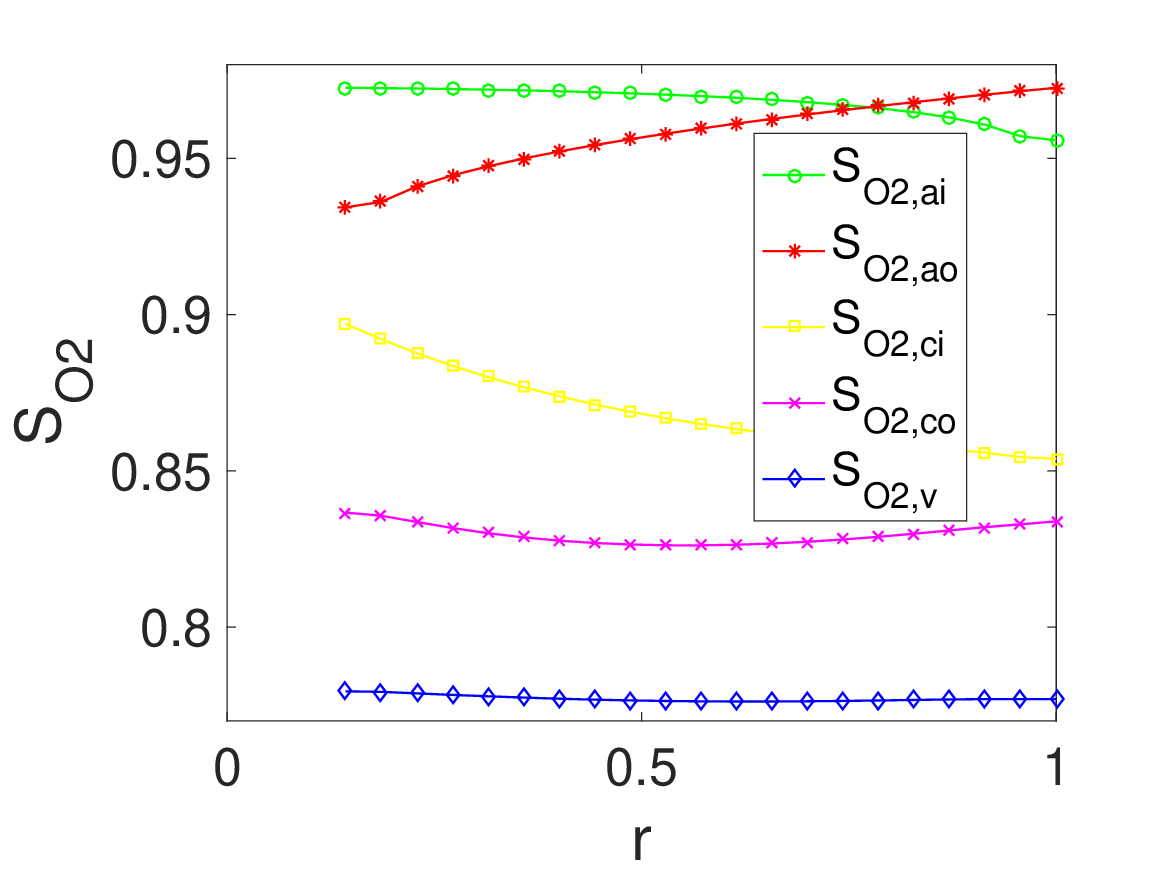}
\caption{The steady-state profiles of (a) oxygen binding capacity $H_k$ and (b) oxygen saturation $S_{O2,k}={C_{k}^{n}}/(C_{k}^{n}+ \tilde{C}_{50}^{n})$ in (\ref{eq47}) in the five vascular domains}
\label{fig9}
\end{center}
\end{figure}

Figure \ref{fig9}(a) shows the profiles of oxygen binding capacity  $H_k$ in vascular domains, indicating that they are almost a constant (i.e., $H_0$) for all five vascular domains, which verifies the argument that they can be assumed as a constant under the normal physiological conditions when water leak across blood vessel wall is small. Figure \ref{fig9}(b) shows the profiles of oxygen saturation $S_{O_2}$ in the five vascular domains, which are quite similar in shapes to those in Figure \ref{fig7}(d) since they almost differ by a scaling factor $H_0$.  At a fixed location, there is oxygen saturation drop (about 20 \%) from artery to vein through the capillary, due to capillary oxygen exchanges by diffusion (see Figure \ref{fig8}(c)).  For example, at $r=0.53$, the oxygen saturation drops by 19\%  (from 97\% to 78\%) from artery $\Omega_{ai}$  to vein $\Omega_v$ through capillary $\Omega_{ci}$. When it is multiplied by $H_{ci}$, the bound-oxygen drops by 12.8, more than 20 times the drop of the dissolved oxygen 0.55  (from 0.97 to 0.42). The oxygen supply is mainly released from the bound oxygen stored in the hemoglobin in RBCs.

In summary, for the reference case with uniform distribution of resting volume fractions, the blood supply and oxygen delivery is sufficient and stable from two sets of vasculature network, and the pathways are as expected. 

%The spatial drop of $S_{O_2}$ in artery domains is also mainly due to the diffusion of oxygen through blood vessel wall. 

\section{Effects  and sensitivity of parameters}

\subsection{Effects of resting volume fractions}

The resting volume fractions $\eta_j^{re}(r)$ in (\ref{eq43}) depends on the structural information of blood vessels. Without detailed information, we assumed uniform (constant) profile for $\eta_j^{re}(r)$ in the previous section. We now consider two cases with different volume fraction profiles.

In the first case, we assume a Gaussian profile for $\eta_j^{re}(r)$, called Gaussian case 1 later. On the one hand, when the main branch of blood vessel divides into two sub-branches, the total cross sectional area of blood vessel will increase (e.g., 1.2 fold), so the volume fractions will increase according to the level of branches; on the other hand, some branches will terminate at certain length, so the volume fractions will decrease after certain branching level.  We assume that the two artery resting volume fractions follow Gaussian profile with different mean and standard deviation
\begin{equation}
\label{eq54}
\begin{aligned}
&\eta_{ai}^{re}(r)\sim \mathcal{N}(0.4,0.2), \quad \eta_{ao}^{re}(r)\sim \mathcal{N}(0.7,0.2),
\end{aligned}
\end{equation}
roughly speaking the maximum reaches at around the 1/3 of the domain from the start of artery.  Suppose the capillaries and the vein at resting state are responding to the profiles of the two sets of arteries
\begin{equation}
\label{eq55}
\begin{aligned}
& \eta_{ci}^{re}\sim \mathcal{N}(0.4,0.2), \quad \eta_{co}^{re} \sim \mathcal{N}(0.7,0.2),\quad \eta_v^{re} \sim \eta_{ci}^{re} + \eta_{co}^{re}.
\end{aligned}
\end{equation}
A scaling constant will be multiplied on the profiles of each volume fraction $\eta_j$ ($j=ai,ao,ci,co,v$)  to ensure the weighted average value over the whole domain is the same as those in the uniform case.

\begin{figure}[h]
\begin{center}
\includegraphics[width=0.45 \textwidth]{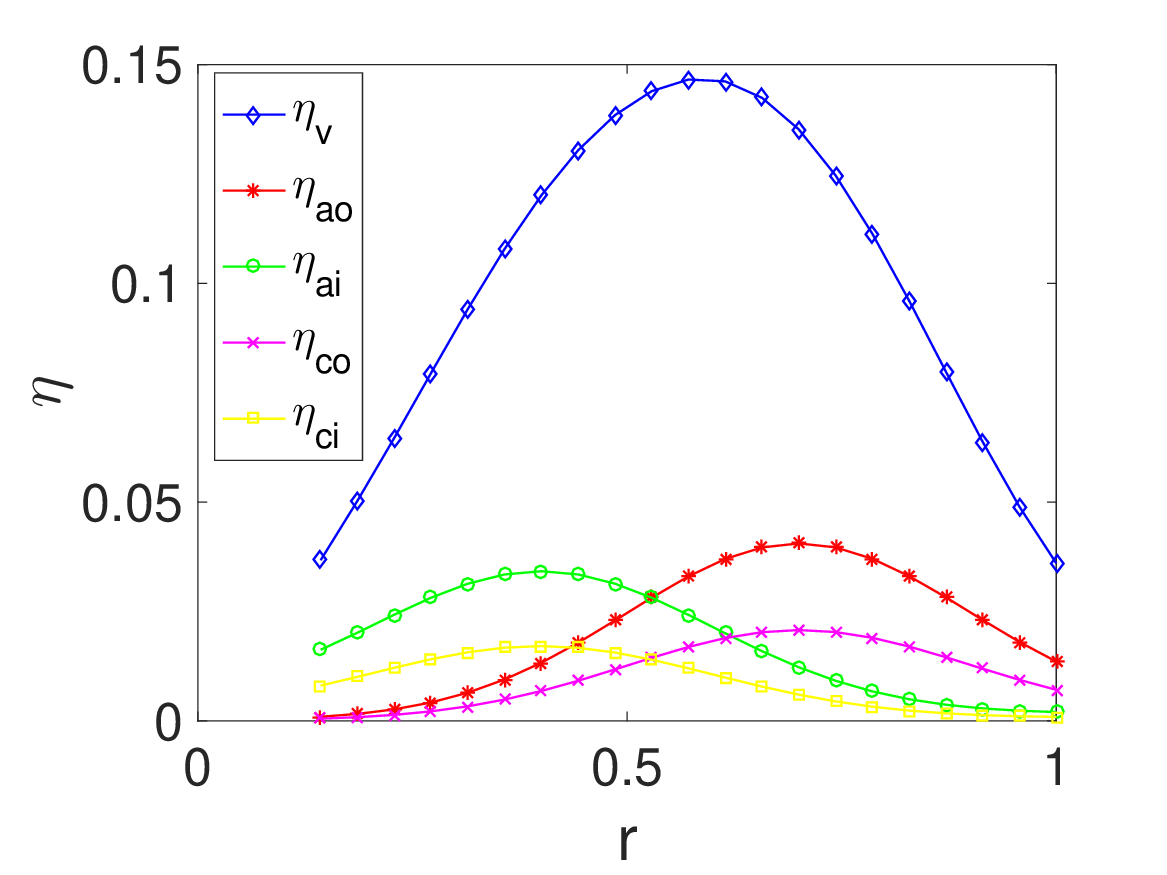}
\includegraphics[width=0.45 \textwidth]{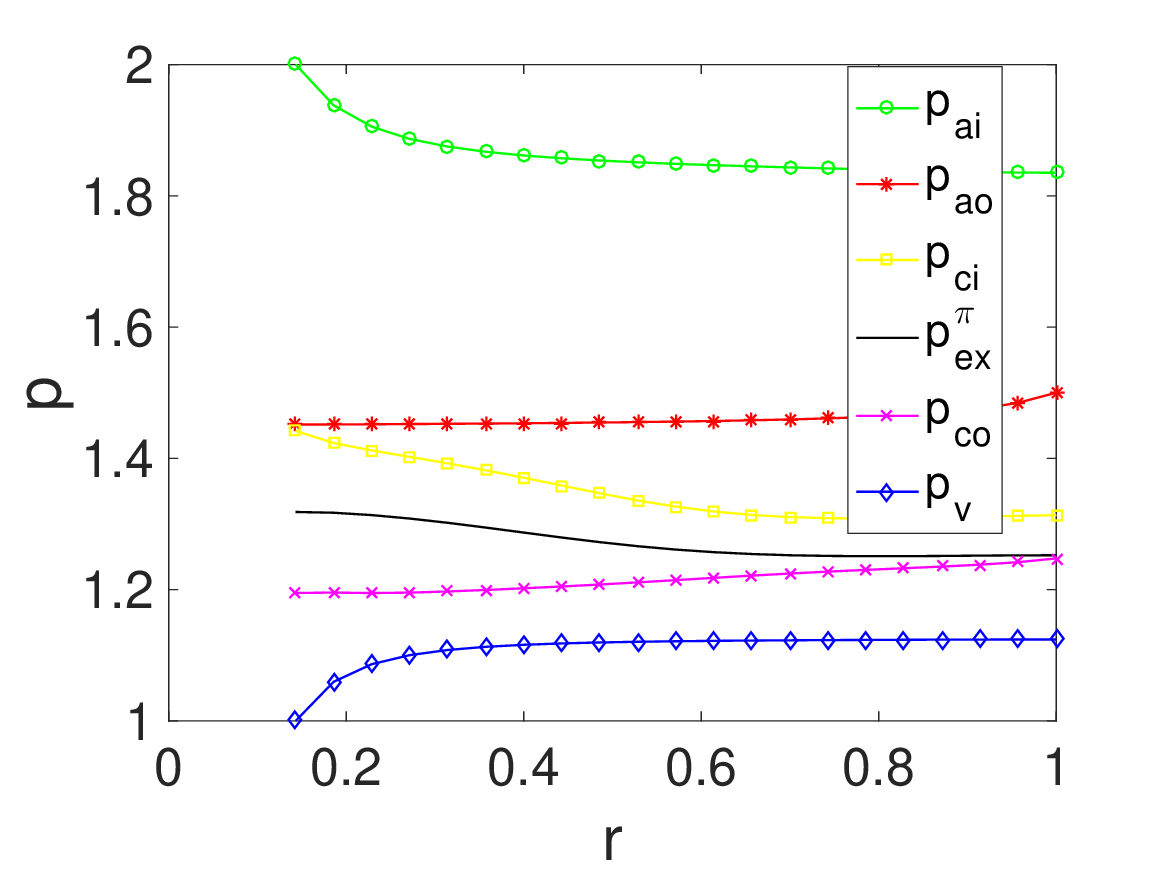}
\caption{Profiles of volume fractions and pressures at steady state in Gaussian case 1.}
\label{fig10}
\end{center}
\end{figure}

\begin{figure}
\begin{center}
\includegraphics[width=0.45 \textwidth]{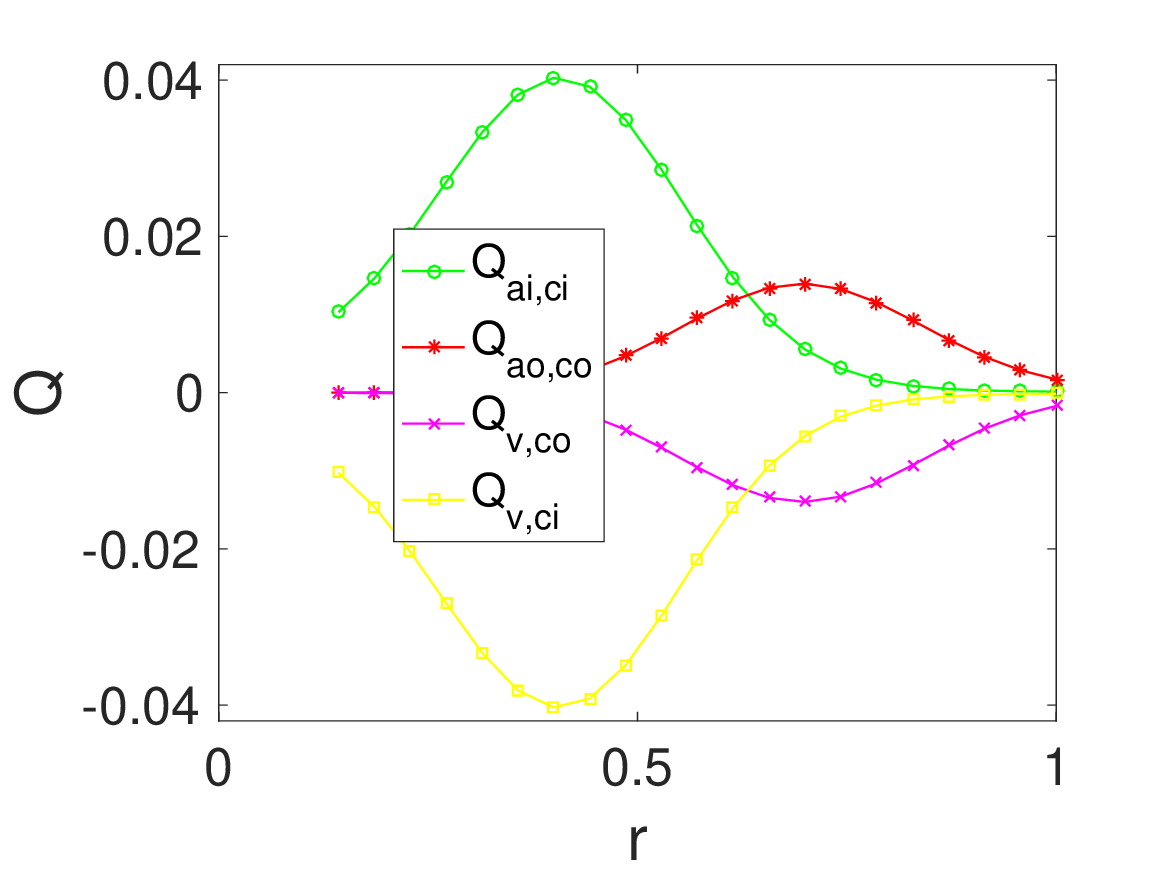}
\includegraphics[width=0.45 \textwidth]{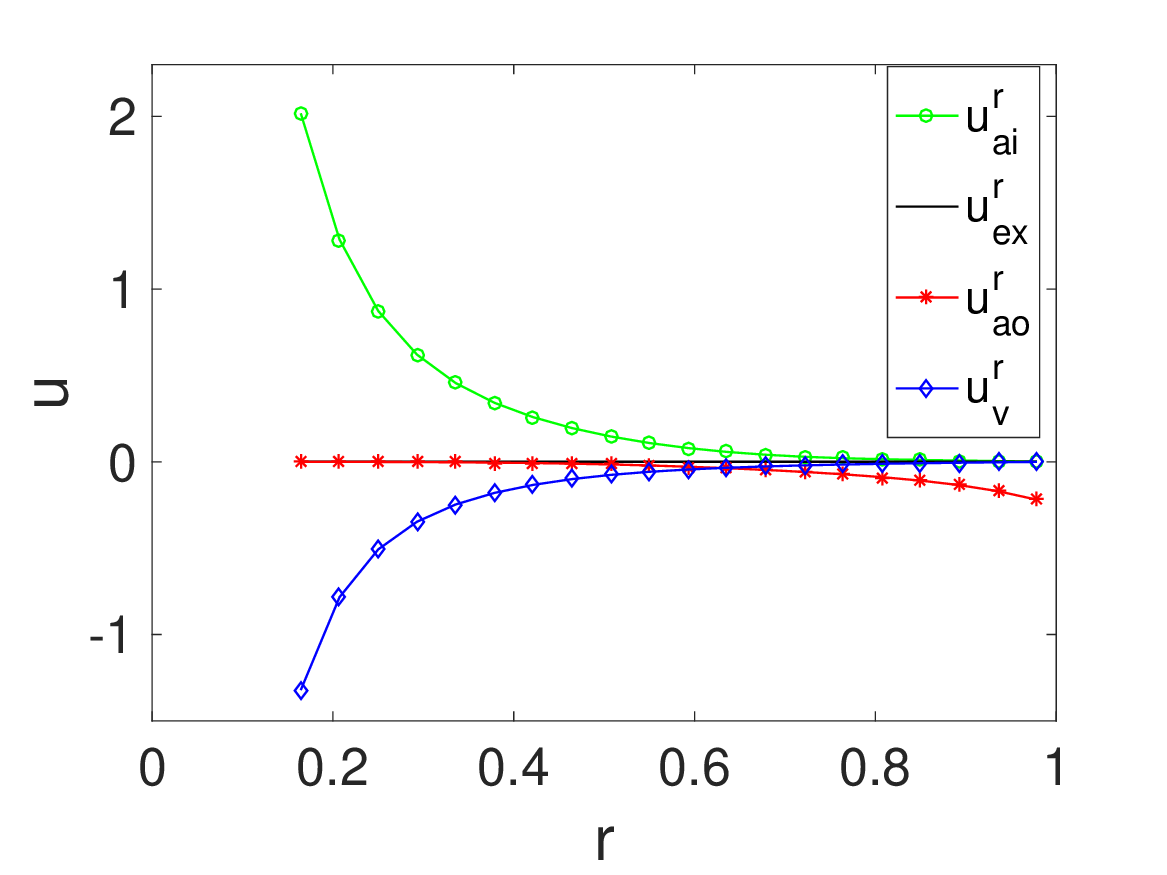}
\caption{The flow rates and velocities in the Gaussian case 1}
\label{fig11}
\end{center}
\end{figure}

The volume fractions $\eta_j$ at steady state (at $t=200$) are shown in Figure \ref{fig10}(a), which follow similar resting Gaussian profiles for $\eta_j^{re}$, since the difference between them is quite small (with max around $3*10^{-3}$) due to the large moduli $\tilde{\lambda}_j$. Figure \ref{fig10}(b) shows the pressure profiles, where one major difference from the uniform case is that the total pressure drop in artery domain $\Omega_{ai}$ is smaller but that in vein domain $\Omega_{v}$ is larger.

Figure \ref{fig11}(a) shows the flow rates between capillary domains and other vascular domains, which follow similar Gaussian profiles as the volume fractions in each capillary domain, e.g., the maximum values are also aligned at similar spatial locations $r=0.4,0.7$.  The in-domain velocities in Figure \ref{fig11}(b) follow similar trend as the uniform case, but the velocity in vein $\Omega_v$ is larger at $r=R_0$ since the volume fraction (equivalently total cross-sectional area) $\eta_v$ is much smaller at $r=R_0$ than that in the uniform case.  Table \ref{table2} shows the average values of these flows rates in Figure \ref{fig11}(a) and the two artery boundary flow rates defined in (\ref{eq43_1}). The change in average values is partly due to the fact that $\Omega_{ao}$ contributes more to inner region (small $r$) while $\Omega_{ai}$ contributes more to outer region. The blood supply from the $\Omega_{ai}$ boundary (i.e., the CRA) is significantly increased,  while that from the $\Omega_{ao}$ boundary (i.e., the PCA) stays almost the same. Therefore, the resting volume fractions strongly affects the blood supply and exchanges in the vasculature.

% The total supply of blood flow from the two arteries at the boundary are at similar level....

\begin{table}[h]
\begin{center}
\begin{tabular}{|c|c|c|c|c|}
   \hline
 &  $Q_{ai,ci}, |Q_{v,ci}|$& $Q_{ao,co}, |Q_{v,co}|$ & $Q_{ai}^\ast$ & $Q_{ao}^\ast$\\
 \hline
Uniform Case & 0.0076& 0.0072& 0.0034& 0.0034\\
\hline
 Gaussian case 1 & 0.0164 & 0.0055 & 0.0060 & 0.0034\\
 \hline
  Gaussian case 2 & 0.0332 & 0.0068 & 0.0090 & 0.0038\\
 \hline
\end{tabular}
\end{center}
\caption{Average values of capillary flows rates over the whole region and the two artery boundary flow rates in three cases. }
\label{table2}
\end{table}

\begin{figure}[h]
\begin{center}
\includegraphics[width=0.45 \textwidth]{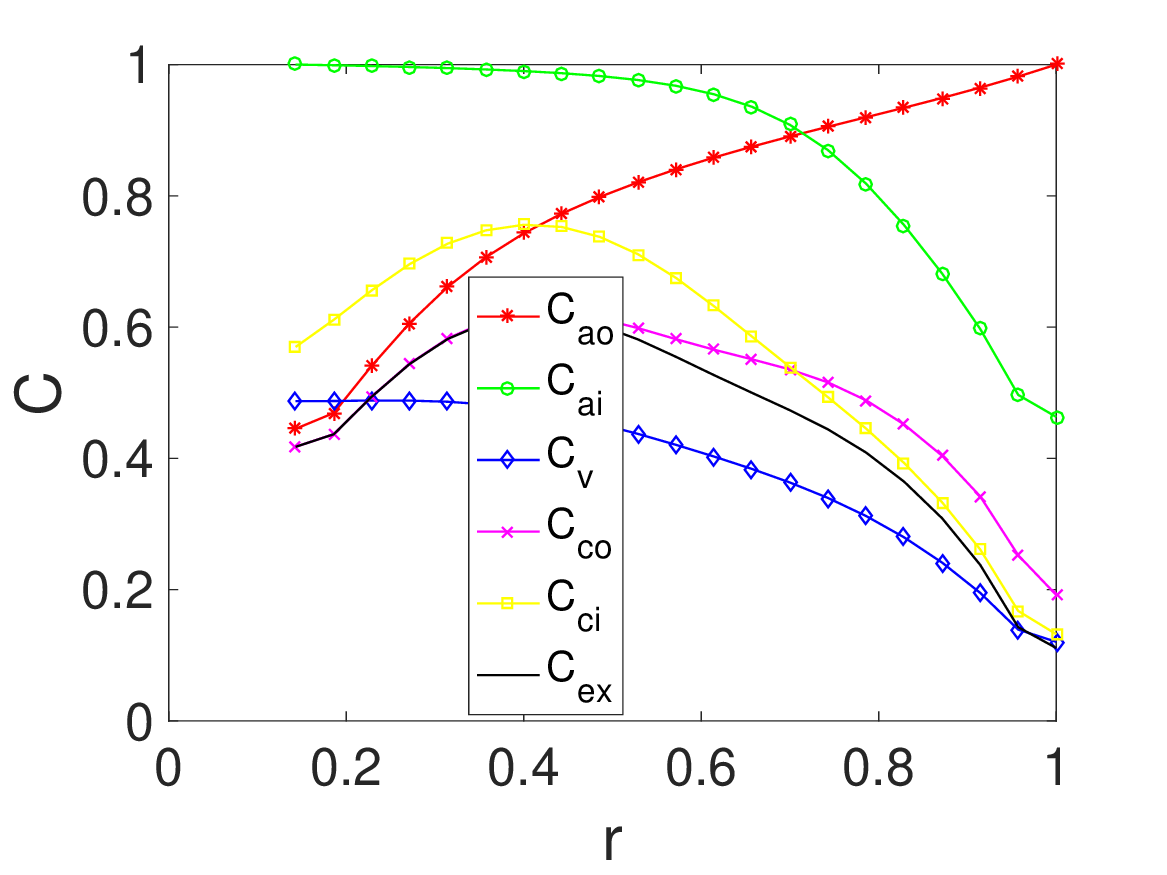}
\includegraphics[width=0.45 \textwidth]{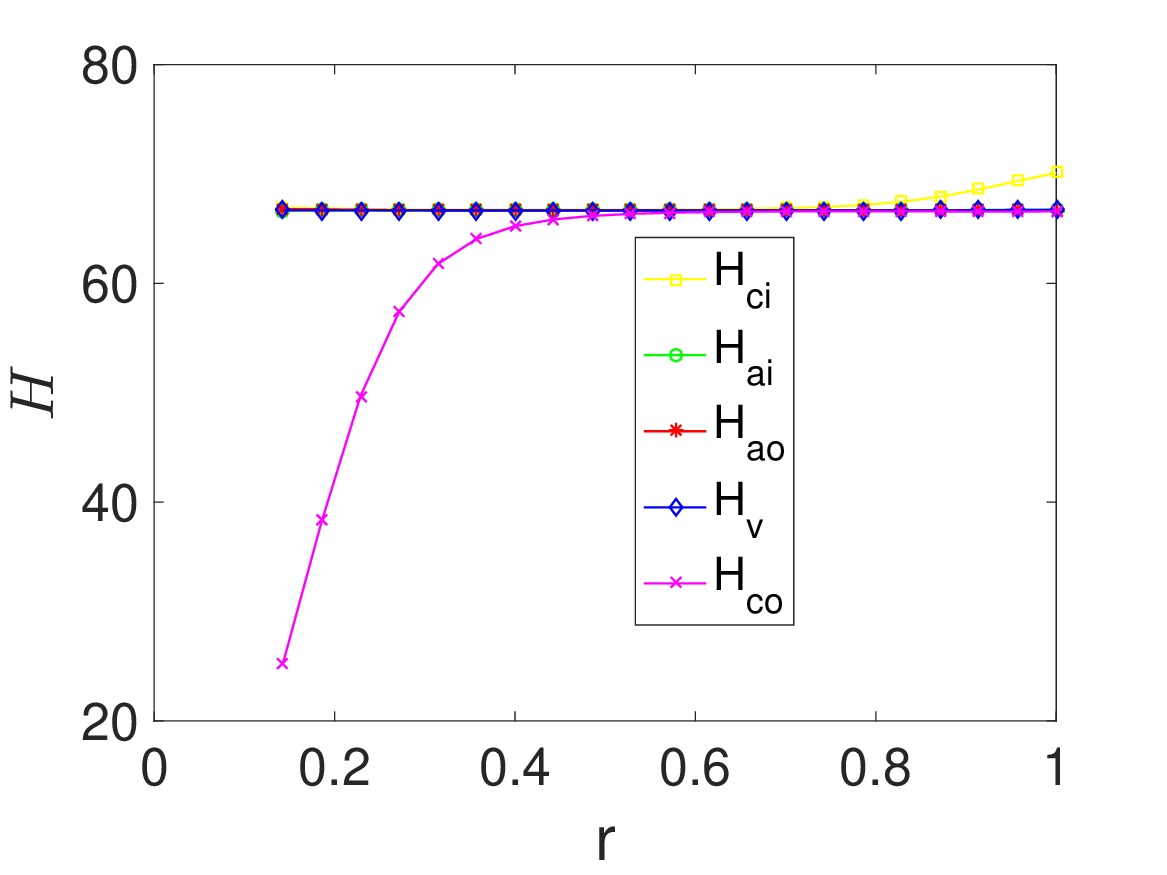}
\caption{The profiles of $C_k,H_k$ at steady state $t=200$, with Gaussian case 1.}
\label{fig12}
\end{center}
\end{figure}

For oxygen transport part, Figure \ref{fig12} shows the profiles of of $C_k$ and $H_k$ at steady state ($t=200$). Figure \ref{fig12}(a) shows larger variations for the concentrations compared with the uniform case, where the low concentration of $C_{ex}$ near $r=1$ indicate insufficient supply of oxygen there. Figure \ref{fig12}(b) shows that $H_k$ in arteries and vein are almost the normal constant, but the $H_{ci},H_{co}$ in capillary domains near boundary has significant variations since the water leak is comparable with the small blood exchanges there.
% It reaches the maximum at the intersection point between the curves of $C_{ex}$ and $C_{v}$. 
% The total oxygen concentrations $\bar{C}_k$ and the oxygen saturations $S_{O2,k}$ in Figure \ref{fig12}(b,d) follow similar trend as $C_k$ in Figure \ref{fig12}(a).

\begin{figure}[h]
\begin{center}
\includegraphics[width=0.45 \textwidth]{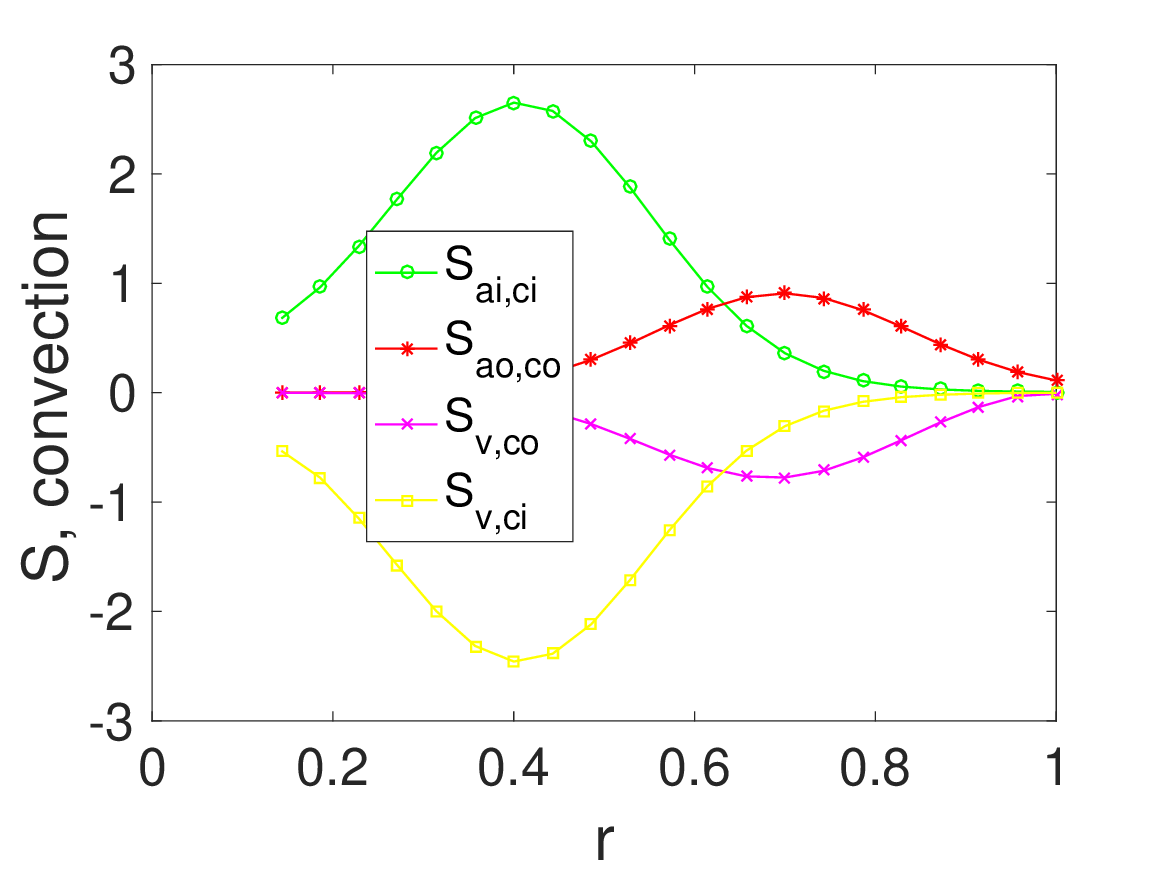}
\includegraphics[width=0.45 \textwidth]{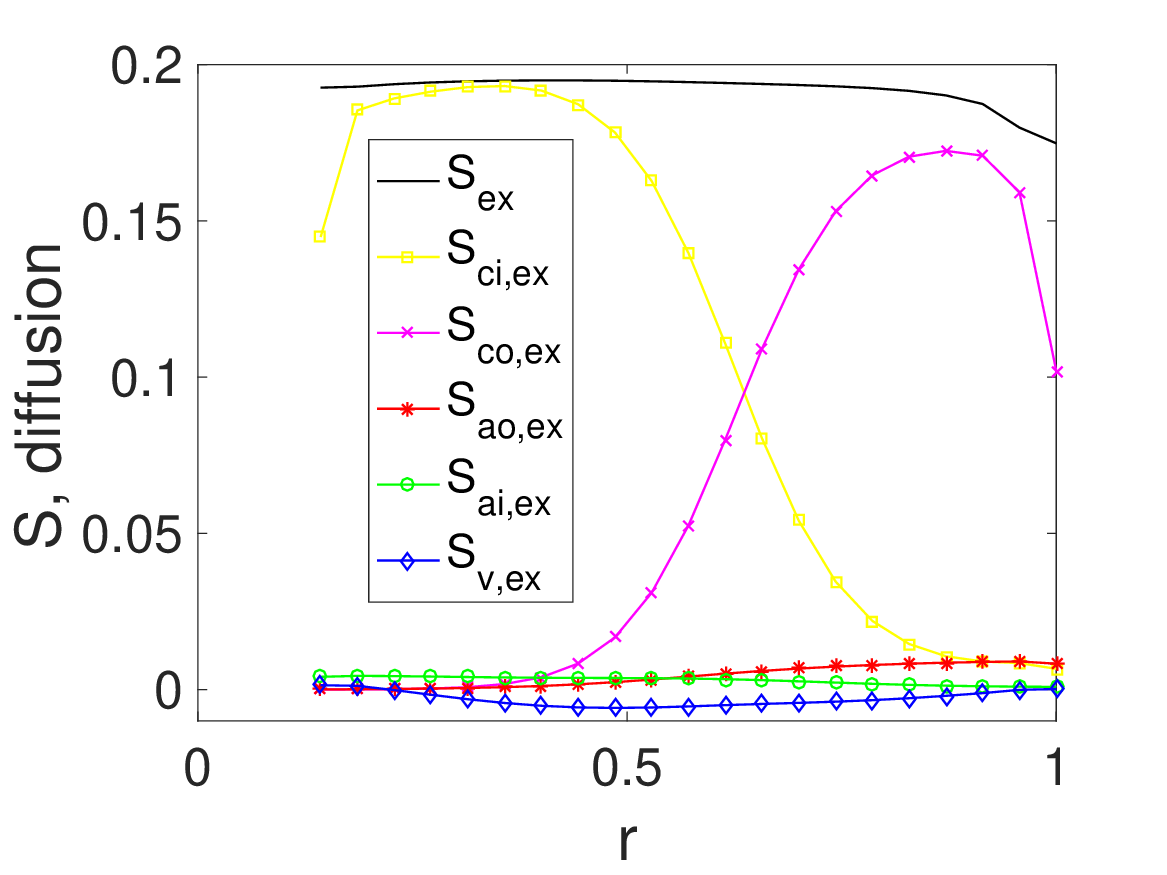}
\caption{The oxygen exchange rates in Gaussian case 1. }
\label{fig13}
\end{center}
\end{figure}

The pathways of oxygen supply to the extravascular domain are similar to the previous uniform case, but Figure \ref{fig13} shows larger spatial variations for oxygen exchange rates in different domains.  The oxygen exchange in Figure \ref{fig13}(a) between capillary and other vascular domains follow Gaussian-like profiles, since it is convection dominant (see Figure \ref{fig11}(a)). Figure \ref{fig13}(b) shows  that the two capillary sets due to terms $S_{ci,ex},S_{co,ex}$ serve different regions, one for inner and the other for outer region, and their sum is almost the total oxygen consumption $S_{ex}$. The total oxygen consumption $S_{ex}$ has a dip of about 10\% near $r=1$, which means insufficient supply of oxygen in that region, consistent with small $C_{ex}$ in Figure \ref{fig12}(a).

\begin{figure}
\begin{center}
\includegraphics[width=0.45 \textwidth]{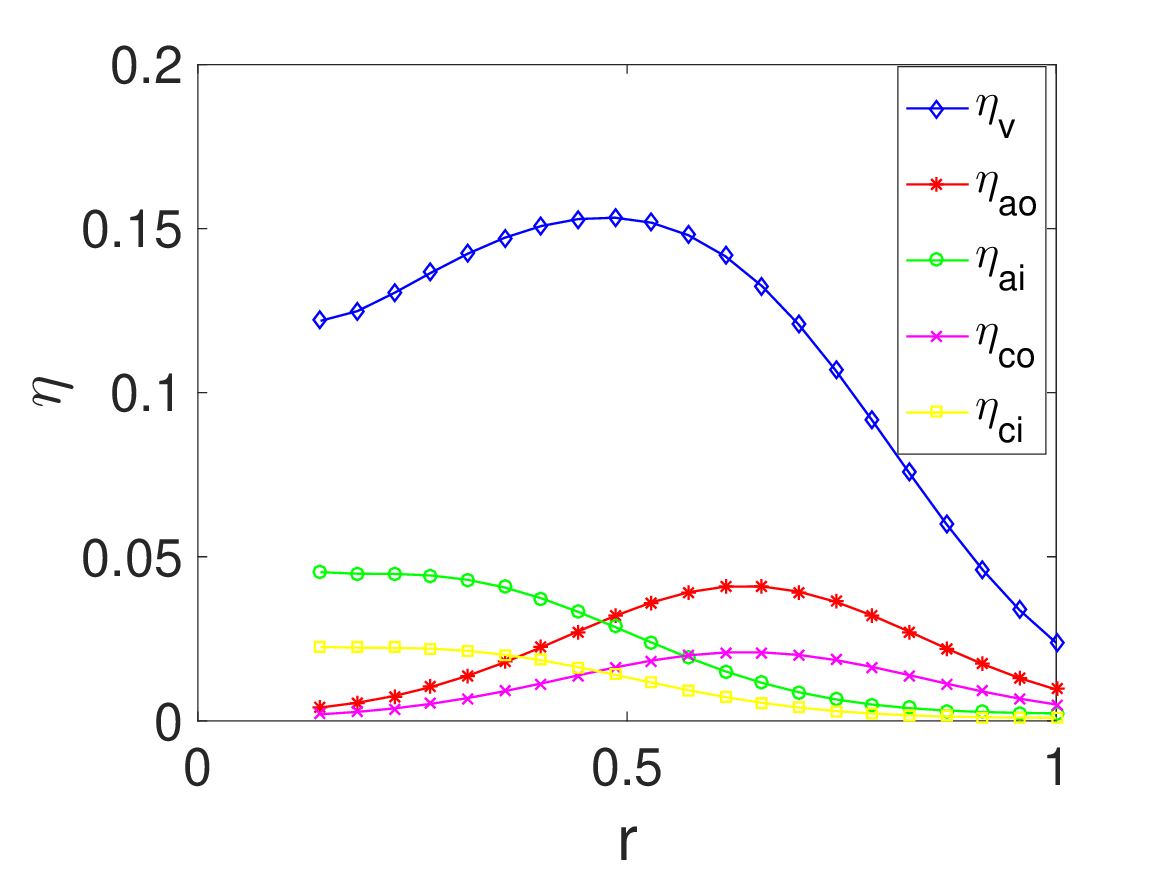}
\includegraphics[width=0.45 \textwidth]{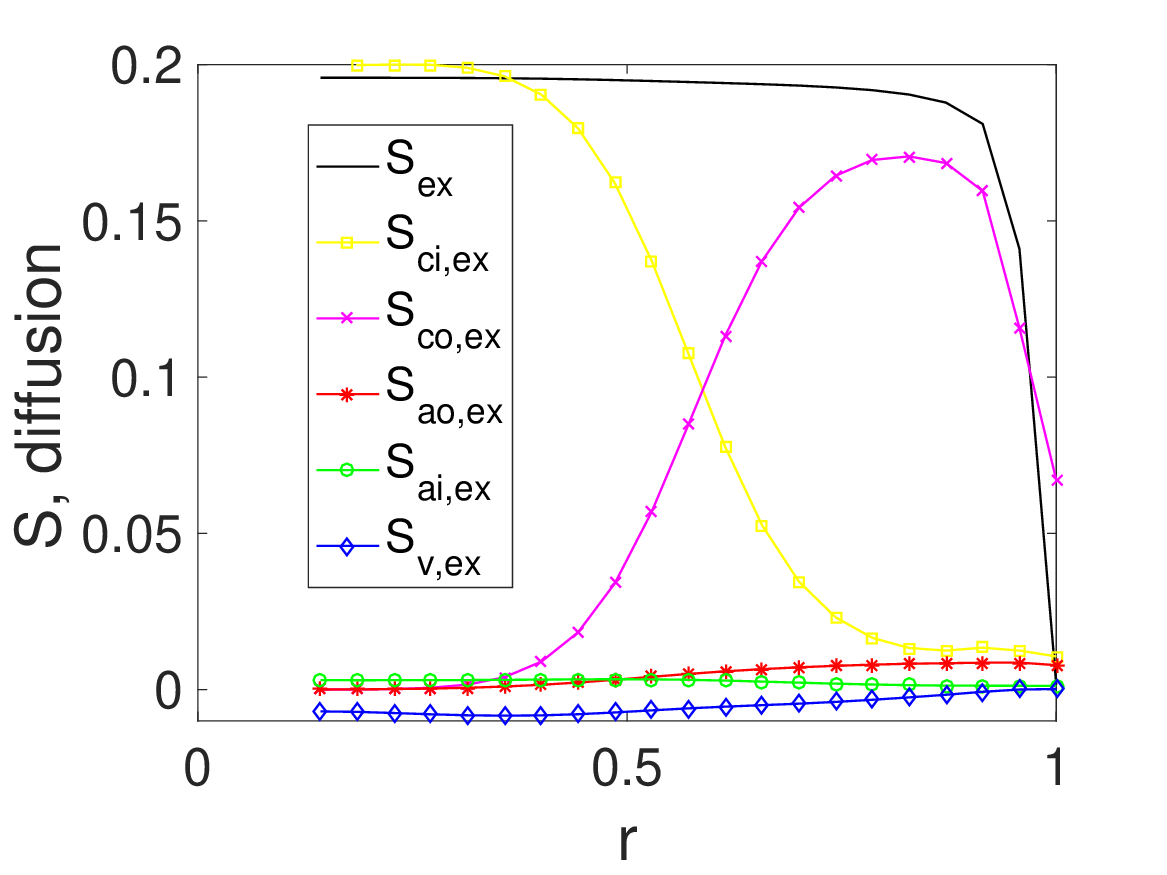}
\caption{The profiles of volume fractions and oxygen exchange rates in Gaussian case 2. }
\label{fig13_1}
\end{center}
\end{figure}

In the second case, called Gaussian case 2, we consider
\begin{equation}
\label{eq54_1}
\begin{aligned}
&r\eta_{ai}^{re}(r)\sim \mathcal{N}(0.4,0.2), \quad r \eta_{ao}^{re}(r)\sim \mathcal{N}(0.7,0.2),\\
&r \eta_{ci}^{re}\sim \mathcal{N}(0.4,0.2), \quad r \eta_{co}^{re} \sim \mathcal{N}(0.7,0.2),\quad \eta_v^{re} \sim \eta_{ci}^{re} + \eta_{co}^{re}.
\end{aligned}
\end{equation}
which means the absolute values of cross-sectional areas of blood vessels follow the Gaussian profiles, where the extra factor $r$ is due to the polar coordinates used. The volume profiles at steady state are shown in Figure \ref{fig13_1}(a). The blood supply from boundary is also increased as shown in Table \ref{table2}, particularly for the $\Omega_{ai}$ boundary. The insufficiency of oxygen supply near $r=1$ is more severe in this case, shown in Figure \ref{fig13_1}(b). 

In the above two cases, we find that (1) the structural profiles of vasculature play essential roles in the overall blood supply and distribution of exchanges,  and (2)  although the total blood supply from boundary is increased with Gaussian cases, the supply of oxygen could still be insufficient in some local region due to the uneven exchanges. To simulate more practical situations, more information on the blood vessel sizes and distributions is needed.  Also, the maximum consumption parameter $S_{ex}^{max}$ could vary in space, e.g., depending ion channel/pump distributions, so the distribution of resting capillary volume fractions could relate to the biological need to ensure stable oxygen supply.

% The third case is ... two sets of capillary vs one set of capillary?? modeling need for two sets?

\subsection{Effects of leak coefficients}

In this subsection, we consider the effects of the leak coefficients $L_{j,ex}$ in (\ref{eq7}). The changes of these coefficients could be due to the changes of properties of blood vessels walls and could relate to damages from pathological conditions \cite{moyseyenko2023,lamb2007,guidoboni2020}. Since the capillary change/damage could occur more easily, we mainly focus on the coefficients $L_{ci,ex}, L_{co,ex}$ for illustration.

\begin{table}[h]
\begin{center}
\begin{tabular}{|c|c|c|}
   \hline
Parameters $L_{ci,ex}, L_{co,ex}$&  $40$ fold & $500$ fold \\
\hline
 \% change of $\Delta p_{ai}$, $Q_{ai}^\ast$  &1.5\%,1.6\% & 10.1\%,10.7\% \\
\hline
\% change of $\Delta p_{ao}$, $Q_{ao}^\ast$ & -0.8\%, -0.4\% & 3.6\%, 5.9\% \\
\hline
\% change of $\Delta p_{v}$, $Q_{v}^\ast$   &  0.5\%, 0.6\% & 7.7\%, 8.3\% \\
\hline
average $Q_{ci,ex},Q_{co,ex}$  & 32 fold  & 118 fold\\
\hline
\end{tabular}
\end{center}
\caption{Changes of pressure drops $\Delta p_{j}$, boundary flow rates $Q_{j}^\ast$ ($j=ai,ao,v$), and average water exchange rates by changing the parameters $L_{ci,ex}, L_{co,ex}$ simultaneously.}
\label{table3}
\end{table}

\begin{figure}[h]
\begin{center}
\includegraphics[width=0.45 \textwidth]{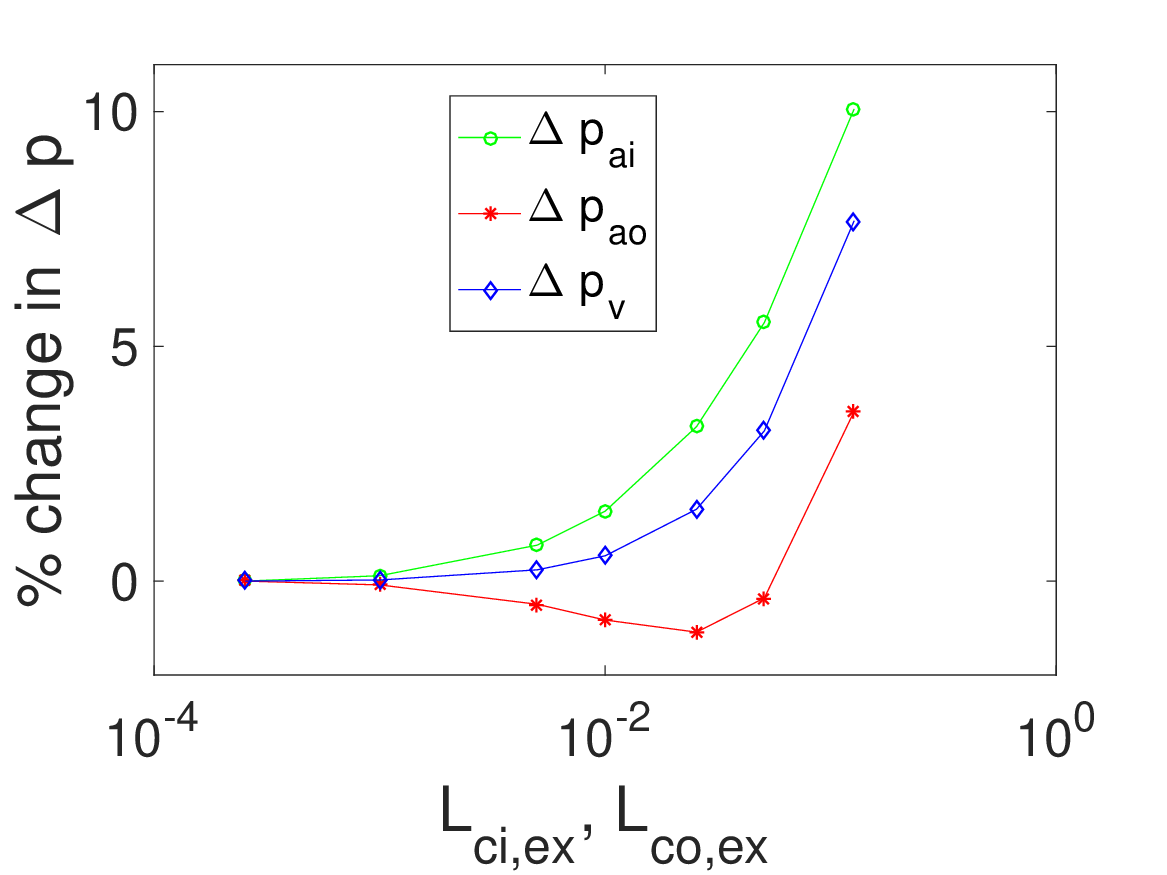}
\includegraphics[width=0.45 \textwidth]{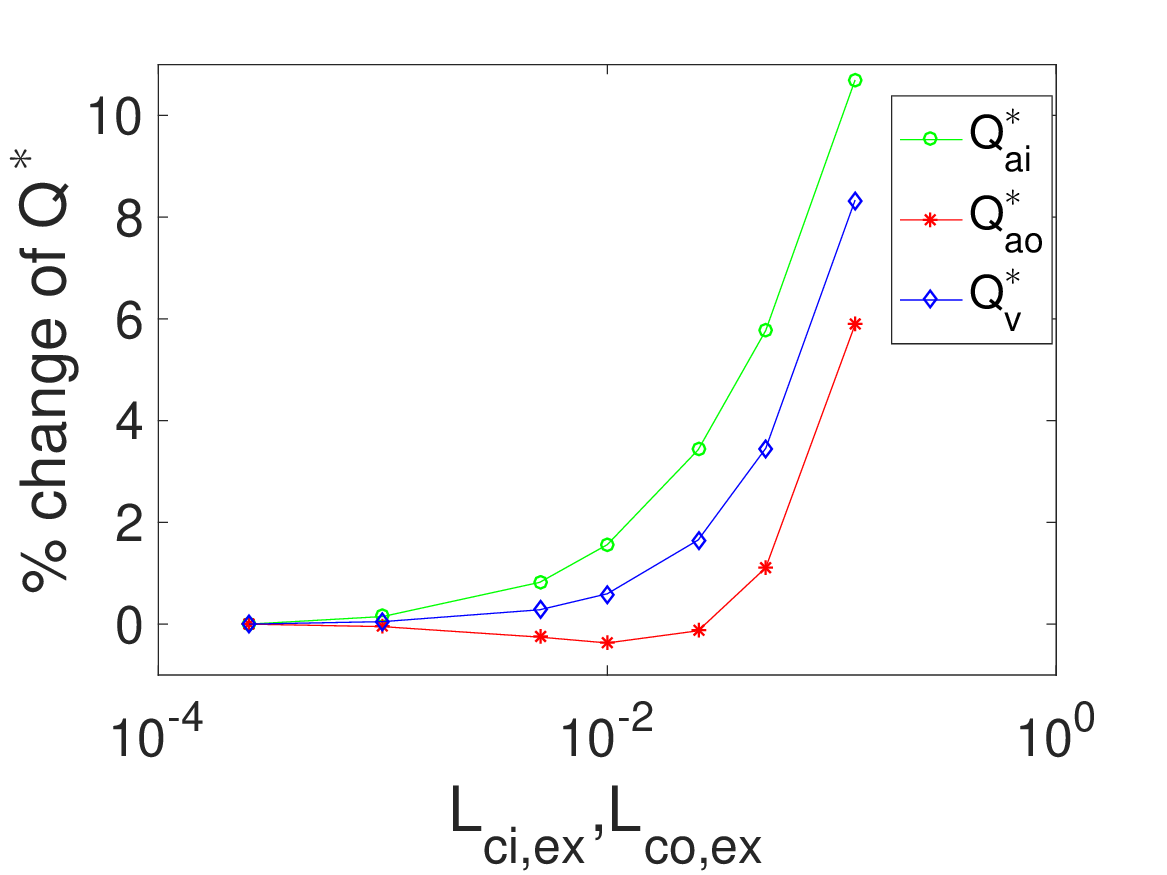}
\includegraphics[width=0.45 \textwidth]{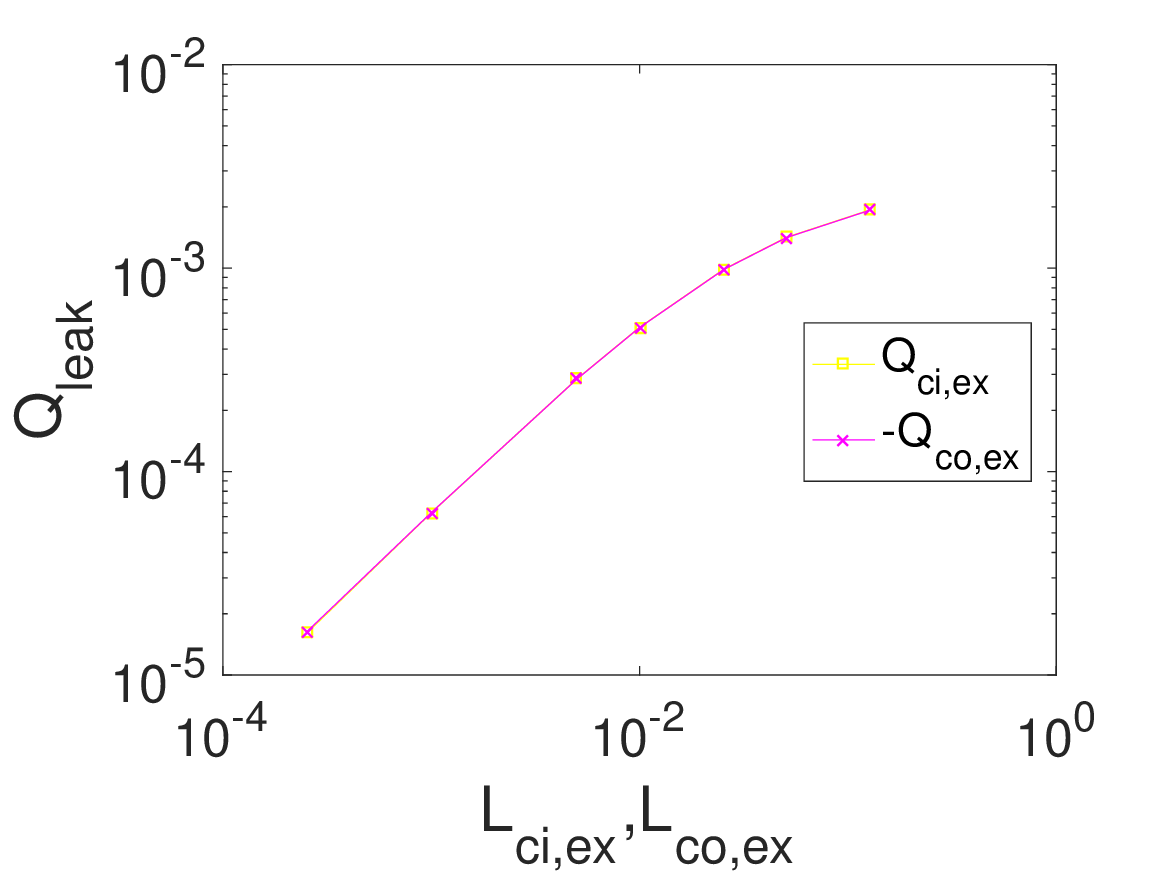}
\includegraphics[width=0.45 \textwidth]{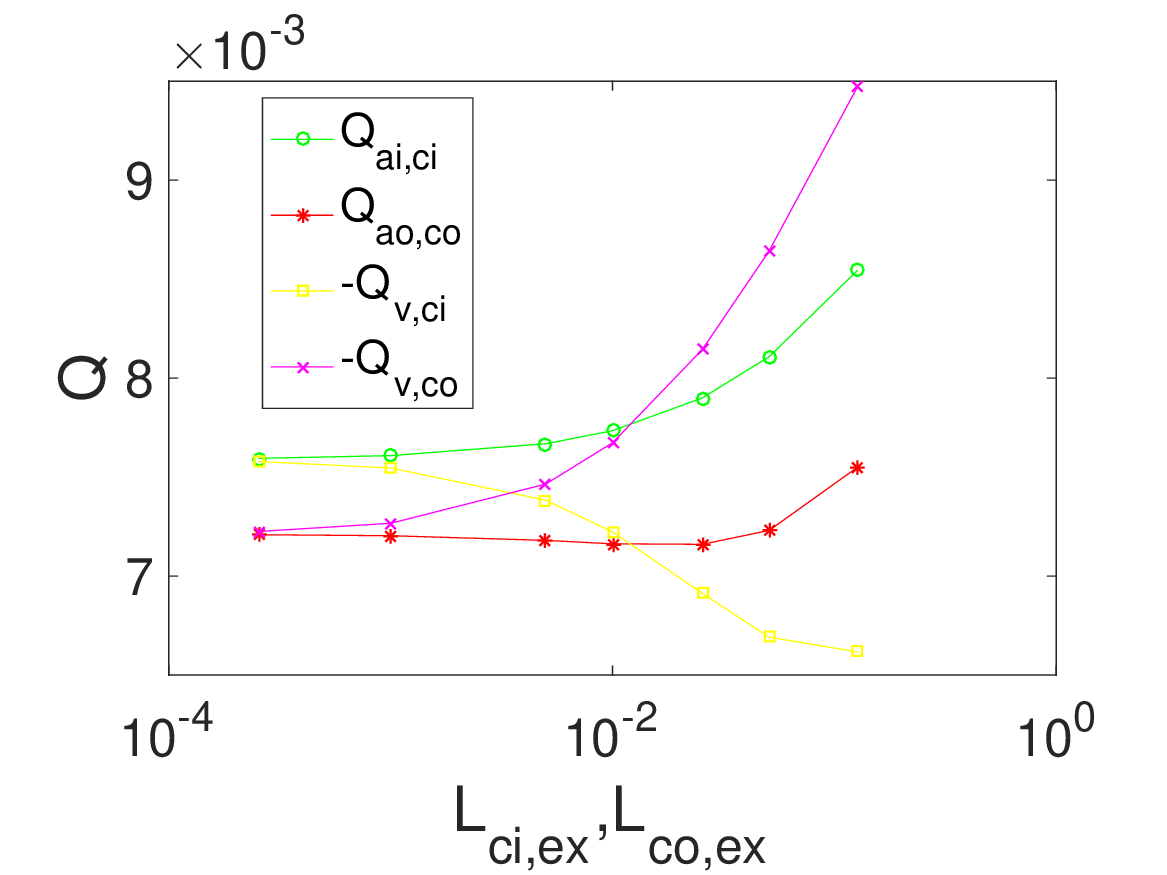}
\caption{The dependence of pressure drops, boundary flow rates, and average exchange rates on the leak coefficients $L_{ci,ex},L_{co,ex}$ with unit $\mu$m/(Pa s). }
\label{fig14}
\end{center}
\end{figure}

For water/blood circulation part, Table \ref{table3} and Figure \ref{fig14} show the changes of pressure drop and water flow rates by gradually increasing $L_{ci,ex},L_{co,ex}$ simultaneously. We find
\begin{itemize}
\item There are more significant changes for pressure drops $\Delta p_k= |p_k(R_0)-p_k(R_1)|$ and boundary fluxes $Q_k^\ast$ ($k=ai,ao,v$) after the parameters $L_{ci,ex},L_{co,ex}$ reach some threshold (e.g., at the value of $10^{-2}$ $\mu$m/(Pa s); about 40 fold of original value), shown in Figure  \ref{fig14}(a,b). This is when the water exchange rates $Q_{ci,ex},Q_{co,ex}$ become comparable with other flow rates between different domains as shown in Figure \ref{fig14}(c,d).  For example, Table \ref{table3} shows that the boundary flux $Q_{ai}^\ast$ from CRA increases by 1.6 \% with 40 fold parameter increase, but increases by 10.7\% with 500 fold parameter increase.
\item There is an additional major pathway for water circulation after the threshold
\begin{equation}
\label{eq56}
\begin{aligned}
\Omega_{ci} \rightarrow \Omega_{ex} \rightarrow \Omega_{co},\quad  Q_{ci,ex}>0, \quad Q_{co,ex}<0,
\end{aligned}
\end{equation}
which is based magnitudes and signs of quantities in Figure \ref{fig14}(c,d), e.g, $Q_{ci,ex}>0, |Q_{ai,ci}|>|Q_{v,ci}|, |Q_{ao,co}|<|Q_{v,co}|$. Table \ref{table3} and Figure \ref{fig14}(c) show that there is almost proportional change of $Q_{ci,ex},Q_{co,ex}$ with respect to the parameter change initially before the threshold, since they are small and unaffected by other flow rates.  After the threshold when they are comparable with other flow rates in Figure \ref{fig14}(d),  their changes will interact with other flow rates and the additional pathway will play important roles on the blood/water circulation.
\end{itemize}

%Because the pressures $p_{ci},p_{ex}^\pi,p_{co}$ are closer to each other with large leak parameter, so the increase of $Q_{ci,ex},Q_{co,ex}$ are less significant. 

\begin{figure}[h]
\begin{center}
\includegraphics[width=0.45 \textwidth]{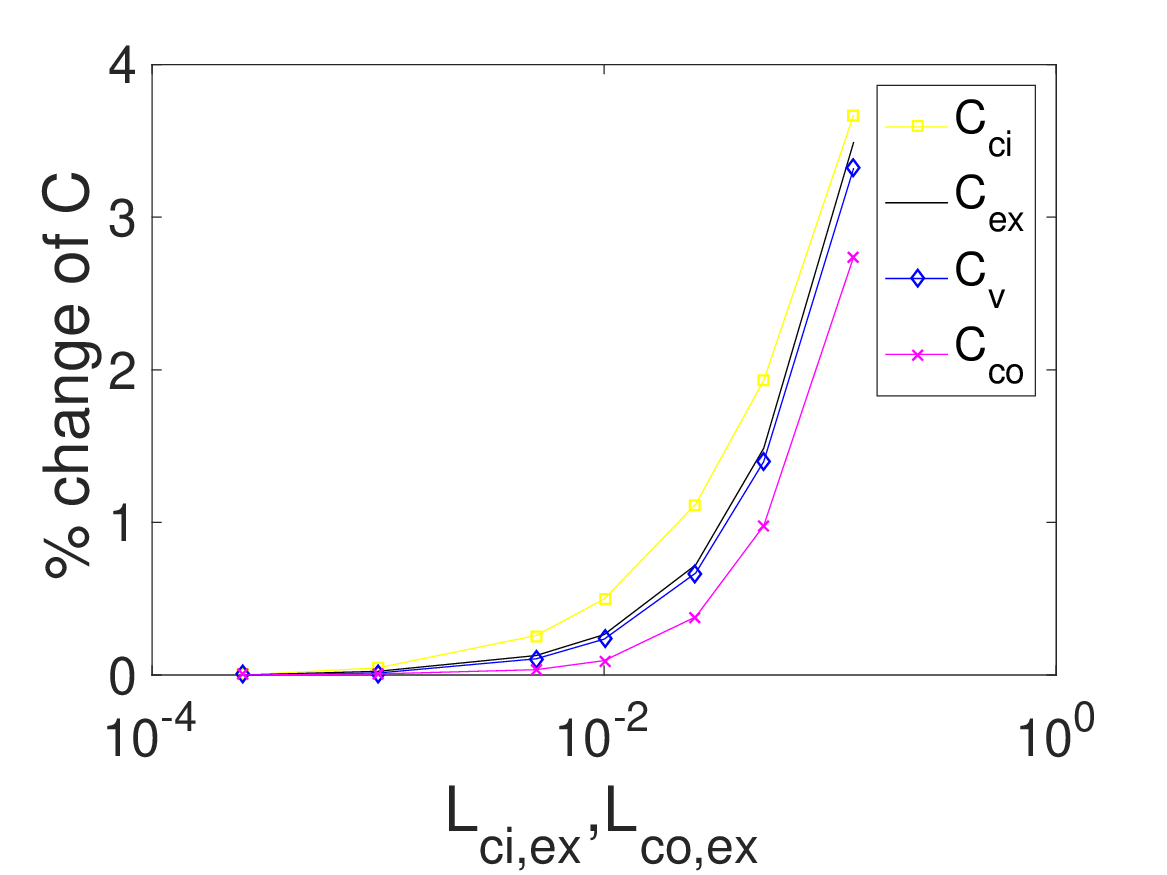}
\includegraphics[width=0.45 \textwidth]{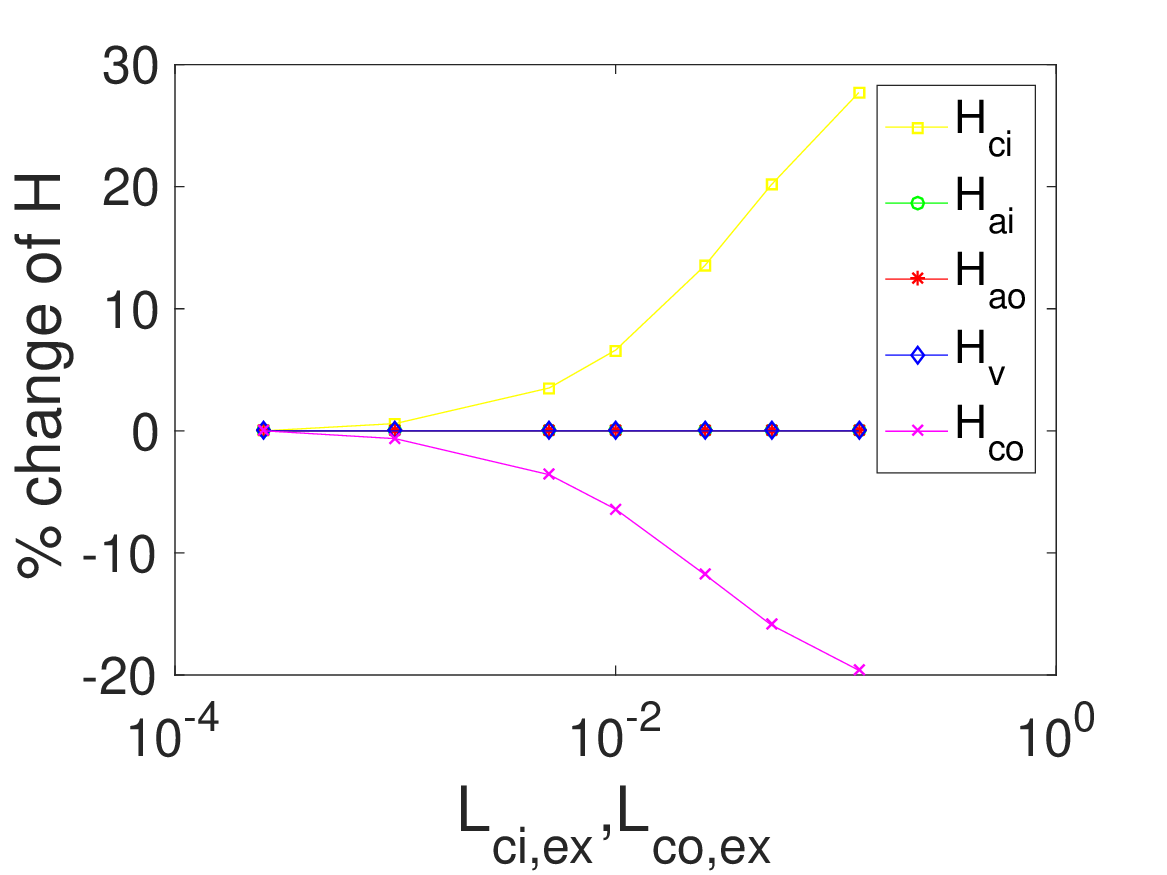}
\caption{Percentage changes of average concentrations $C_k$ and average oxygen binding capacities $H_k$ by changing the leak coefficients $L_{ci,ex},L_{co,ex}$. }
\label{fig15}
\end{center}
\end{figure}

\begin{figure}[h]
\begin{center}
\includegraphics[width=0.45 \textwidth]{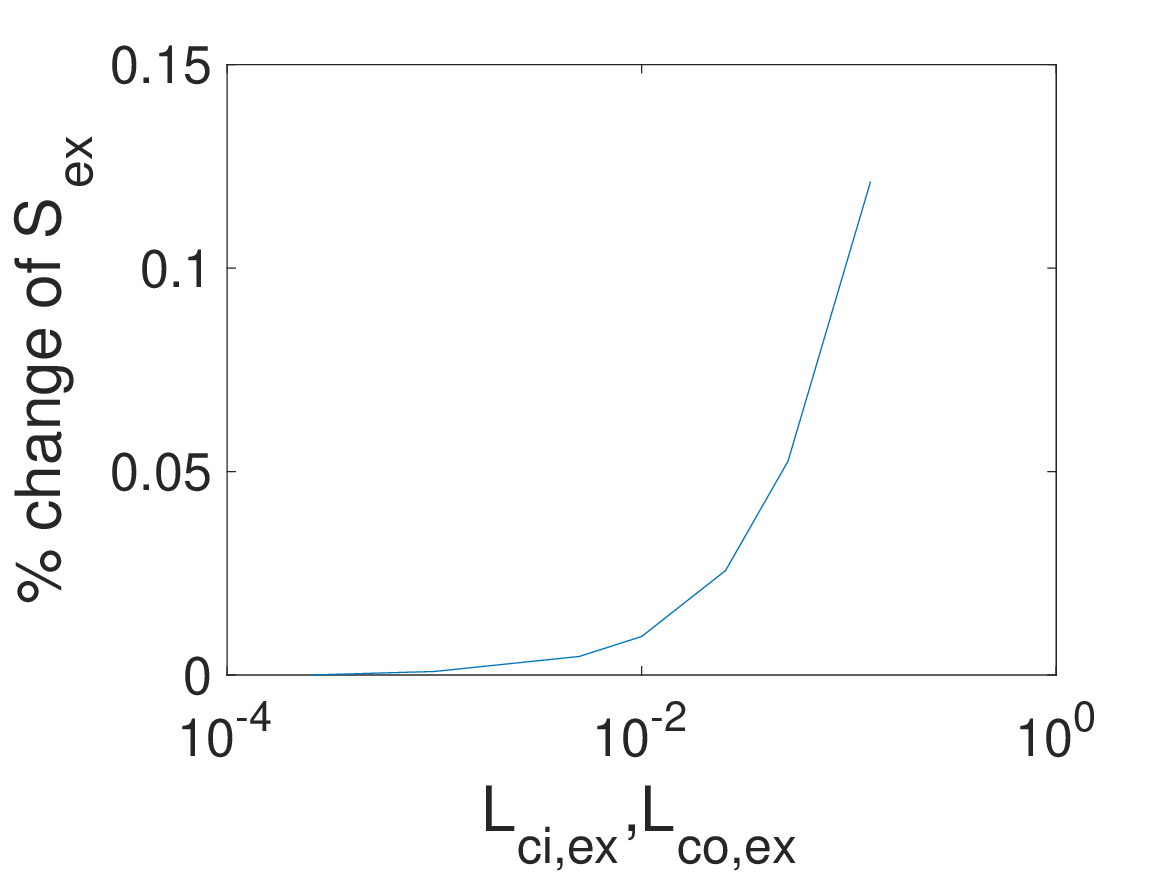}
\includegraphics[width=0.45 \textwidth]{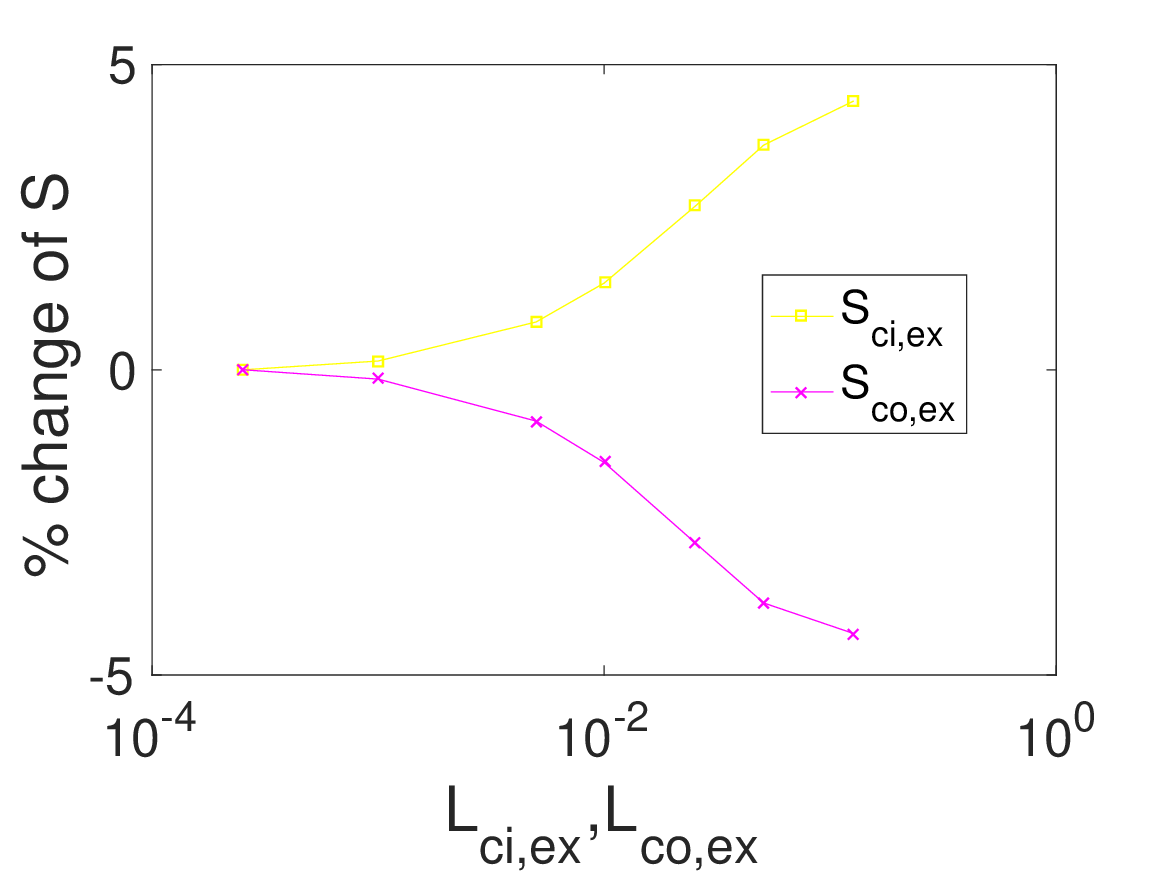}
\caption{Percentage changes of average oxygen consumption $S_{ex}$ and the oxygen supply from two capillary domains by changing the leak coefficients $L_{ci,ex},L_{co,ex}$.}
\label{fig16}
\end{center}
\end{figure}

For the oxygen transport part, we observe the following:
\begin{itemize}
\item Figure \ref{fig15}(a) shows that there are significant increases for the average concentrations $C_{k}$ ($k=ci,co,ex,v$) after some threshold for the parameters $L_{ci,ex},L_{co,ex}$ (e.g., $10^{-2}$ $\mu$m/(Pa s)), which is a consequence of the increased boundary flow rates in Figure \ref{fig14}(b).   Figure \ref{fig15}(b) shows that oxygen binding capacity $H_{ci}$ increases and $H_{co}$ decreases significantly after the threshold, as a consequence of the additional pathway of water circulation in (\ref{eq56}).
\item Figure \ref{fig16}(a) shows that the oxygen consumption is kept at steady level, with less than $0.15\%$ change over the studied range of parameters $L_{ci,ex},L_{co,ex}$. Figure \ref{fig16}(b) show that the oxygen supply from capillary domain $\Omega_{ci}$ increases while that from capillary domain $\Omega_{co}$ decreases (so the sum kept almost constant), as a consequence of the additional pathway of water circulation in (\ref{eq56}).
%This is because $S_{ex}$ is close to maximum value $S_{ex}^{max}$ with relatively stable $C_{ex}$, as in Figure \ref{fig6}(b). 
\end{itemize}

In summary, the water leak coefficients through capillary wall  will have strong impacts on blood circulation after it passes a threshold, due to additional pathway of water circulation. This in turn affects oxygen delivery, with redistributed supply from the two
sets of capillaries.

{\bf  Remark:} The effects of leak coefficients $L_{ai,ex}, L_{ao,ex},L_{v,ex}$ in arteries and vein have very similar features as the above case, except the following differences. With simultaneous changes of these three parameters, the threshold is smaller and at about $10^{-3}$ $\mu$m/(Pa s) since the pressure differences across blood vessel wall of artery and vein is larger. The additional pathway for water circulation is
\begin{equation}
\label{eq57}
\begin{aligned}
\Omega_{ai},\Omega_{ao}\rightarrow \Omega_{ex} \rightarrow \Omega_{v}, \quad Q_{ai,ex}>0, ~~ Q_{ao,ex}>0, ~~ Q_{v,ex}<0.
\end{aligned}
\end{equation}
For the oxygen transport part, the quantities $H_k$ in different domains are affected due to  (\ref{eq57}).

\subsection{Effects of blood viscosity $\mu_b$}

\begin{figure}[h]
\begin{center}
\includegraphics[width=0.45 \textwidth]{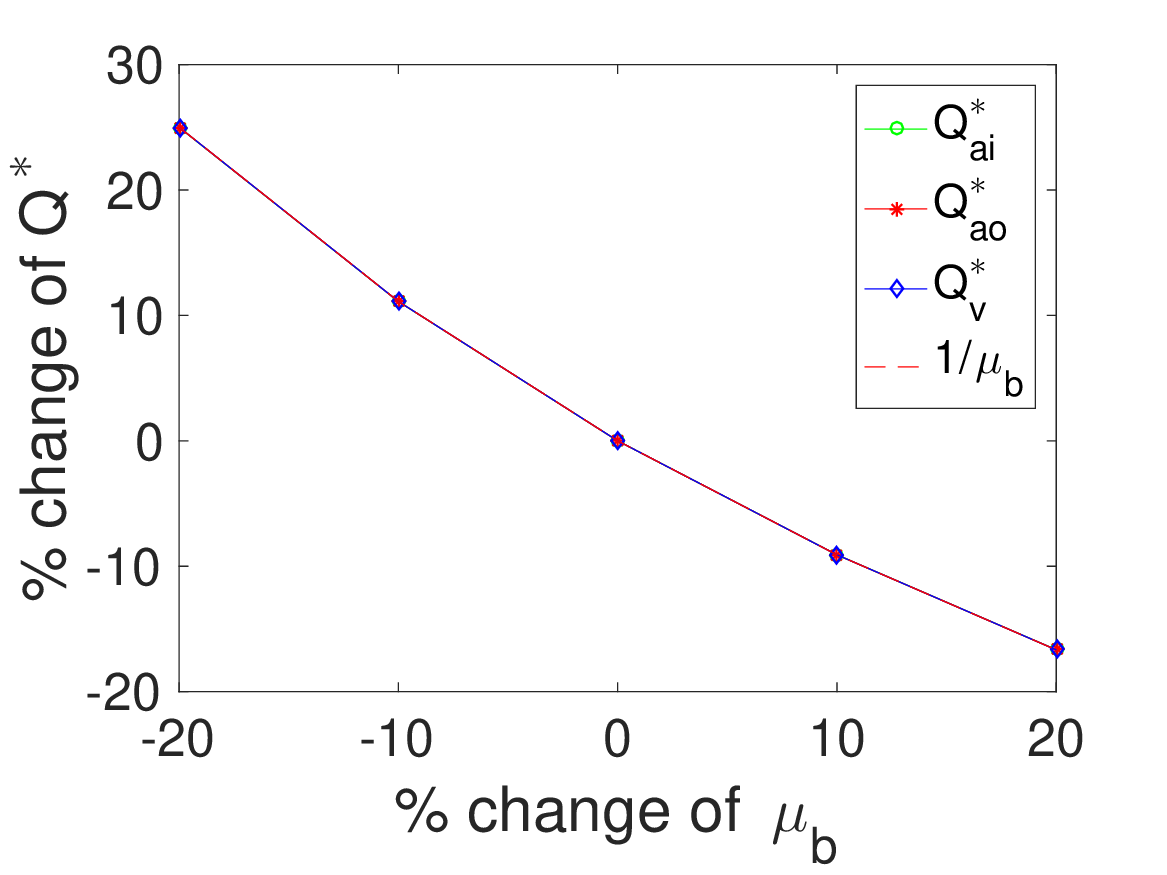}
\caption{Percentage change of boundary blood supply with changing $\mu$.}
\label{fig17}
\end{center}
\end{figure}

In this subsection, we focus on the effects of blood viscosity $\mu_b$, which could increase due to aging or other disease progress.  For the blood/water circulation, by gradually changing $\mu_b$ by $\pm 20 \%$, we observe that the parameter $\mu_b$ has almost a scaling effect for the water/blood flow. The pressure profiles in Section \ref{section3_1} stay the same, but all the flow rates change by a factor of $1/\mu$. Figure \ref{fig17} shows the percentage changes of  boundary flow rates $Q_{k}^\ast$ ($k=ai,ao,v$) with changes of $\mu$, which align with the change of $1/\mu_b$ for illustration of scaling effect. This scaling effect can also be seen from the governing equations and parameter relations. 

\begin{figure}[h]
\begin{center}
\includegraphics[width=0.45 \textwidth]{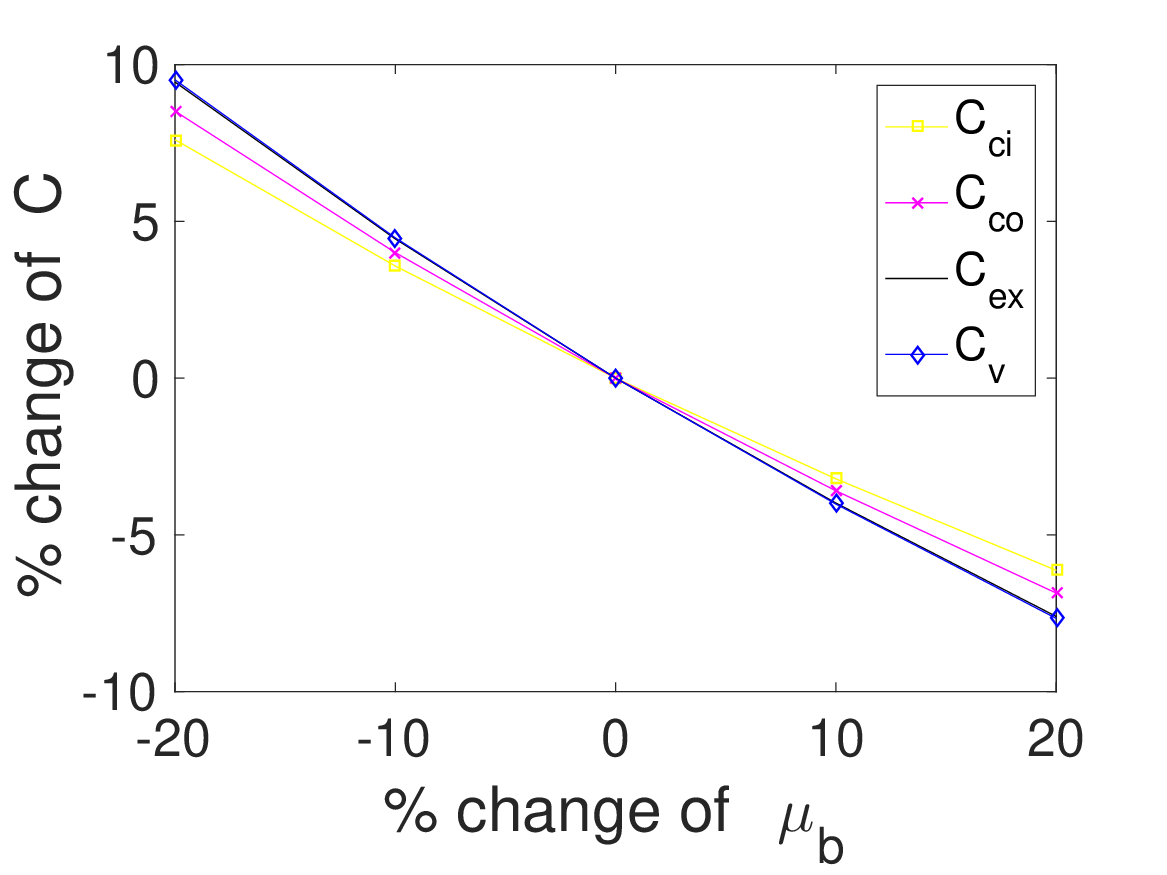}
\includegraphics[width=0.45 \textwidth]{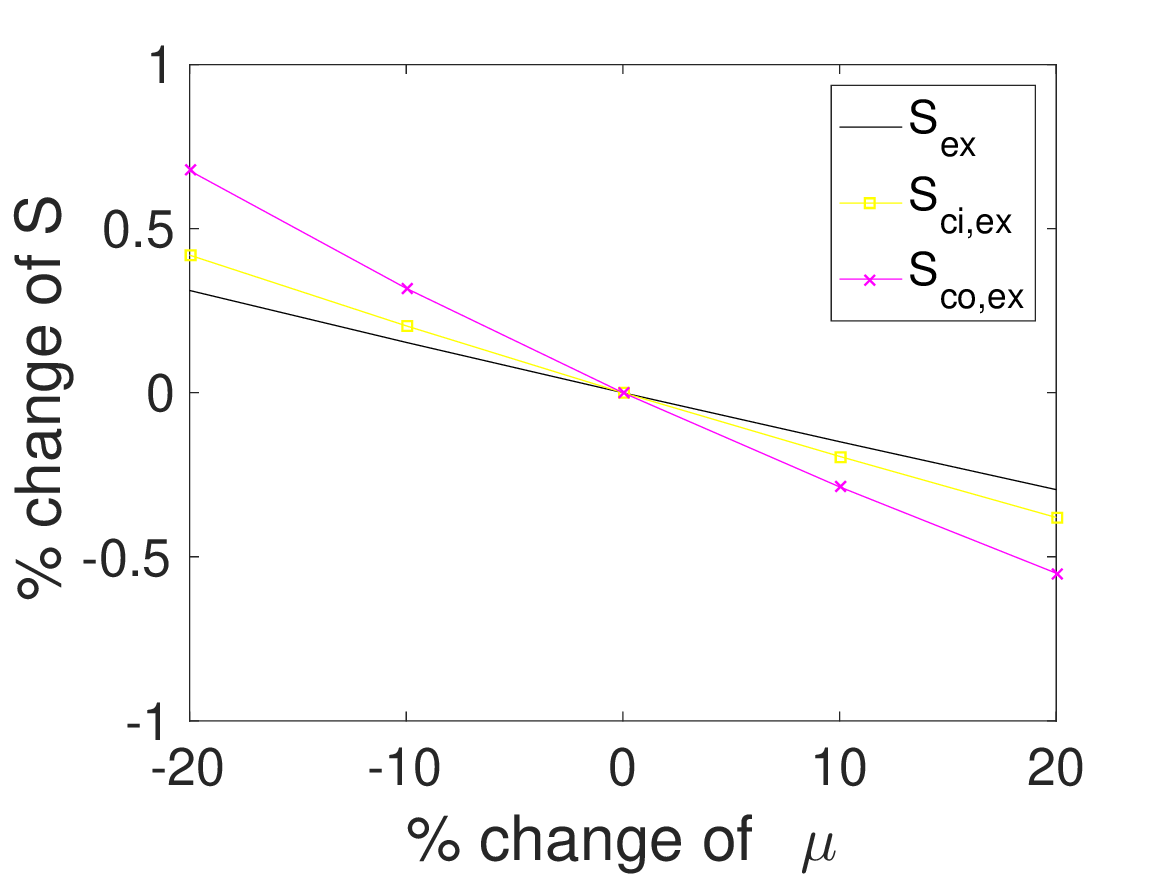}
\caption{oxygen with changing $\mu$}
\label{fig18}
\end{center}
\end{figure}

For the oxygen transport, Figure \ref{fig18}(a) shows that the average concentrations for $C_j$ ($j=ci,co,v,ex$) over the region decrease as $\mu$ increase, for example, they decrease by 6-8\% with $\mu$ increased by 20 \%. It is a consequence of decreased blood supply in Figure \ref{fig17}. But the difference between the capillary concentrations and extravascular concentrations is relatively stable, so the oxygen exchanges from capillary to the extravascular space only drops slightly ($<1\%$ with $\mu$ increased by 20 \%), as shown in Figure \ref{fig18}(b). The total consumption is also kept relatively stable in Figure \ref{fig18}(b), with uniform case for resting volume fractions. 

{\bf Remark}: With the Gaussian case 1, compared with the values in Figure \ref{fig13}(b),  the average $S_{ex}$ is decreased from 0.192 to 0.189 with $\mu$ increased by 20 \%, and locally near $r=1$, $S_{ex}$ is impacted significantly, decreasing from 0.175 in  to 0.146. This is expected since less boundary blood supply will have more severe effects on the local region with limited vasculature. 

Next, we investigate how much the pressure at artery boundary should be adjusted to maintain the normal blood supply when the $\mu$ increases. At the boundary, fixed flow rates $Q_{ai}^\ast, Q_{ao}^\ast$ in (\ref{eq43_1}) are used to replace the pressure boundary conditions, and they are set to be the same as the reference case. Table \ref{table4} shows the pressure adjustments at artery boundary to maintain blood supply with increase of $\mu$. Pressures have to increase by 8.2\% (3.3 mmHg) for $P_{ai,0}$ and by 6.5\% (2  mmHg) for $P_{ao,1}$ when $\mu$ is increased by 20\%. In the uniform case, the oxygen delivery is not affected much, but for the Gaussian cases the oxygen supply in local regions will be affected.

In summary, increased viscosity has a scaling effect on blood circulation according to the decreasing factor $1/\mu_b$. This in turn affects the oxygen concentrations, and could worsen the local oxygen delivery for non-uniform cases. Biologically, when  $\mu$ is increased under various factors, as a remedy for sufficient blood supply and oxygen delivery , the blood pressure in artery has to increase correspondingly.

\begin{table}[h]
\begin{center}
\begin{tabular}{|c|c|c|c|}
   \hline
 increase of $\mu$ &  0 & $10\%$ & $20\%$\\
 \hline
$P_{ai,0}$ & 2& 2.0834, (4.2\%,~ 1.7 mmHg),& 2.1648, (8.2\%,~ 3.3 mmHg)\\
\hline
$P_{ao,1}$ & 1.5 & 1.5488, (3.3\%,~ 1 mmHg)&  1.5974, (6.5\%,~ 2  mmHg) \\
 \hline
\end{tabular}
\end{center}
\caption{Pressure adjustments at artery boundary to maintain blood supply with increase of $\mu$.}
\label{table4}
\end{table}

\subsection{Effects of periodic arterial pressure}

\begin{figure}[h]
\begin{center}
\includegraphics[width=0.4 \textwidth]{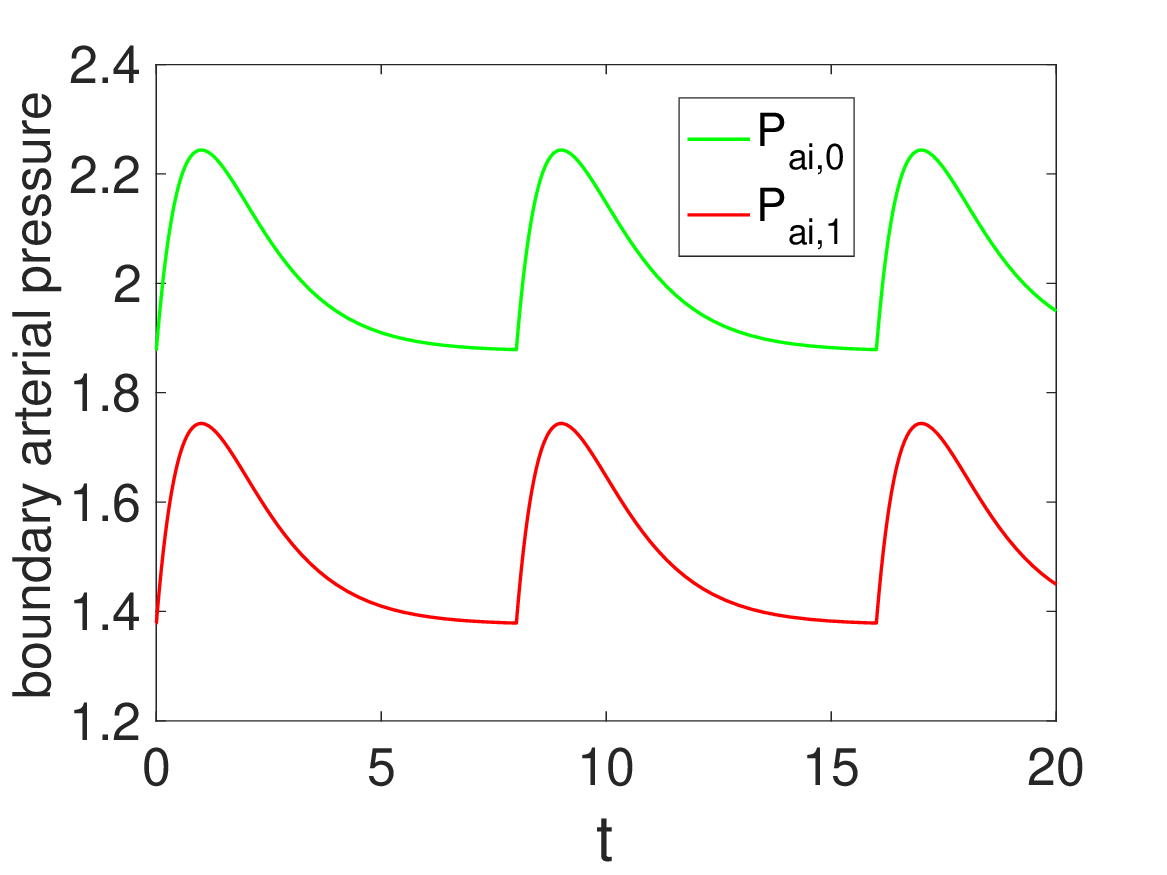}
\caption{Periodic profiles of boundary arterial pressures for $P_{ai,0},P_{ao,1}$.}
\label{fig31}
\end{center}
\end{figure}

In this subsection, we consider the case of periodic changes of arterial pressure due to the pulse pressure. Figure \ref{fig31} shows the periodic profiles of the dimensionless boundary arterial pressures $P_{ai,0}(t)$ and $P_{ao,1}(t)$ in (\ref{eq12},\ref{eq14}) used as boundary conditions of the following simulation. The profile of variation is constructed roughly based on a shifted function $t*e^{-t}$ in the first period, which has a shape similar to a pulse profile. The period is T=8, which is 0.8 seconds with units.   The total variation within one period is about 0.365, i.e., the pulse pressure between the systolic and diastolic pressures is about 7.3 mmHg with units. The dimensionless average values  (mean arterial pressures, MAP) for $P_{ai,0},P_{ao,1}$ over one period are the same as the reference values 2 and 1.5 (i.e., 40 mmHg and 30 mmHg with units), which are located at 1/3 of total variation and therefore consistent with the common definition. For illustration, we chose our curves in Figure  \ref{fig31} based on profiles of pressures and flow rates in \cite{julien2023,Olufsen2014,kenfack2004,lekakis2005,Anatomy2013,nagaoka2006} and the knowledge of MAP in \cite{prada2016,guidoboni2019,Anatomy2013}. More specific references may be available for more accurate curves that we have not yet found.

\begin{figure}[h]
\begin{center}
\includegraphics[width=0.3 \textwidth]{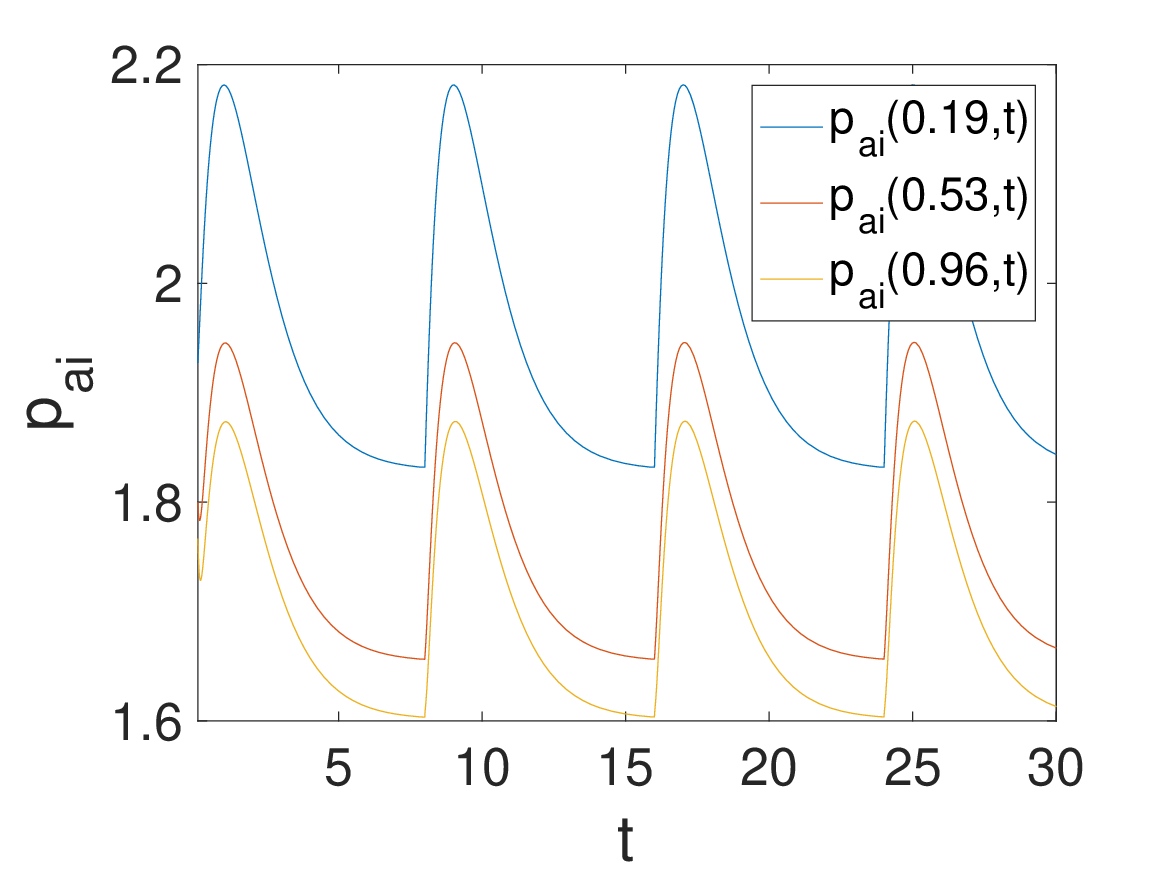}
\includegraphics[width=0.3 \textwidth]{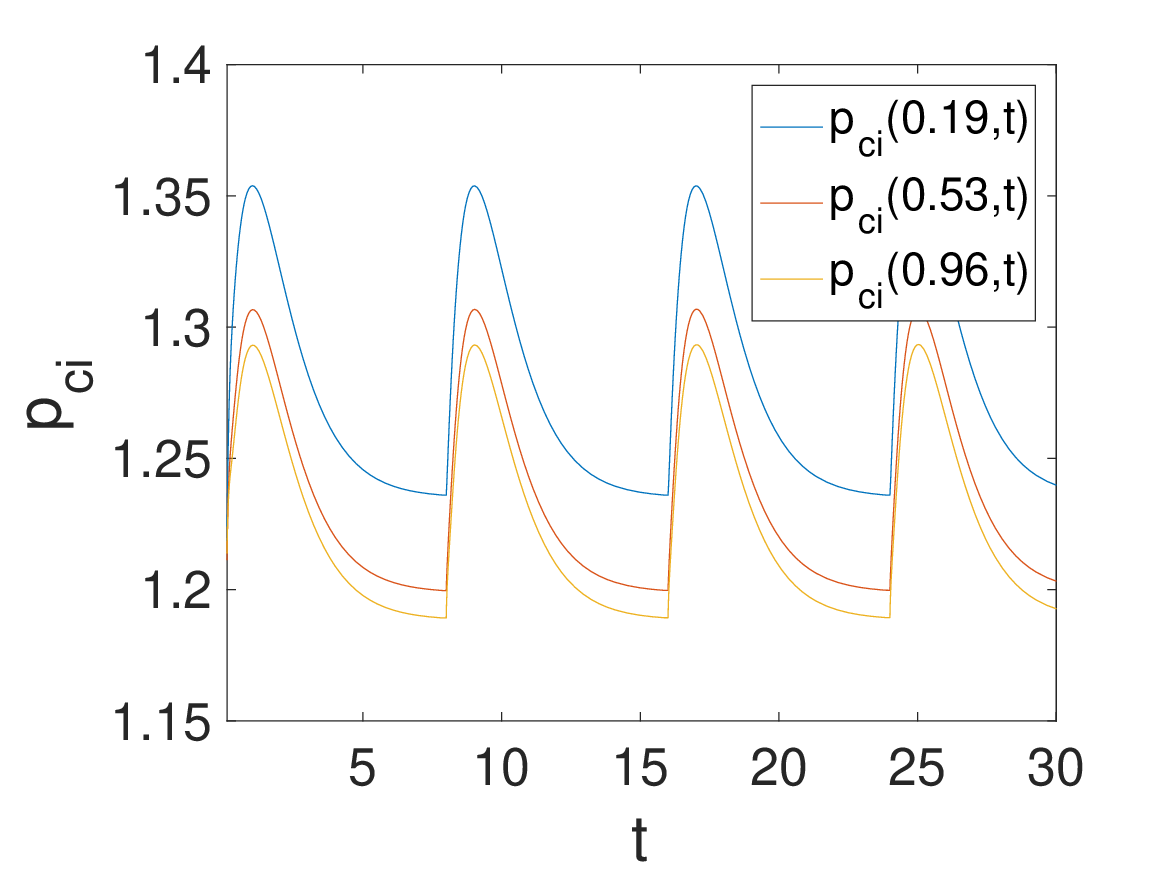}
\includegraphics[width=0.3 \textwidth]{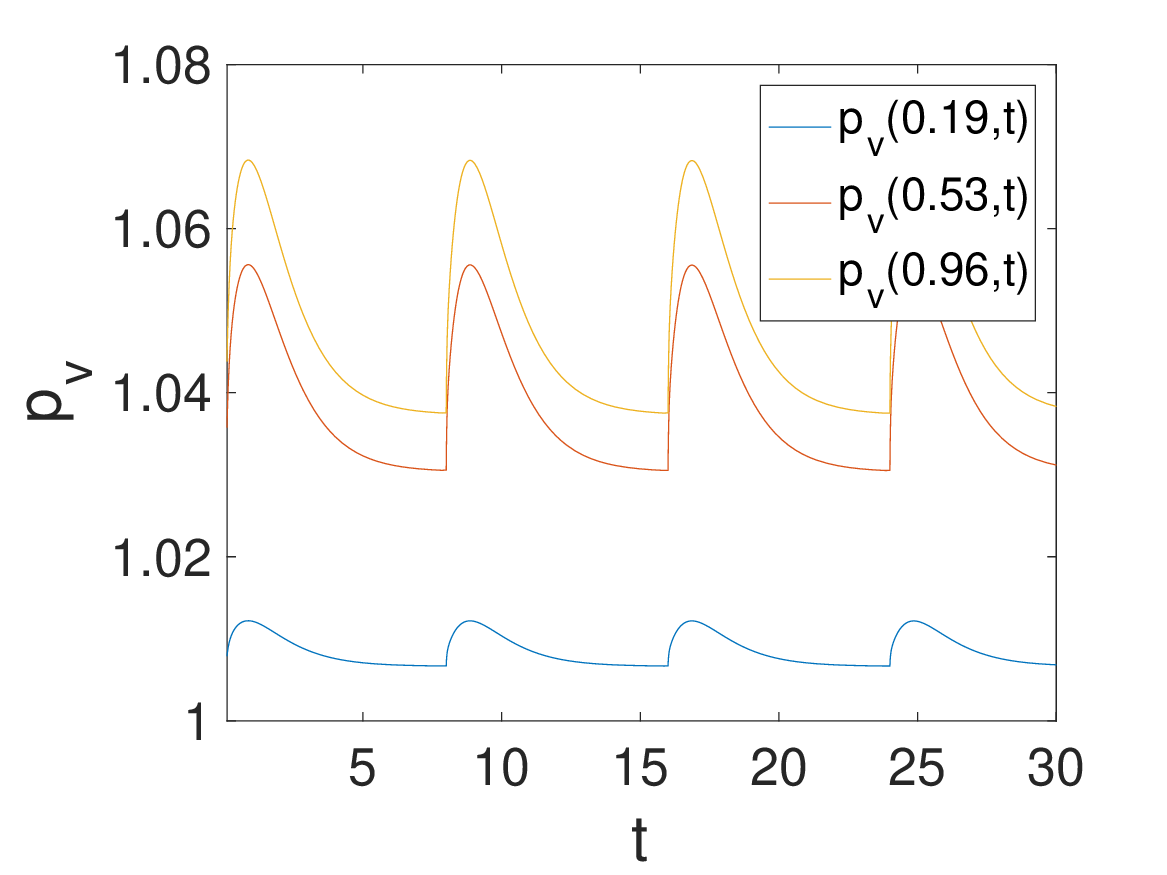}
\caption{dynamics of the three pressures  $p_{ai},p_{ci},p_v$ at three locations $r=0.19, 0.53,0.96$.}
\label{fig32}
\end{center}
\end{figure}

Figure \ref{fig32} shows the dynamics of the three three pressures  $p_{ai},p_{ci},p_v$ at three typical locations $r=0.19, 0.53,0.96$ (one in the middle, two close to two ends). They all show periodic oscillations similar to the given profile in Figure \ref{fig31}, but with different magnitude of variations. The periodic variations in arteries are the largest, and the variations in veins are the smallest in magnitude. The other pressures show similar trend and variations. The volume fractions show small periodic adjustments due to the force balance and changes in pressures.

\begin{figure}[h]
\begin{center}
\includegraphics[width=0.4 \textwidth]{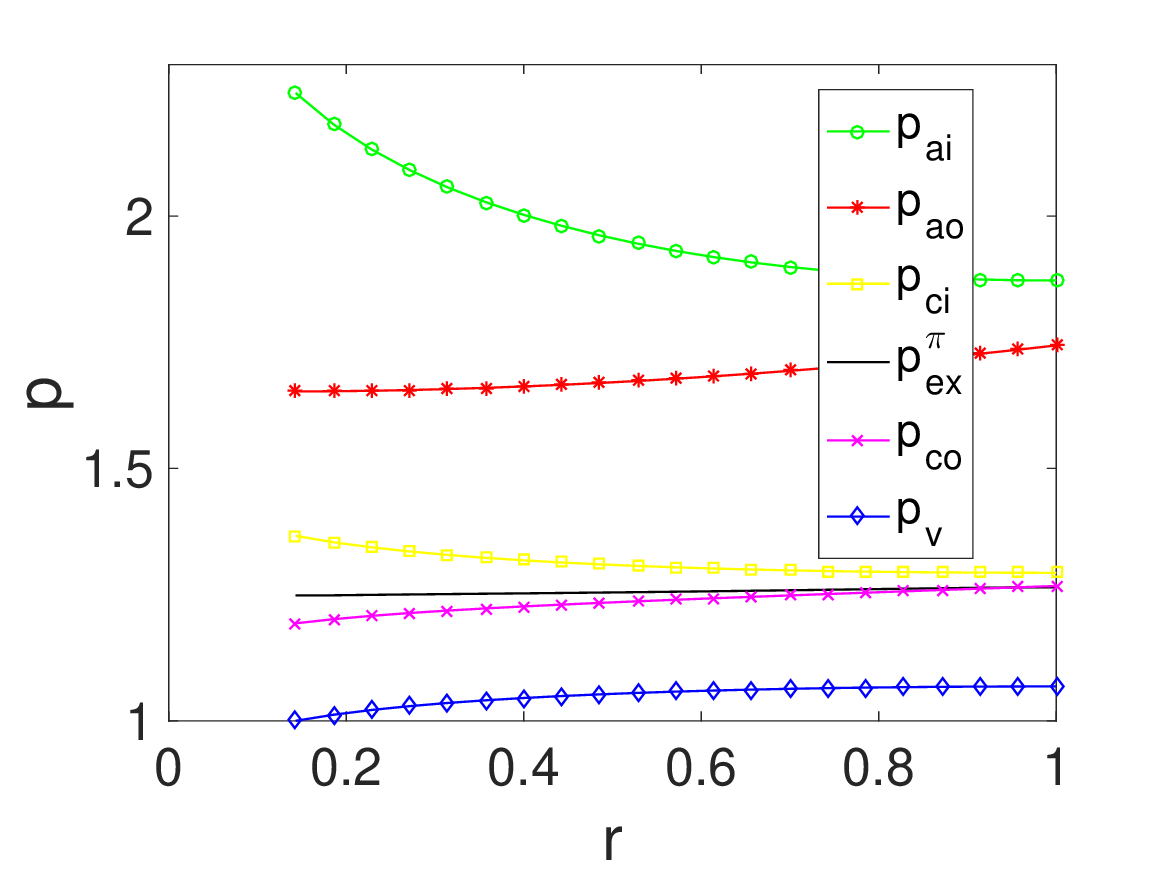}
\includegraphics[width=0.4 \textwidth]{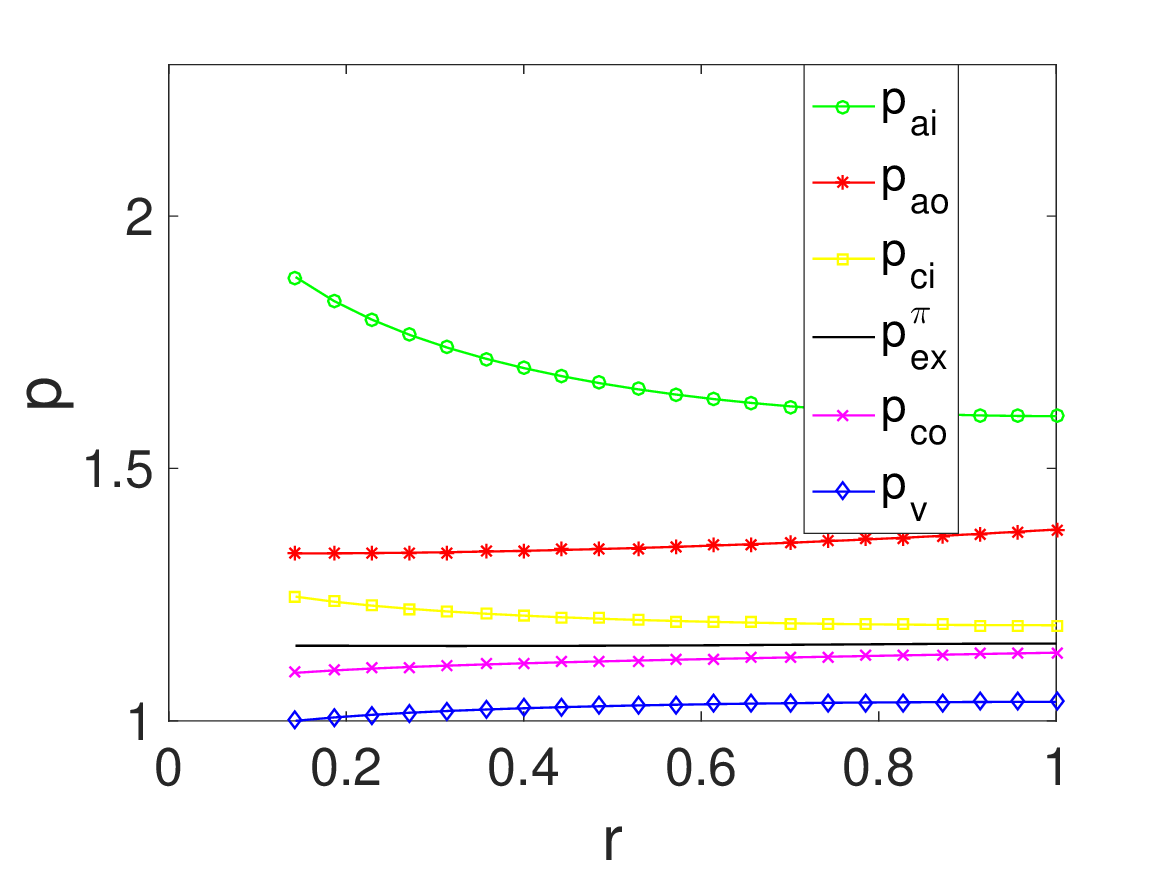}
\caption{spatial  profiles of all pressures in the whole region at two particular times (i.e., systolic and diastolic pressures respectively).}
\label{fig33}
\end{center}
\end{figure}

Figure \ref{fig33} shows the spatial profiles of all pressures in the whole region at two particular times when the boundary arterial pressures are maximum and minimum (systolic and diastolic pressures respectively). The spatial profiles and their relative positions for pressures are quite similar to that in Figure \ref{fig3}, but move up and down simultaneously and stably according to dynamic changes in boundary arterial pressures in Figure \ref{fig31}.

\begin{figure}[h]
\begin{center}
\includegraphics[width=0.4 \textwidth]{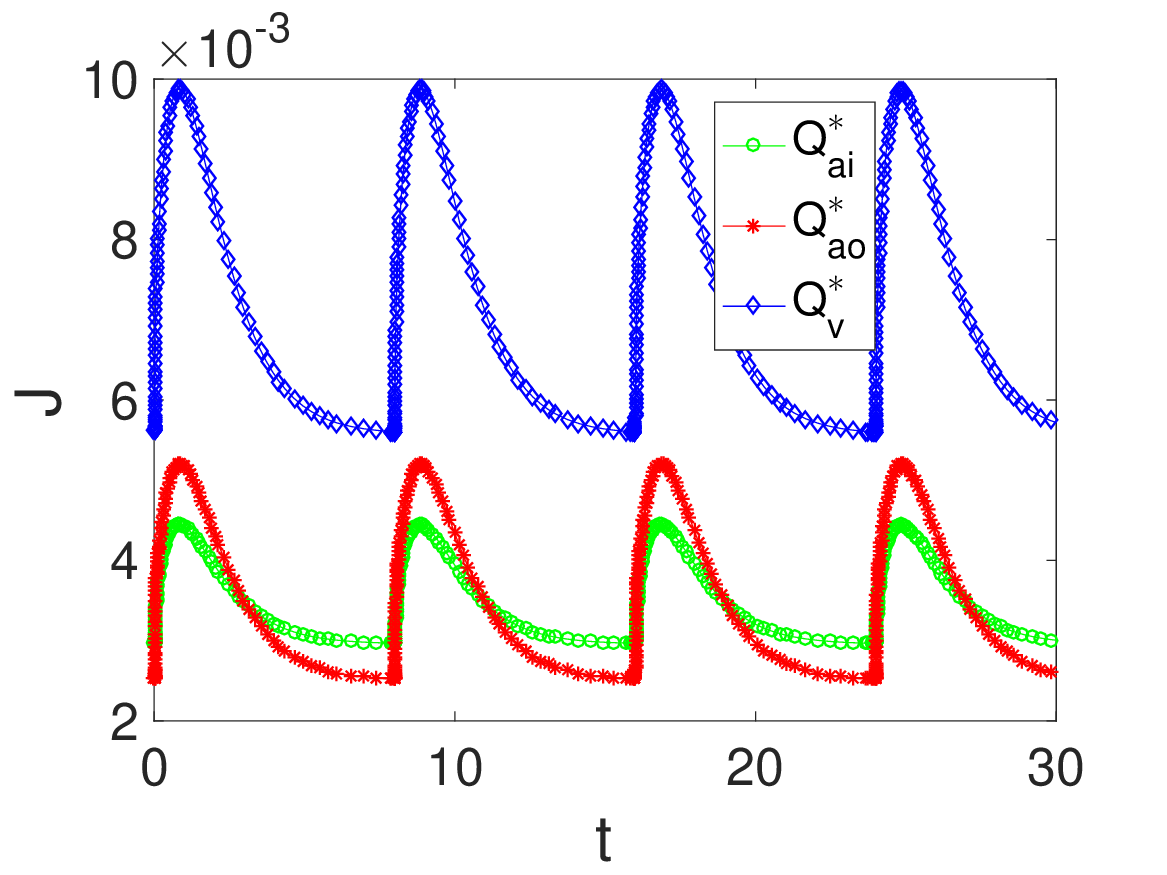}
\caption{The dynamics of the boundary blood flow rates in the periodic case.}
\label{fig34}
\end{center}
\end{figure}

\begin{figure}[h]
\begin{center}
\includegraphics[width=0.4 \textwidth]{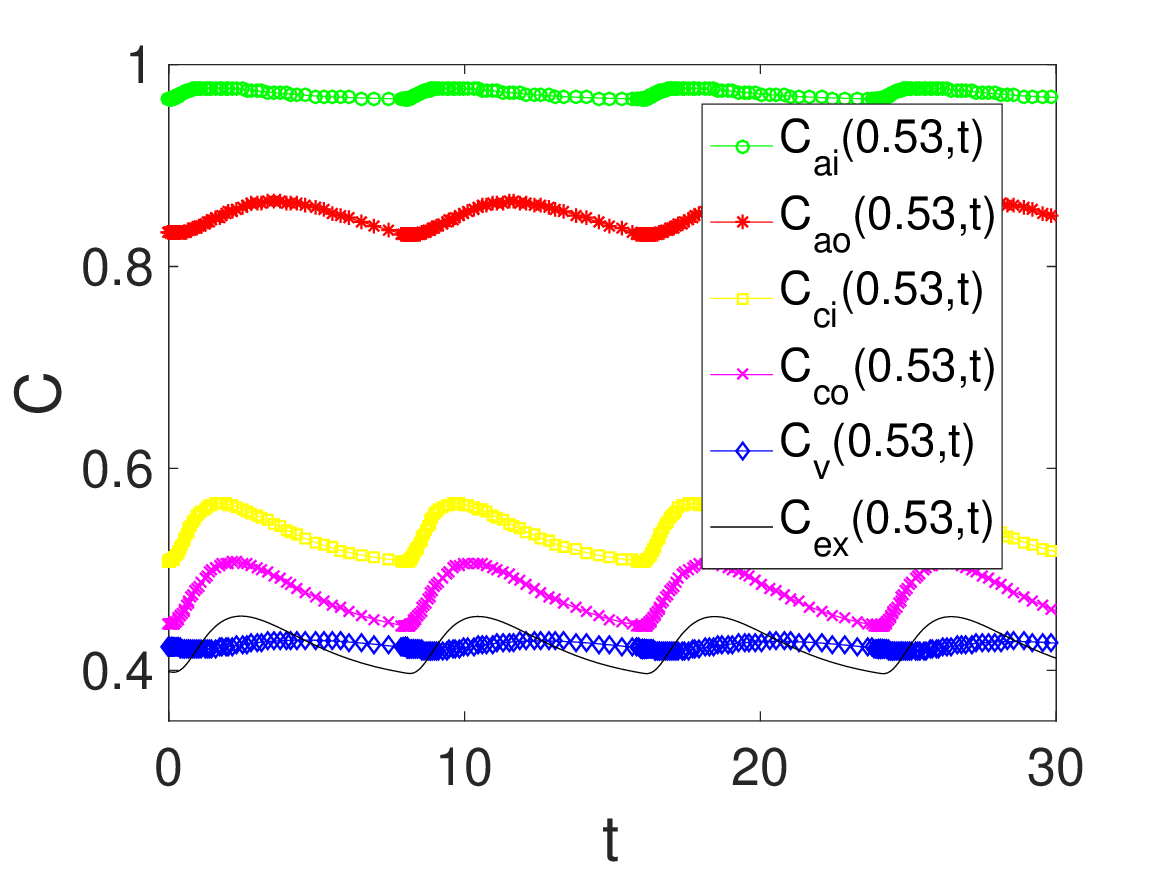}
\includegraphics[width=0.4 \textwidth]{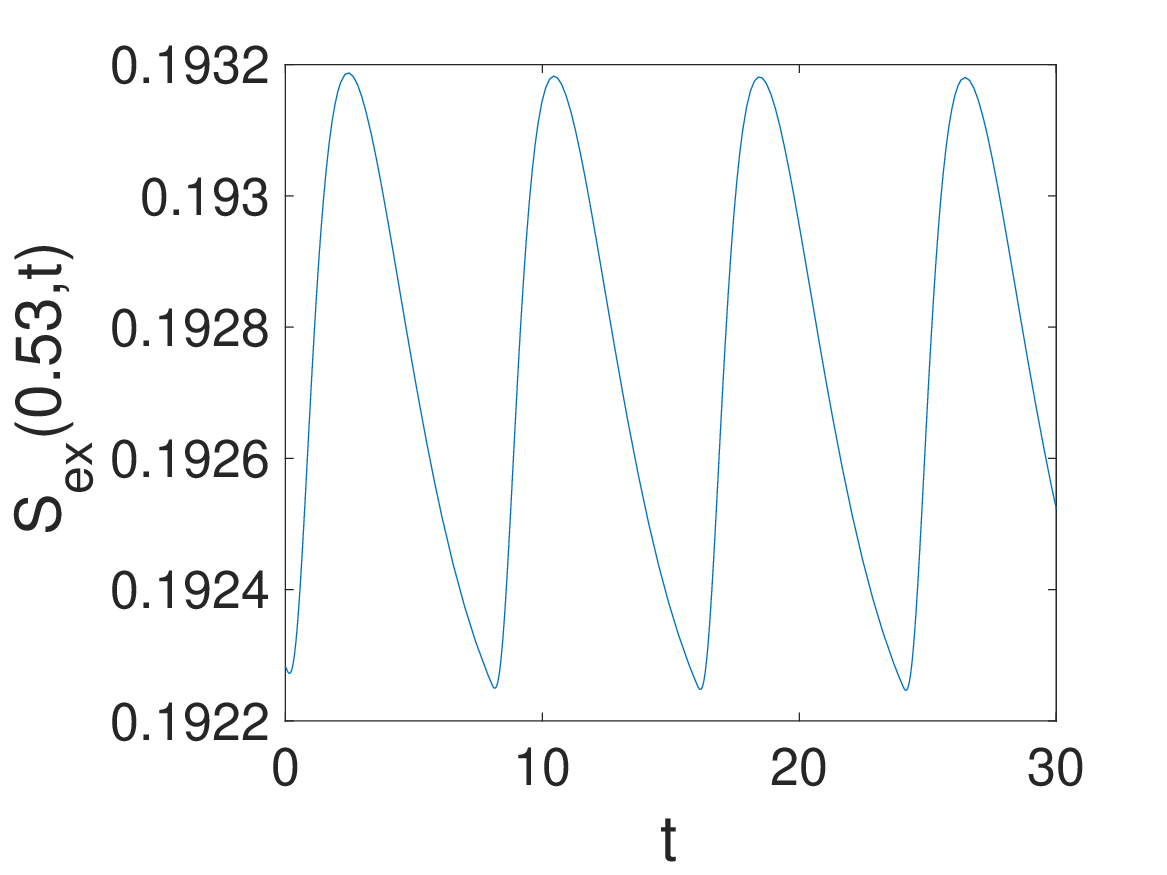}
\caption{The dynamics of the concentrations and oxygen consumption at $r=0.53$.}
\label{fig35}
\end{center}
\end{figure}

Figure \ref{fig34} shows the dynamics of the boundary blood flow rates, which has a similar periodic profiles as the given boundary arterial pressures. The average value of blood flow rates over one period is computed to be $Q_{ai}^{\ast}=Q_{ao}^\ast = 0.0034$, the same as those in Table \ref{table5}. So for the blood circulation, the uniform case in  Section \ref{section3_1} can be considered as the averaged version of the case with periodic arterial pressure. 

Figure \ref{fig35} shows the dynamics of the concentrations and oxygen consumption at $r=0.53$. The concentrations show periodic variations but the total oxygen consumption is relatively stable with very small changes. The small change for oxygen consumption is due the relation $S_{ex}(C_{ex})$ used in Figure \ref{fig6}, because even with changes of $C_{ex}$ (say in [0.4,0.5]), the consumption $S_{ex}$ stays at almost stable constant level. Then, as a consequence of stable consumption in Figure \ref{fig35}(b) and varying (but sufficient) flow rates in Figure \ref{fig34}, the concentrations will have more significant variations, i.e., with more blood flow (and RBCs) from boundary, the concentrations do not need to drop too much to satisfy the stable oxygen consumption. We also notice there is a delay in the timing of peak values of concentrations compared with the timing of peak values of blood flow rates (i.e., the timing of systolic pressure).

In summary, with periodic pressure conditions, the blood circulation and oxygen concentrations show similar periodic variations in time and stable spatial profiles. The uniform case is almost the averaged version of periodic case.

\subsection{Effects of other parameters}

In this subsection, we study the effects of a few other model parameters, including $\lambda_j, \beta_j,\delta_j$ in blood circulation part and $S_{ex}^{max}, C_{1/2},C_{ai,0},H_0$ in the oxygen delivery part.
 
 %The distribution $\eta$ seems more important.
% $ P_j^{re}$ in Yi's paper?
% If $ P_j^{re}=0$, the effect of $\lambda$ will be more significant on permeability.
 
First, we study the effects of the modulus $\lambda_j$ ($j=ai,ao,v,ci,co$) of blood vessel wall in the force balance (\ref{eq11}). Since the diameters of blood vessels could change in response to the environment changes (or stimulus like oxygen)  \cite{guidoboni2019,physiology2009,arciero2013},  from modeling perspective, this could be reflected by the response of blood wall property to the environment changes. For example, in the present framework, the feedback effect of oxygen on the blood flow could be incorporated through the dependence of $\lambda_j$ on oxygen concentrations, since the volume fractions (reflecting the diameter of blood vessels) will change accordingly and influence the blood flow. Here, we illustrate the effect of $\lambda_j$ ($j=ai,ao,v,ci,co$) by reducing them simultaneously.  Table 5 shows the comparison of the reference uniform case and the case when all $\lambda_j$ are reduced by 50\%, showing that the volume fractions for $\eta_{j}$, particularly for $\eta_{ai},\eta_{ci}$, are increased, which in turn increases the blood supply from the boundary, particularly for $Q_{ai}^\ast$. The main reason is that the permeabilities $\bar{\kappa}_j, \bar{K}_j$ in (\ref{eq41}) between vascular domains will increase with the increased volume fractions (or diameters of blood vessels), leading to more blood flow. The influence on oxygen delivery is not as significant, since in the uniform case, the blood supply and oxygen supply is quite sufficient and stable, so it does not need the increased blood flow to help out. We also tested the Gaussian case 1, when $\lambda_j$ are increased by 50\%, the minimum value of $S_{ex}$ near $r=1$ in Figure \ref{fig13}(b) increases from 0.1748 to 0.1794, so in that case the change of $\lambda_j$ can help the local oxygen delivery.

\begin{table}[h]
\begin{center}
\begin{tabular}{|c|c|c|c|c|c|c|}
   \hline
  &  $\eta_{ai}$ &   $\eta_{ao}$ & $\eta_{ci}$ & $\eta_{co}$ & $Q_{ai}^\ast$ & $Q_{ao}^\ast$\\
 \hline
with $\lambda_j$ in Uniform case  & 0.0143 & 0.0257 & 0.0070 & 0.0131 &0.0034 & 0.0034 \\
\hline
reduce $\lambda_j$ by $50\%$  & 0.0160 & 0.0262 & 0.0077 & 0.0135 & 0.0042 & 0.0035\\
 \hline
\end{tabular}
\end{center}
\caption{Effects of $\lambda_j$ in the uniform case. }
\label{table5}
\end{table}

\begin{figure}[h]
\begin{center}
\includegraphics[width=0.45 \textwidth]{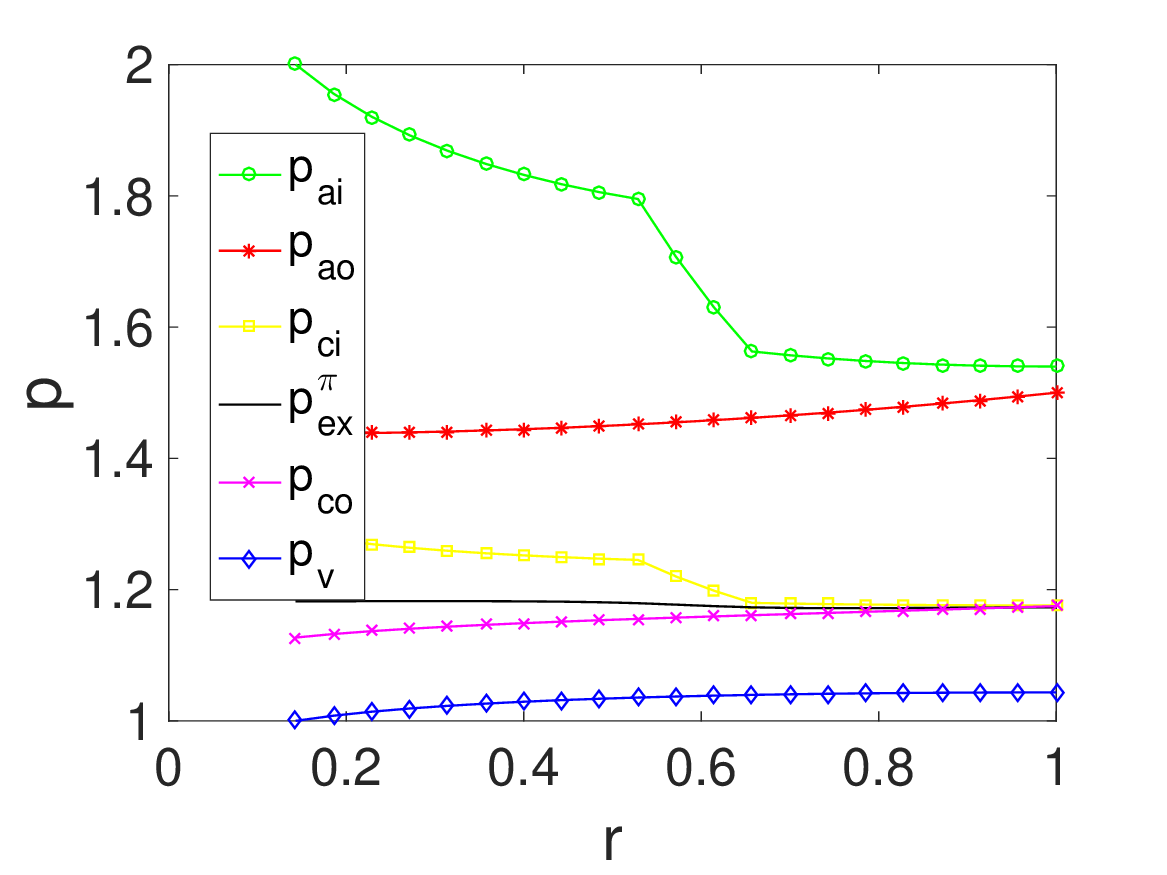}
\includegraphics[width=0.45 \textwidth]{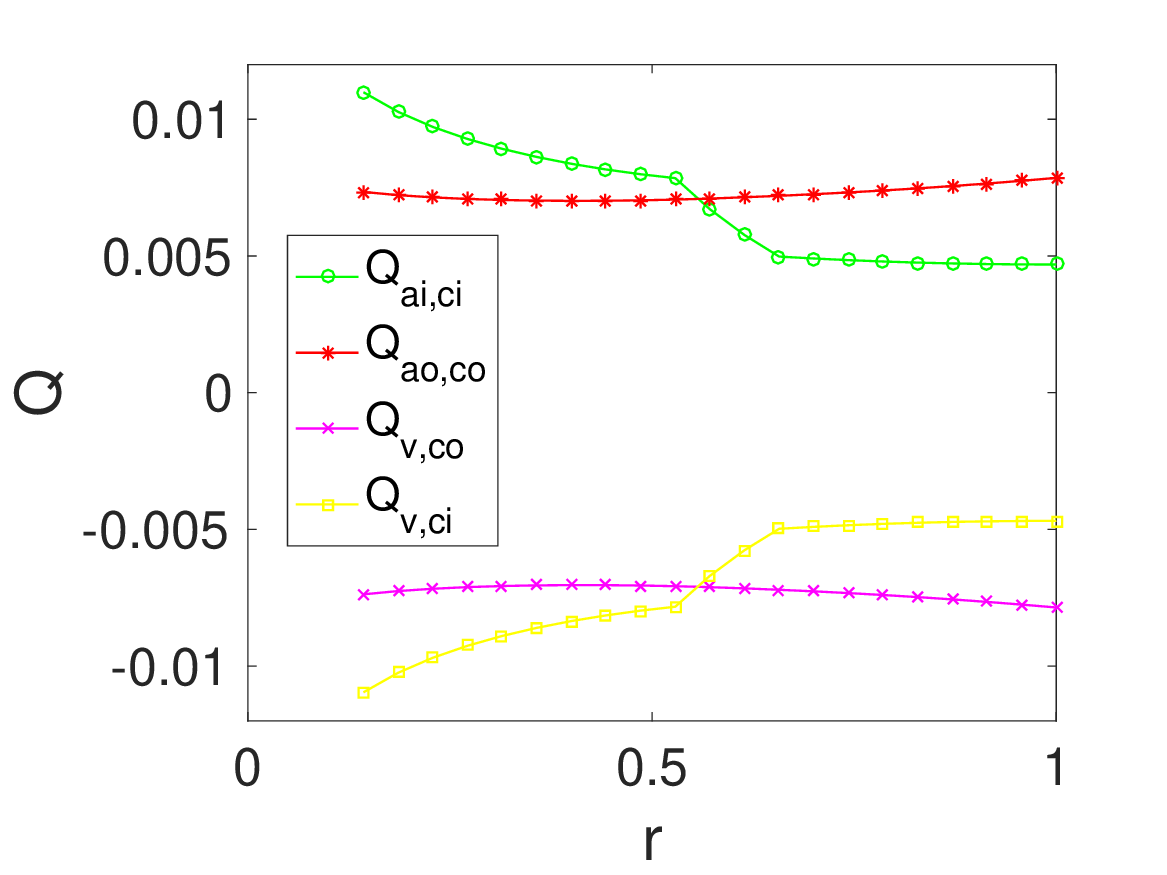}
\caption{Pressure profiles and blood exchange rates with local change of $\beta_{ai}$, which is reduced to 10\% of original value near $r=0.6$.}
\label{fig20}
\end{center}
\end{figure}

\begin{figure}[h]
\begin{center}
\includegraphics[width=0.45 \textwidth]{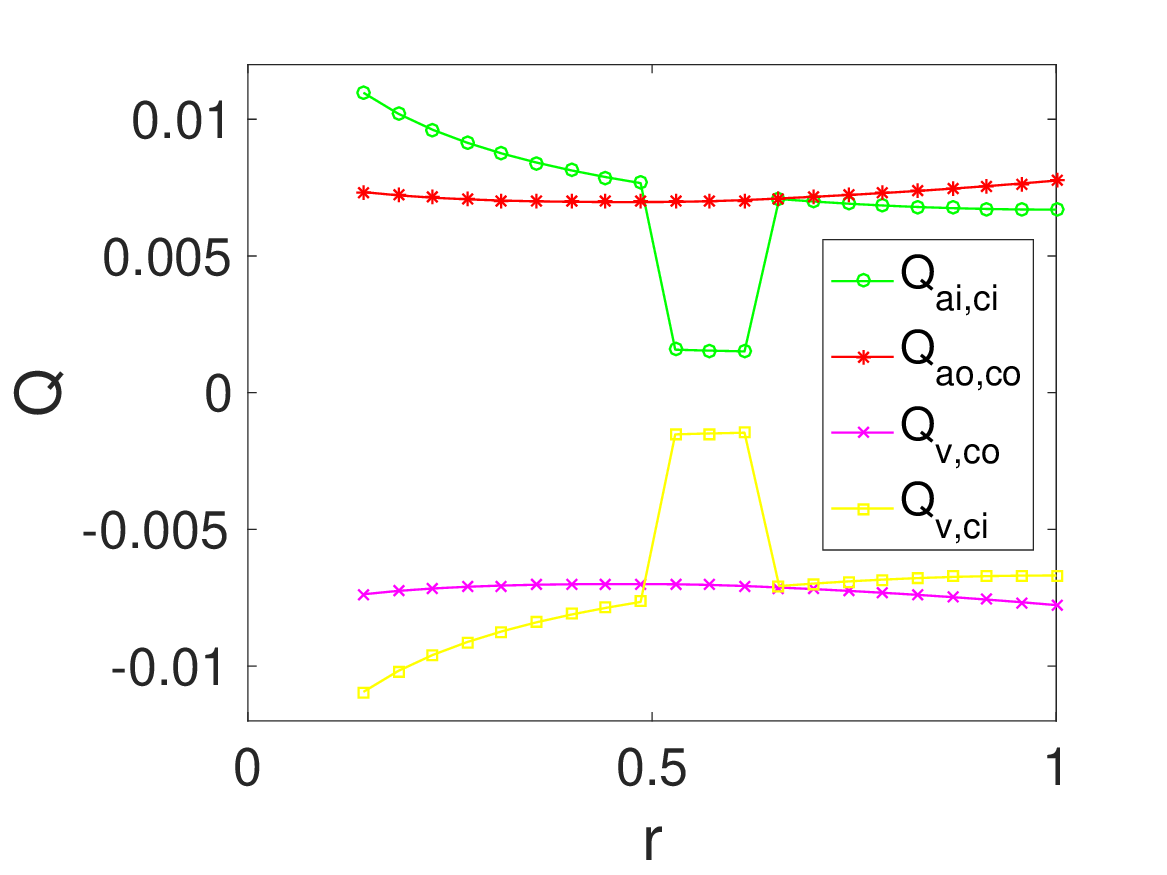}
\includegraphics[width=0.45 \textwidth]{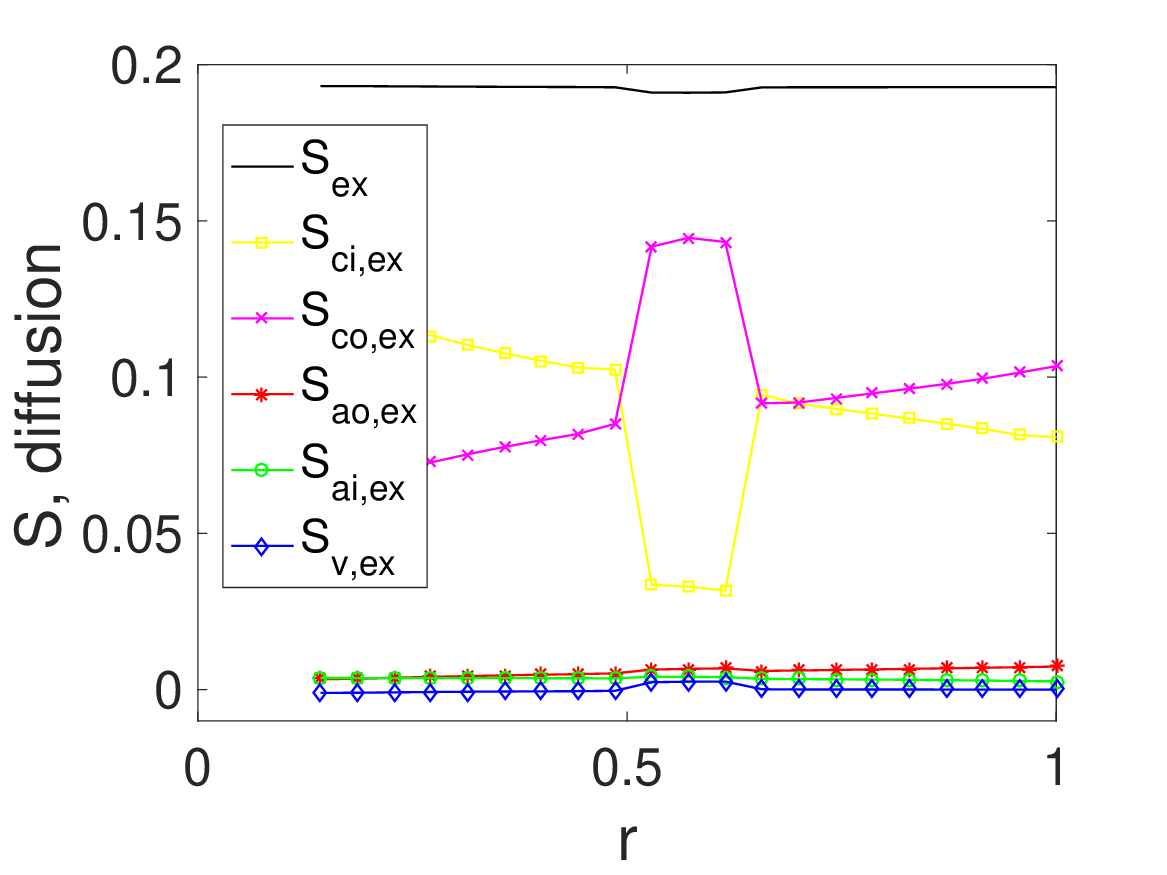}
\caption{Blood exchange rates and oxygen supply with local change of $\delta_{ci}$, which is reduced to 10\% of original value near $r=0.6$.}
\label{fig21}
\end{center}
\end{figure}

Next, we analyze the effects of partial blockage of blood vessels in some local region in either artery or capillary, which could be caused by pathological conditions such as arterial stenosis \cite{julien2023,guidoboni2019}. The blockage leads to decreased permeability, and for illustration we take $\beta_{ai}$ and $\delta_{ci}$ as the effective parameter for the blockage of artery and capillary. In Figure \ref{fig20}, we set $\beta_{ai}$ to be 10\% of original value (i.e., 90\% decrease) in a small interval near $r=0.6$. Figure \ref{fig20}(a) shows that the pressures $p_{ai}$ and $p_{ci}$ have a more significant drop near $r=0.6$ to counterbalance some blockage effect. The average value of blood flow rates $Q_{ai,ci}$ and $Q_{v,ci}$ after the blockage location (the downstream regions) are reduced by 26\% (from  0.0065 to 0.0048), shown in Figure \ref{fig20}(b). The boundary input $Q_{ai}^\ast$ is reduced  by 15\%  (from normal value 0.0034 to 0.0029). For the oxygen delivery part, the overall oxygen delivery/consumption is still quite stable without observable change, since the overall blood supply is still sufficient and there is redistribution of oxygen supply in the affected region, i.e., more supply is from the other capillary domain $\Omega_{co}$.  We have also tested the case when $\beta_{ai}$ is reduced by 50\% locally, and the changes for $Q_{ai,ci},Q_{v,ci}$  and  $Q_{ai}^\ast$ are quite small ($< 3\%$ decrease). In Figure \ref{fig21}, we set $\delta_{ai}$ to be 10\% of original value in a small interval near $r=0.6$. The flow rates $Q_{ai,ci}$ and $Q_{v,ci}$ are reduced by about 80\% (from 0.0071 to 0.0015) locally near $r=0.6$, while other regions are almost unaffected.  The oxygen consumption only drops slightly  ( about 1\%) in the local region near $r=0.6$, since it is compensated by more oxygen supply from the undamaged capillary $\Omega_{co}$. We have also tested the case when both $\delta_{ci},\delta_{co}$ are reduced by 90\%, the reduction in flow rates are similar and at about 80\%, but the oxygen consumption only drops by about 8.4\%. In brief, the system is not very sensitive to permeability changes mainly because of the present of two capillary networks and the relation $S_{ex}(C_{ex})$ with small $C_{1/2}$ in Figure \ref{fig6}(b).

Next, we analyze the effects of the parameters $S_{ex}^{max}$ and  $C_{1/2}$ in the Michaelis-Menten formula (\ref{eq25}) for oxygen consumption.  Due to lack of data, these two parameters are estimated for different tissues in appendix \cite{causin2016, arciero2008,secomb2004}, so they do not necessarily reflect the real situation for optic nerve. These parameters are biologically important, for example, the max demand $S_{ex}^{max}$ could change because of high metabolism or neuron activities. We examine a case when $S_{ex}^{max}$ is increased by $50\%$. By the first two rows in Table \ref{table6} for Uniform case, the consumption of oxygen also increases  correspondingly by roughly $50\%$ on average and keeps stable spatial profile, but the concentrations $C_{ex},C_v$ are decreased accordingly as more oxygen from RBCs has to release to meet the increased need. For  Gaussian case 1, the last two rows in Table \ref{table6} show similar increased average consumption and decreased average concentrations $C_{ex},C_v$.  But locally near $r=1$, the minimum oxygen consumption is affected significantly and reduced to 13\% of average value, since the mismatch between demand and supply is worsened as the blood supply is insufficient there. The parameter $C_{1/2}$ influence the shape of the consumption curve in Figure \ref{fig6}(b), with increased $C_{1/2}$ the curve will have a smoother transition and lie below the original curve. So with increased $C_{1/2}$, the average consumption will be smaller for uniform case, and more local regions will be affected for oxygen delivery  in the Gaussian case 1 due to insufficient blood supply and more sensitive changes of $S_{ex}$ with $C_{ex}$ in certain range (say [0.2, 0.4]). In brief, increased $S_{ex}^{max}$ or $C_{1/2}$ will worsen
the oxygen supply for local regions with insufficient blood supply.

\begin{table}[h]
\begin{center}
\begin{tabular}{|c|c|c|c|c|}
   \hline
  &  average $C_{ex}$ &   average $C_{v}$ & average $S_{ex}$ & min $S_{ex}$ \\
 \hline
reference Uniform case  & 0.424 & 0.429 & 0.193 & 0.193  \\
\hline
increase $S_{ex}^{max}$ by $50\%$  & 0.357 & 0.351 & 0.287 & 0.287 \\
 \hline
 reference Gaussian case 1  & 0.456 & 0.378 & 0.192 & 0.175  \\
\hline
increase $S_{ex}^{max}$ by $50\%$  & 0.379 & 0.308 & 0.272 & 0.035 \\
 \hline
\end{tabular}
\end{center}
\caption{Effects of $S_{ex}^{max}$ by increasing it by 50\%. }
\label{table6}
\end{table}

Finally, we study the effects of parameters $C_{ai,0}$ (we set $C_{ai,0} = C_{ao,1}$) and $H_0$ in (\ref{eq35},\ref{eq36}), which are related to the supply of oxygen from artery boundary. Biologically, the oxygen content from boundary artery could drop due to various conditions like anaemia (i.e, low $H_0$) and high altitude or carbon monoxide poisoning (i.e., low $C_{ai,0}$).  For the uniform case, when $H_0$ or $C_{ai,0}$ is decreased by 20\%, there is only negligible impact on the oxygen delivery, e.g., the consumption rate $S_{ex}$ has negligible change (deceased by $<1\%$), because the supply of blood flow and hence oxygen is still sufficient. For the Gaussian case 1, Table \ref{table7} shows the comparison when the $C_{ai,0}$ or $H_0$ is decreased by 20\%. With $C_{ai,0}$ decreased by 20\%, the profile and values of $S_{ex}$ have negligible change ($<1\%$ for mean and minimum values), while the concentrations decrease by moderate percentage (e.g., 5-7\%). That means the overall profiles of concentrations are shifted downward to maintain relatively stable oxygen consumption. With $H_{0}$ decreased by 20\%, the oxygen consumption/supply in some local region will be affected more significantly, as the minimum $S_{ex}$ drops from 0.175 to 0.146 near $r=1$. This is because the oxygen in hemoglobin (in RBCs) is the main source of oxygen supply, and if hemoglobin concentration is decreased (here reflected by $H_0$), the supply of oxygen is worsened in local regions with insufficient blood supply. The concentrations like $C_{ex},C_v$ will be decreased accordingly (by about 10\%) since this allows more oxygen release from hemoglobin to maintain roughly stable $S_{ex}$. Overall, decreased $H_0$ has more impact on local regions compared with decreased $C_{ai,0}$.

\begin{table}[h]
\begin{center}
\begin{tabular}{|c|c|c|c|c|}
   \hline
  &  average $C_{ex}$ &   average $C_{v}$ & average $S_{ex}$ & min $S_{ex}$ \\
 \hline
 reference Gaussian case 1  & 0.456 & 0.378 & 0.192 & 0.175  \\
\hline
 decrease $C_{ai,0}$ by $20\%$  & 0.421 & 0.359 & 0.191 & 0.172 \\
 \hline
decrease $H_0$ by $20\%$  & 0.414 & 0.339 & 0.189 & 0.146 \\
 \hline
\end{tabular}
\end{center}
\caption{Effects of $H_0$ and $C_{ai,0}$ in the Gaussian case 1. }
\label{table7}
\end{table}
 
%how about demand is spatially varying?

%How the percentage (average percentage, distribution) of $\eta$ changes affect the water flow/O2 level; 
%

\section{Conclusions}

In this work, we have developed a multi-domain model for blood circulation and oxygen transport in optic nerve, with biological structures and various physical mechanisms incorporated. The arteries, veins and capillaries for vasculature are treated as different domains in the model for the same geometric region. Simulated baseline results show mechanisms and scales consistent with literature and intuition. Then, the effects of various important model parameters (relevant to pathological conditions) are investigated in detail, and the model provide insights into the possible implications from those parameter changes.  Vasculature distribution (or resting volume fractions here), leak coefficients after a threshold, blood viscosity, maximum oxygen demand, and hemoglobin concentrations have significant impacts on the blood circulation and oxygen delivery, particularly for local regions with insufficient blood supply. 

There are limitations for the current model and possible generalizations from this framework. This work focuses on a 1D case based on optic nerve geometry and simplicity considerations, but it can be extended to high-dimensional cases when more biological structural information is available. It could be extended for retina as well, where more image/experimental data are available.  The coupling here in the model is only between blood circulation and oxygen transport, but could include ion transport mechanisms in different subdomains of by the tissue domain. 

% One main coupling is the convection effect from fluid to oxygen transport, and the feedback coupling could be added into the model for the autoregulation. 

 \appendix

\section{Parameter values and estimates}

\subsection{Parameters in blood/water circulation}

\begin{table}
\caption{\label{table8}  The parameters in blood circulation}
\vspace{-0.6cm}
\begin{center}
\begin{tabular}{|c|c|c|}
\hline
parameter & value & references\\
\hline 
Radius of central vessels $R_0$ & 113 $\mu $m & \cite{prada_thesis}\\
\hline 
Radius of optical nerve $R_1$ & 790 $\mu $m  & \cite{prada_thesis}\\
\hline
Tissue/Extravascular pressure $P_{ex,1}$  &  2.92 mmHg & \cite{prada_thesis}\\
\hline
CRA pressure $P_{ai,0}$ & 40 mmHg & \cite{prada_thesis}\\
\hline
PCA pressure $P_{ao,1}$ & 30 mmHg & \cite{prada_thesis}\\
\hline
CRV pressure $P_{v,0}$ & 20 mmHg & \cite{prada_thesis}\\
\hline
 osmotic pressure constant $\pi_{ex}$ & 8 mmHg &  \cite{physiology2009}\\
    \hline
  osmotic pressure constant $\pi_{j}$ in vessel & 28 mmHg &    \cite{physiology2009}\\
\hline
vascular volume fraction $\sum \eta_k^{re}$ & 15.67 \% & \cite{prada_thesis}\\
\hline
capillary, averaged $\eta_{co}^{re}+\eta_{ci}^{re}$ & 1.9 \% =1.27+ 0.63 \% & \cite{Tim2011}\\
\hline
volume fraction, average $\eta_v^{re}$ &  10 \% & \cite{prada_thesis,physiology2009}\\
\hline
volume fraction, average $\eta_{ao}^{re} + \eta_{ai}^{re}$ &   3.77 \%= 2.51+1.26 \%  & \cite{prada_thesis,physiology2009}\\
\hline
  area coefficient $M_{v,ex}^0$ & $0.021$ /$\mu m$ &  estimate\\
 \hline
   area coefficient  $M_{ao,ex}^0$ & $0.016$ /$\mu m$ &  estimate\\
 \hline
   area coefficient  $M_{ai,ex}^0$ & $0.011$ /$\mu m$ &  estimate\\
 \hline
   area coefficient  $M_{co,ex}^0$ & $0.075$ /$\mu m$ &  estimate\\
  \hline
   area coefficient  $M_{ci,ex}^0$ & $0.053$ /$\mu m$ &  estimate\\
   \hline
 permeability coefficient $\beta_{v}$ & 9000 $(\mu m)^2$ &  estimate\\
    \hline
 permeability coefficient $\beta_{ao}$ & 16000 $(\mu m)^2$ &  estimate\\
     \hline
 permeability coefficient $\beta_{ai}$ & 31500 $(\mu m)^2$ &  estimate\\
      \hline
 permeability $\kappa_{ex}$ & $4\times 10^{-4} (\mu m)^2$ &  \cite{zhu2021}\\
    \hline
% \end{tabular}
%\end{center}
%\end{table}%
%
%
%\begin{table}
%\caption{\label{table9}  The parameters in blood circulation}
%\begin{center}
%\begin{tabular}{|c|c|c|}
%   \hline
tortuosity $\tau_{ex}$ & 0.9 &  estimate\\
\hline
tortuosity $\tau_{ai},\tau_{ao},\tau_{v}$ & 0.5&  estimate\\
\hline
leak coefficient $L_{ao,ex},L_{ai,ex},L_{v,ex}$ & $1\times 10^{-6}$ $\mu m$/(Pa s) &  \cite{zhu2021}\\
\hline
water leak coefficient  $L_{ci,ex},L_{co,ex}$ & $2.54\times 10^{-4}$ $\mu m$/(Pa s) & \cite{antcliff2001}\\
\hline
permeability coefficient $\delta_{ai}$ & $0.39$ /(Pa s)  & estimate\\
\hline
permeability coefficient $\delta_{ao}$ & $0.2$ /(Pa s)  & estimate\\
\hline
permeability coefficient $\delta_{v}$ & $0.11$ /(Pa s)  & estimate\\
\hline
permeability coefficient $\delta_{ci}, \delta_{co}$ & $3.25, 1.61$ /(Pa s)  & \cite{Tim2011}\\
\hline
viscosity $\mu_b$, $\mu_{ex}$ & 0.011  Pa s  & \cite{prada_thesis,causin2016,roux2020}\\
\hline
modulus $\lambda_{j}, j=ai,ao,ci,co$ &$7.8 \times 10^5$ Pa & \cite{prada_thesis} \\
\hline
modulus $\lambda_{v}$ &$1.5 \times 10^5$ Pa & estimate \\
\hline
resting pressure $P_{ex}^{re}$ & $2.92$ mmHg & estimate \\
\hline
resting pressure $P_{ai}^{re},P_{ao}^{re}$ & $25$ mmHg & estimate \\
\hline
resting pressure $P_{v}^{re},P_{ci}^{re},P_{co}^{re}$ & $20$ mmHg & estimate \\
\hline
\end{tabular}
\end{center}
\end{table}%

%{\bf Units conversion: }
%
%1 mm Hg = 133.3 Pa; 
%
%1dyn = $10^{-5}$ N= 10 $\mu$ N
%
%1Pa s = 1 Kg/m/s = 10 g/cm/s

% diffusion of oxygen: $3*10^{-5} cm^2/s$?

% Filtration coefficient:  0.01 ml/min/mmHg/100g
%\hline
%velocity in veins (averaged) & 19.3 mm/s & Wang  et al 2009\\
%\hline
%velocity in capillary (averaged) & 0.88 mm/s & Tim David 2011\\
%\hline
%flow rate in capillary (averaged) & 1.77 nl/min & Tim David 2011\\

{\bf Estimate of $\beta_j$ and $M_{j,ex}^0$}

First, we estimate the values of $\beta_k$ ($k=ai,ao,v$) in (\ref{eq3}), and we omit the subscript $k$ in some derivation below. In Poiseuille's flow through a cylindrical blood vessel of radius $r_{bv}$, the permeability $\kappa$ (the $\kappa_j$ in the formula of (\ref{eq2})) is derived as \cite{prada_thesis,physiology2009}
$$
\kappa= \frac{1}{8} r_{bv}^2.
$$
The volume fraction $\eta$ is given by \cite{prada_thesis}
$$
\eta = \pi r_{bv}^2 N_{bv} L/V,
$$
where $V$ is a control volume,  $L$ and $N_{bv}$ are the length and number of parallel blood vessels in the control volume. So, we can write them as 
$$
r_{bv}^2 = \beta \eta, \quad \beta=\frac{V}{\pi N_{bv} L}, \quad \kappa = \frac{1}{8} \beta \eta,
$$
where the last one is the formula (\ref{eq3}) used in the maintext, and the first formula will be used to estimate the coefficient $\beta$. 
Here we assume constant $\beta$ for simplicity (of course, $\beta$ could be varying as it characterizes the structural information about distribution of branches) and estimate it by choosing an average radius of blood vessel for each domain. For example, we choose average $r_{bv} = 30 \mu m$ for vein domain and get 
$$
\beta_v \approx \frac{r_{bv}^2}{\eta_v^{re}} =\frac{(30 \mu m)^2}{0.1} = 9000 (\mu m)^2. 
$$
Similarly we choose $r_{bv} = 20 \mu m$ for two artery domains and get $\beta_{ai},\beta_{ao}$ respectively as in Table \ref{table8}.  We also choose $r_{bv} = 3 \mu m$ for capillary domains to get $\beta_{ci},\beta_{co}$ (although not directly used in maintext), which will be used to estimate $M_{ci,ex}^0,\delta_{ci}, M_{co,ex}^0,\delta_{co}$.

Next, we estimate $M_{j,ex}^0$ defined in (\ref{eq6}), and omit $j,ex$ in the general formula below. By definition, the area of blood vessel wall per unit control volume is
$$
M= 2\pi r_{bv} N_{bv} L/V = 2 \sqrt{1/\beta} \sqrt{\eta} = M^0  \sqrt{\eta} \quad \Rightarrow \quad M^0 = 2 \sqrt{1/\beta},
$$
where the definitions of $\eta,\beta$ are used in the second equality. Then, for each domain, we have the estimate
$$
 M_{j,ex}^0 = 2 \sqrt{1/\beta_j},\quad j=ai,ao,v,ci,co.
$$

{\bf Estimate of $\delta_j$}

The estimate of $\delta_j$ is also based on $\beta_j$. For example, $\delta_{ai}$ is estimated as
$$
\delta_{ai} = \frac{1}{8} \frac{\beta_{ai} {\tau}_{ai} }{\mu_b} \frac{1}{(\Delta r)^2} \approx 0.39 /(\mathrm{Pa}\, \mathrm{s})
$$
where $\Delta r= R_1-R_0 =677  \mu m$ is used. Similarly the same $\Delta r$ is used for estimates of $\delta_{ao},\delta_v$. For capillary domains, we have used $\Delta r=50 \mu m$ to estimate $\delta_{ci},\delta_{co}$, because capillary network is local and connects the artery and vein domains. The estimated $\delta_{ci},\delta_{co}$ have similar values as those calculated from \cite{Tim2011}.

% The diameter ratio of daughter to parent is about $d_1/d_0=0.8$, the area ratio is about $1.2$ from data, consistent with $0.8^2 *2 = 1.28$, and the Marray formula $d_0^3 = d_1^3 + d_2^3$ is also consistent $0.8^3 *2 =1$

{\bf Other estimates}

The total resting volume fractions for two arteries are taken from \cite{prada_thesis,physiology2009}. We split it as two parts, 2/3 for the domain $\Omega_{ao}$ and 1/3 for the domain $\Omega_{ai}$, because it is believed that the artery from PCA is the primary component \cite{prada_thesis,onda1995}. Similarly, we adopted the split  the total capillary resting volume fraction (from \cite{Tim2011}) as two parts, 2/3 and 1/3 respectively for the two capillary domains $\Omega_{co},\Omega_{ci}$. The modulus $\lambda_j$ for arteries are taken from \cite{prada_thesis} (see also \cite{nagaoka2006,camasao2021}),  and the $\lambda_v$ is chosen to be 5 times smaller than that, because the compliance (related to the inverse of $\lambda_j \eta_j$ here) is about 24 times larger in veins \cite{physiology2009}. The tortuosity in blood vessels are chosen as 0.5 because this is related to the relative orientation/angle of blood vessels to the radial direction, while it is set as 0.9 for extracellular space since it is almost connected in every direction.

\subsection{Parameters in oxygen transport}

Since we directly used the concentration of oxygen instead of partial pressure of oxygen in our model, the values from the following references will be converted by multiplying oxygen solubility coefficient $\alpha_{O2}$. After conversion, the max consumption rate is $6*10^{-4}$ ml O2/ml/s in \cite{causin2016} for retina, and some consumption rate in the range $ [1,42]*10^{-4}$ ml O2/ml/s  is used in  \cite{arciero2008}, the value $ 23*10^{-4}$ ml O2/ml/s is used in \cite{secomb2004} for the brain. Based on some simulation, in order to be compatible with estimates of blood flow velocity and normal oxygen concentrations (about 40 mmHg partial pressure multiplied by $\alpha_{O2}$) in vein, we choose a relatively larger one $6*10^{-3}$ ml O2/ml/s for the optical nerve. The values of $C_{1/2}, C_{50}$ etc are taken from  \cite{causin2016} after conversion.

The permeability of oxygen in capillary is estimated by the formula \cite{causin2016}
$$l_{ci,ex}=l_{co,ex} = \frac{D_w}{t_{cap}} =\frac{1*10^{-9} \textrm{m}^2/\textrm{s}}{0.5 \mu \textrm{m}}= 0.002 \textrm{m/s}$$ 
where $D_w $ is the diffusivity and $t_{cap}$ is the thickness for capillary vessel wall. The values are similar to those in \cite{goldman2008}. For artery and vein, we used the same formula, but with a much smaller diffusion constant and a larger vessel wall thickness \cite{physiology2008} %[page 465]
$$l_{ai,ex}=l_{ao,ex}= l_{v,ex} = \frac{D_w}{t_{ai}}= \frac{1*10^{-10} \textrm{m}^2/\textrm{s}}{2 \mu \textrm{m}}= 5*10^{-5} \textrm{m/s}.$$ 
For the estimates, we also referred to \cite{secomb2004}.

For the effective diffusion constants between vascular domains, we take $D_{ai,ci}$ for example, which is estimated by harmonic average
$$
D_{ai,ci}= \frac{D_{ai}^{*} D_{ci}^{*} }{D_{ai}^{*}  + D_{ci}^{*} }\approx 3.6*10^{-5} /s
$$
where
$$D_{ai}^{*} = \frac{\eta_{ai} D_{ai} \tilde{\tau}_{ai}}{(\Delta r)^2} \approx 3.6*10^{-5} , \quad D_{ci}^{*} = \frac{\eta_{ci} D_{ci} \tilde{\tau}_{ci}}{(\Delta \tilde{r})^2} \approx 3.3*10^{-3}
$$
where $\Delta r = 667 \mu$m and $\Delta \tilde{r} = 50 \mu$m are used for artery and capillary. The other three $D_{ao,co},D_{v,ci},D_{v,co}$ are estimated similarly.

\begin{table}
\caption{Parameters in oxygen transport}
\begin{center}
\begin{tabular}{|c|c|c|}
\hline
max consumption rate $S_{ex}^{max}$ &$6 \times 10^{-3}$ ml O2/ml/s & estimate\\
\hline
half-max parameter $C_{1/2}$ & $4.8 \times 10^{-5}$ ml O2/ml &  \cite{causin2016}\\
\hline
solubility coefficient in blood $\alpha_{O2}$  & $3 \times 10^{-5}$ ml O2/ml /mmHg & \cite{causin2016} \\
\hline
boundary concentration $C_{ai,0},C_{ao,1}$  & $3 \times 10^{-3}$ ml O2/ml & \cite{causin2016} \\
\hline
average concentration of $[Hb]$ in blood  & $0.15$ g/ ml  & \cite{pittman2016,popel1989} \\
\hline
binding-capability of Hb, $C_{Hb}$  & $1.34$ ml O2/g  &  \cite{pittman2016} \\
\hline
average binding-capability of blood, $H_0$  & $0.2$ ml O2/ml  & \cite{pittman2016,arciero2008} \\
\hline
Hill exponent $n_{Hill}$  & 2.7 & \cite{causin2016} \\
\hline
Half saturation constant in Hb, $C_{50}$  &  $8 \times 10^{-4}$ ml O2/ml & \cite{causin2016} \\
\hline
diffusion constant $D_{k},k= ai, ao, v, ci, co$  &  $2.18 \times 10^{-9}$ m$^2$/s & \cite{causin2016} \\
\hline
diffusion constant  $D_{ex}$  &  $1 \times 10^{-9}$ m$^2$/s & \cite{causin2016} \\
\hline
oxygen permeability, $l_{ci,ex},l_{co,ex}$  &  $0.002 $ m/s & \cite{causin2016} \\
\hline
oxygen permeability, $l_{ai,ex},l_{ao,ex}, l_{v,ex}$  &  $5\times 10^{-5} $ m/s & \cite{causin2016} \\
\hline
diffusion constant $D_{ai,ci},D_{ao,co}$  &  $3.6\times 10^{-5} $ /s,$7.3 \times 10^{-5} $ /s & estimate \\
\hline
diffusion constant $D_{v,ci},D_{v,co}$  &  $2.8 \times 10^{-4} $ /s & estimate \\
\hline
\end{tabular}
\end{center}
\label{default}
\end{table}%

\section*{Acknowledgments}
This work was partially supported by the 
National Natural Science Foundation of China (no. 12231004, 12071190)  and Natural Sciences and Engineering Research Council of Canada (NSERC).


\section*{References}
\bibliographystyle{elsarticle-num.bst} 
\bibliography{reference.bib}


%\begin{thebibliography}{00}
%%% Text of bibliographic item
%
%\bibitem{}
%
%\end{thebibliography}
\end{document}